\pdfoutput=1

\documentclass[12pt,twoside]{mitthesis}
\usepackage{lgrind}
\usepackage{cmap}
\usepackage[T1]{fontenc}
\usepackage{nicematrix}
\usepackage{epigraph}
\usepackage{mathtools}
\usepackage{graphicx}
\usepackage{dcolumn}
\usepackage{xcolor}
\usepackage{bm}
\usepackage[utf8]{inputenc} 
\usepackage{url}
\usepackage{amsmath}
\usepackage{amssymb}    
\usepackage{graphicx}
\usepackage{listings}
\usepackage{hyperref} 
\usepackage{nicematrix}

\renewcommand{\v}[1]{\ensuremath{\mathbf{#1}}} 
\newcommand{\gv}[1]{\ensuremath{\mbox{\boldmath$ #1 $}}} 
\newcommand{\grad}[1]{\nabla #1} 
\renewcommand{\div}[1]{\nabla \cdot #1} 
\newcommand{\pd}[2]{\frac{\partial #1}{\partial #2}} 
\newcommand{\uv}[1]{\ensuremath{\mathbf{\hat{#1}}}} 
 
\renewcommand{\d}[2]{\frac{d #1}{d #2}} 
\newcommand{\delpar}[1]{\nabla_{\parallel} #1} 
\newcommand{\gradperp}[1]{\grad{}_{\perp} {#1}} 

\newcommand{\curv}[1]{{C}_{\left({#1}\right)}}

\newcommand{\pderiv}[2]{
\frac{\partial #1}{\partial #2}
}
\newcommand{\pderivInline}[2]{
\partial #1/\partial #2
}

\newcommand{\cs}{c_s}

\newcommand{\bhatZ}{\v{b_0}}

\newcommand{\cur}{j_{\parallel}}

\newcommand{\vpi}{v_{\parallel i}}						
\newcommand{\vpe}{v_{\parallel e}}					
\newcommand{\gvort}{\omega}		

\newcommand{\momSrce}{S_{\mathcal{M}\parallel e}}							
\newcommand{\momSrci}{S_{\mathcal{M}\parallel i}}							
\newcommand{\nSrcN}{S_n}									
\newcommand{\enerSrceN}{S_{E,e}}							
\newcommand{\enerSrciN}{S_{E,i}}							

\newcommand{\LB}{L_B}

\newcommand{\alphad}{\alpha_d}

\newcommand{\epa}{\epsilon_R}
\newcommand{\epv}{\epsilon_v}
\newcommand{\epg}{\epsilon_G}
\newcommand{\epge}{\epsilon_{Ge}}

\newcommand{\kappai}{\kappa^i}
\newcommand{\kappae}{\kappa^e}
\newcommand{\BRef}{B_0}

\newcommand{\tRef}{t_0}

\newcommand{\nRef}{n_0}

\newcommand{\TeRef}{T_{e0}}
\newcommand{\TiRef}{T_{i0}}

\newenvironment{eqnal}{\equation\aligned}{\endaligned\endequation}

\newcommand{\fortranl}[1]{\lstinline[language={[90]Fortran}]{#1}}

\let\originalepigraph\epigraph 
\renewcommand\epigraph[2]{\originalepigraph{\textit{#1}}{\textsc{#2}}}

\def\all{all}
\ifx\files\all  \else 

\begin{document}

\title{Physics-informed machine learning techniques for edge plasma turbulence modelling in computational theory and experiment}

\author{Abhilash Mathews}
\prevdegrees{B.Sc. Physics, Western University (2017)}
\department{Department of Nuclear Science and Engineering}

\degree{DOCTOR OF PHILOSOPHY IN APPLIED PLASMA PHYSICS}

\degreemonth{May}
\degreeyear{2022}
\thesisdate{May 11, 2022}


\supervisor{Jerry W. Hughes, Ph.D.}{Principal Research Scientist, MIT Plasma Science and Fusion Center} 

\reader{Anne E. White, Ph.D.}{Professor and Head of Nuclear Science and Engineering}
\chairman{Ju Li, Ph.D.}{Battelle Energy Alliance Professor of Nuclear Science and Engineering \\Chairman, Department Committee on Graduate Theses}

\maketitle



\cleardoublepage
\setcounter{savepage}{\thepage}
\begin{abstractpage}
%
%
%
Edge plasma turbulence is critical to the performance and operation of magnetic confinement fusion devices. Drift-reduced Braginskii two-fluid theory has for decades been widely applied to model boundary plasmas with varying success. Towards better understanding edge turbulence in both theory and experiment, a custom-built physics-informed deep learning framework constrained by partial differential equations is developed to accurately learn turbulent fields consistent with the two-fluid theory from partial observations of electron pressure. This calculation is not otherwise possible using conventional equilibrium models. With this technique, the first direct quantitative comparisons of turbulent field fluctuations between electrostatic two-fluid theory and electromagnetic gyrokinetic modelling are demonstrated with good overall agreement found in magnetized helical plasmas at low normalized pressure.

To translate these computational techniques to experimental fusion plasmas, comprehensive 2-dimensional diagnostics operating on turbulent time scales are necessary. For this purpose, a novel method to translate brightness measurements of HeI line radiation into local plasma fluctuations is demonstrated via a newly created deep learning framework that integrates neutral transport physics and collisional radiative theory for the $3^3 D - 2^3 P$ transition in atomic helium. Using fast camera data on the Alcator C-Mod tokamak, this thesis presents the first 2-dimensional time-dependent experimental measurements of the turbulent electron density, electron temperature, and neutral density in a fusion plasma using a single spectral line. With this experimentally inferred data, initial estimates of the 2-dimensional turbulent electric field consistent with drift-reduced Braginskii theory under the framework of an axisymmetric fusion plasma with purely toroidal field are calculated. The inclusion of atomic helium effects on particle and energy sources are found to strengthen correlations between the electric field and electron pressure while broadening turbulent field fluctuation amplitudes which impact ${\bf E \times B}$ flows and shearing rates.
\end{abstractpage}


\cleardoublepage

\section*{Acknowledgments}

\setlength{\epigraphwidth}{0.8\textwidth}
\epigraph{I suppose in the end, the whole of life becomes an act of letting go, but what always hurts the most is not taking a moment to say goodbye.}{Yann Martel, Life of Pi}


It is only as I stitch together the final pieces of my PhD that I am realizing it is nearly over. A trek that once almost felt endless just a short while ago is now just about finished. And at this end, there are countless people to thank. 

To start, this thesis and myself owe immense gratitude to J.W. Hughes. He provided me an example every day of what it means to be a good scientist and an even better human. Freedom is not worth having if it does not include the freedom to make mistakes\footnote{Mahatma Gandhi}, and Jerry aptly provided ample freedom to find my own meaningful problems and make my own mistakes. There may be no greater gift as a student, and Jerry's wit---in life and experiment---made this journey a joy. I thank A.E. White for making this work all possible beginning on the first day I walked into the PSFC as a Cantabrigian by bringing Jerry into her office. While the project evolved over the years, Anne knew how to start it all. And ever since our first serendipitous encounter in Austin, D.R. Hatch was a source of constant encouragement when exploring unknown territory and truly beginning this research. I am also grateful to my thesis defence members D.G. Whyte and J.D. Hare for their time in crafting this dissertation. At every stage, from my NSE admission letter to navigating a pandemic abroad, B. Baker was incredibly supportive throughout the entirety of graduate school.


Upon first meeting Mana on campus, I didn't realize how instrumental he would be to this work as a mentor and a friend, but this document would not exist without him. I can also say essentially all the same about Jim, who truly made these last chapters of the thesis possible. And while he may not be a formal co-author listed on publications, I wholeheartedly thank Ted for always being there at the beginning of my PhD. I could not have asked for a better neighbour. I am also indebted to the contributions of all my collaborators: M. Francisquez taught me to conduct the two-fluid simulations presented in Chapter 2 using the \texttt{GDB} code run on MIT's Engaging cluster and co-developed with B. Zhu and B.N. Rogers; N. Mandell developed and ran the electromagnetic gyrokinetic simulations described in Chapter 3, which were performed on the Perseus cluster at Princeton University and the Cori cluster at NERSC, and based upon the \texttt{Gkeyll} framework led by A. Hakim and G.W. Hammett; B. LaBombard and D. Brunner operated the mirror Langmuir probe utilized in Chapter 4; A.Q. Kuang and M.A. Miller assisted with probe data analysis in Chapter 4; S.G. Baek ran DEGAS2 simulations to inform neutral modelling in Chapters 4 and 5; J.L. Terry and S.J. Zweben operated the GPI diagnostic in Chapters 4 and 5; M. Goto, D. Stotler, D. Reiter, and W. Zholobenko developed the HeI collisional radiative codes employed in Chapters 4 and 5. The content of these pages are enabled by their technical support and camaraderie. Any errors in this thesis are my own.


While this may be the end of my PhD at MIT, the sights of Park Street and Killian Court will forever feel warm. I thank all the wonderful people I ran into on Albany Street and across the world over the past years including Nick, Francesco, Pablo, Rachel, Beatrice, Thanh, Eli, Bodhi, Fernanda, Nima (Harvard's P283B taught me to truly view physics questions as constrained-optimization problems), Lt. Reynolds, Erica, Muni, Christian, Patricio, Yu-Jou, Libby, Cassidy, Alex, Lucio, Aaron, Sam, Evan, Anna (from Switzerland), Anna (from Austria), Manon, Eva, Josh, and the never-faraway Peter. Prior to ever setting foot in Cambridge, I am truly lucky to have met a lifelong teacher and advisor in Martin, who pushed me to this point starting from London. I also thank my best friends in Canada for always making me feel at home wherever I am---from Fenway Park with Katie and Sammy, to sailing into the Charles with Donna, to the great lake with George and Maher (or even adventuring on it in Niagara), to escaping the French secret police with Dylan, to Adam's backyards in Toronto and Montr\'eal and Killarney, to living in a van across Valhalla and sleeping in train stations by Berchtesgaden with Nicholas---no place is ever too distant. 

There are also many people without whom this thesis would literally not be possible such as the great folks at Jasper Health Services for helping me to still write, Mississauga Fire Department for allowing me to still breathe, and Procyon Wildlife for inspiring me years ago and still to this day. But beyond all, this PhD is the product of the unconditional love and support I constantly get from my big family---all the way from my dear grandparents to my boisterous cousins who are always full of life. At the core, my parents instilled the values of good work and education in me at an early age. Together with my older brother, they endlessly push me in everything that I do with their time and heart and extraordinary cooking. There can never be enough appreciation for all they've done and for what they mean to me.

And to my littlest brother, Kobe, thank you for teaching me to see the world.
\\
\noindent\rule{15.25cm}{0.4pt}
\\ \\
{\it 
\noindent As you set out for Ithaka\\
hope your road is a long one,\\
full of adventure, full of discovery.\\
Laistrygonians, Cyclops,\\
angry Poseidon---don’t be afraid of them:\\
you’ll never find things like that on your way\\
as long as you keep your thoughts raised high,\\
as long as a rare excitement\\
stirs your spirit and your body.\\
Laistrygonians, Cyclops,\\
wild Poseidon---you won’t encounter them\\
unless you bring them along inside your soul,\\
unless your soul sets them up in front of you.\\
\vspace{-0.7cm}
\\ \\
Hope your road is a long one.\\
May there be many summer mornings when,\\
with what pleasure, what joy,\\
you enter harbors you’re seeing for the first time;\\
may you stop at Phoenician trading stations\\
to buy fine things,\\
mother of pearl and coral, amber and ebony,\\
sensual perfume of every kind---\\
as many sensual perfumes as you can;\\
and may you visit many Egyptian cities\\
to learn and go on learning from their scholars.\\
\vspace{-0.7cm}
\\ \\
Keep Ithaka always in your mind.\\
Arriving there is what you’re destined for.\\
But don’t hurry the journey at all.\\
Better if it lasts for years,\\
so you’re old by the time you reach the island,\\
wealthy with all you’ve gained on the way,\\
not expecting Ithaka to make you rich.\\
\vspace{-0.7cm}
\\ \\
Ithaka gave you the marvelous journey.\\
Without her you wouldn't have set out.\\
She has nothing left to give you now.\\
\vspace{-0.7cm}
\\ \\
And if you find her poor, Ithaka won’t have fooled you.\\
Wise as you will have become, so full of experience,\\
you’ll have understood by then what these Ithakas mean.}\vspace{0.4cm}\\{--- C. P. Cavafy (translated by Edmund Keeley)}

\newpage

\section*{List of Publications}

Part of the content included in this thesis has already been published in peer-reviewed journals or is presently under referee review. Permission to reuse text and figures from these articles has been granted and are listed below:
\begin{itemize}
\item {\bf A. Mathews}, J.W. Hughes, J.L. Terry, S.G. Baek, “Deep electric field predictions by drift-reduced Braginskii theory with plasma-neutral interactions based upon experimental images of boundary turbulence” arXiv:2204.11689
(2022)

\item {\bf A. Mathews}, J.L. Terry, S.G. Baek, J.W. Hughes, A.Q. Kuang, B. LaBombard, M.A. Miller, D. Stotler, D. Reiter, W. Zholobenko, and M. Goto, “Deep modelling of plasma and neutral fluctuations from gas puff turbulence imaging” arXiv:2201.09988 (2022)

\item {\bf A. Mathews}, N. Mandell, M. Francisquez, J.W. Hughes, and A. Hakim, “Turbulent field fluctuations in gyrokinetic and fluid plasmas” Physics of Plasmas {\bf 28}, 112301 (2021) 

\item {\bf A. Mathews}, M. Francisquez, J.W. Hughes, D.R. Hatch, B. Zhu, and B.N. Rogers, “Uncovering turbulent plasma dynamics via deep learning from partial observations” Physical Review E {\bf 104}, 025205 (2021) 

\item {\bf A. Mathews} and J.W. Hughes, “Quantifying experimental edge plasma evolution via multidimensional adaptive Gaussian process regression” IEEE Transactions on Plasma Science {\bf 49}, 12 (2021) 
\end{itemize}

\noindent Funding support came from the Natural Sciences and Engineering Research Council of Canada (NSERC) through the doctoral postgraduate scholarship (PGS D), U.S. Department of Energy (DOE) Office of Science under the Fusion Energy Sciences program by contract DE-SC0014264, Joseph P. Kearney Fellowship, and Manson Benedict Fellowship from the MIT Department of Nuclear Science and Engineering.

\pagestyle{plain}
\tableofcontents
\newpage
\listoffigures
\newpage
\listoftables

\chapter{Introduction}
\setlength{\epigraphwidth}{0.6\textwidth}
\epigraph{It is a profound and necessary truth that the deep things in science are not found because they are useful: they are found because it was possible to find them.}{J. Robert Oppenheimer}

With the neutron discovered less than a century ago \cite{Chadwick_Nobel}, efforts to precisely understand the structure and potential of atomic nuclei have produced breakthroughs in human knowledge and technology. The beginning of World War II launched an era of covert and public research into unlocking the power of the atom, and it is still being widely explored today. One method found to generate power via the interaction of nuclei is through their fusion, where the difference in mass between reactants and products results in net energy. Humanity's advancements are linked with access to power \cite{Earth1,Earth2,Earth3}. As a potential source of abundant electricity across Earth and the depths of space, fusion energy is amongst society's greatest objectives. 

Due to their relatively large terrestrial abundance and/or cross-sections, the fusion reactions with hydrogen isotopes of primary interest in modern experiments include

\begin{equation}
^2_1\text{D} \ + \ ^2_1\text{D} \ \rightarrow \ ^3_2\text{He} \ (0.82 \text{ MeV}) \ + \ ^1_0\text{n} \ (2.45 \text{ MeV})
\end{equation}
\begin{equation}
^2_1\text{D} \ + \ ^2_1\text{D} \ \rightarrow \ ^3_1\text{T} \ (1.01 \text{ MeV}) \ + \ ^1_1\text{H} \ (3.02 \text{ MeV})
\end{equation}
\begin{equation}
^2_1\text{D} \ + \ ^3_1\text{T} \ \rightarrow \ ^4_2\text{He} \ (3.52 \text{ MeV}) \ + \ ^1_0\text{n} \ (14.08 \text{ MeV}),
\end{equation}

but the fusing of these atomic nuclei to yield energy requires the right conditions. Conditions generally reliant upon the hydrogen isotopes being highly energetic themselves since the strong nuclear force can only overcome electrostatic repulsion between nuclei when they are in sufficiently close spatial proximity on the order of femtometres. Such energetic conditions generally necessitate the interacting species exist in ionized states collectively known as plasmas. Past work has demonstrated that fusion reactor concepts with nonequilibrium plasmas are unfeasible (without methods to recirculate power at high efficiencies) \cite{rider_thesis}. Reactors with hot plasmas near thermodynamic equilibrium accordingly require good bulk confinement. 

If defining the circulating power quality factor as $Q = P_f/P_h$, where $P_h$ is the externally applied heating and $P_f$ is the net thermonuclear power, then a simple power balance calculation of an equilibrium plasma (without radiative losses nor electricity recovery from auxiliary sources) finds the following modified Lawson criterion:

\begin{equation}\label{eq:Lawson}
n \tau_E = \frac{12 T}{\langle \sigma v \rangle Y_c}\frac{1}{1 + 5/Q}
\end{equation}

\noindent where $\langle \sigma v \rangle$ is the fusion reaction rate coefficient, $Y_c$ corresponds to the energy yield of charged particles, $\tau_E$ represents the characteristic energy confinement time, and $n$ and $T$ are the volume-averaged plasma density and temperature, respectively, under the approximations of quasineutrality and equilibration between ions and electrons in the plasma. The condition of $Q = 1$ is known as energy breakeven, and fusion power plants ideally seek to operate with $Q \gg 1$ for economic feasibility. If considering $^2_1$D-$^3_1$T plasmas, where $\langle \sigma v \rangle \sim T^2$ in conditions of interest to maximize the equilibrium plasma's fusion reaction rate, then the triple product metric can be cast as

\begin{equation}\label{eq:triple_product}
n T \tau_E \sim \frac{1}{1 + 5/Q},
\end{equation}

\noindent which is directly related to $Q$. Namely, a fusion reactor's efficiency is strongly dependent upon the pressure and energy confinement time of the plasma \cite{twofluid_Lawson}. 

\section{Magnetic confinement fusion}

Decades of worldwide effort on developing fusion concepts have resulted in the creation of numerous experiments, with a magnetic confinement fusion design known as the ``tokamak'' demonstrating values of $Q$ nearing 1 \cite{wurzel2022progress}. Tokamaks confine electrically conductive plasmas in a toroidal chamber using magnetic fields, which aim to insulate the hot ionized gas from the solid walls of the machine. The charged particles continuously swirl along magnetic field lines with toroidal and poloidal components conventionally induced by external electromagnets and transformer coils, respectively, to optimize plasma confinement by accounting for drifts of the gyrating ions and electrons in this curvilinear geometry. But as new tokamaks are being built in attempts to exceed breakeven and demonstrate the viability of fusion energy technology, there are significant uncertainties associated with predictions of $n T \tau_E$ in existing and upcoming experiments from first principles. That is because tokamak plasma transport tends to not only be determined by classical collisional processes diffusing particles and energy across inhomogeneous magnetic fields. Rather, the behaviour of these experimental plasmas are commonly anomalous and strongly influenced by turbulence \cite{Wesson_tokamaks}, which is often termed ``the most important unsolved problem of classical physics.''\footnote{Richard P. Feynman}


To step over this difficulty, the fusion community has in large part relied upon empirical confinement scalings of $\tau_E$ derived from fitting power laws against 0-dimensional observational data and operational parameters across international fusion experiments \cite{scaling1,Physics__2007}. Cross-machine scalings have been similarly developed for estimating the structure of edge pressure profiles \cite{pedestal_scaling2,pedestal_scaling3,pedestal_scaling4} and heat flux widths \cite{Eich_2013,Brunner_2018} since boundary plasma turbulence is difficult to wholly simulate yet strongly influences global profiles and power handling. While these regression efforts are useful for building intuition on physical trends in experiment \cite{Buck-pal,Connor_1977,Petty_2008,Freidberg_elongation}, their ability to extrapolate to novel fusion plasma regimes is not clearly known. Recent progress in integrated modelling efforts have increased the fidelity of tokamak performance calculations \cite{SPARC_confinment}, i.e. to effectively estimate \{$n$, $T$, $\tau_E$\} and thereby $Q$ on future devices, but these core transport simulations still invoke questionable assumptions, particularly in the treatment of boundary conditions which crucially impact reactor performance, core fuelling, and safety of the vessel wall \cite{edge_core_association0,edge_core_association1}. 

Modelling turbulent edge profiles in magnetic confinement fusion devices is especially challenging due to the vast dynamical scales and nontrivial geometry and neutral interactions present whilst going from the hot confined region to the cooler boundary plasmas striking solid material surfaces just a few centimetres away. The sharp plasma gradients extant are potential sources of free energy associated with turbulent edge transport where fluctuation amplitudes can be on the same order as the background quantities. These dynamics can be described by kinetic theory where the evolution of the distribution function of any species $\alpha$, $f_\alpha$, is given by


\begin{equation}\label{eq:kinetic_theory}
\frac{d f_\alpha}{d t} = \frac{\partial f_\alpha}{\partial t} + {\bf \dot{x}}\frac{\partial f_\alpha}{\partial {\bf x}} + {\bf \dot{v}}\frac{\partial f_\alpha}{\partial {\bf v}} = C\{f_\alpha\} + S\{f_\alpha\},
\end{equation}

\noindent where the terms $C\{f_\alpha\}$ and $S\{f_\alpha\}$ symbolize collisions and sources, respectively, in this 6-dimensional phase space. Fully kinetic global simulations for whole device modelling are intractable, though, even with modern computing facilities. Model approximations are thus inevitable, but once approximations are introduced into edge turbulence models, their effects on nonlinear dynamics need to be quantitatively evaluated---against higher fidelity simulations and/or experiment, if possible---for model predictions to be meaningful. An oft-applied set of equations to study the plasma edge is drift-reduced Braginskii fluid theory, which has been studied for decades starting from Braginskii's original formulation \cite{Braginskii1965RvPP}, and remains a highly active research area today \cite{BOUT++,TOKAM3X,GBS,Stegmeir2018,GDB,Thrysoe_2020,giacomin2021gbs,GBS_3fluid_kineticneutrals}. But the underlying model approximations tend to only be marginally satisfied in fusion plasmas and there is a scarcity of clear-cut validation tests of the theory against higher-fidelity simulations or experimental measurements. Even with modern drift-reduced Braginskii codes, diagnostic comparisons tend to be qualitative with considerable uncertainties in quantitative accuracy \cite{2009val,2011val,NIELSEN2015,2016val,2016val_v2,Nielsen_2019,2020val,2021val,code_2022_validations}, although these efforts are improving \cite{Zholobenko_2021_validation}. This challenge in precisely ascertaining the reduced turbulence theory's predictive reliability is in part due to difficulties in aligning nonlinear calculations with numerous free parameters for chaotic systems \cite{chaos_review,chaos_turbulence,chaos_edge}. Unobserved dynamics related to sources and boundaries can be tough to diagnose on turbulent scales in the harsh conditions and geometry of the plasma edge. Quantitative methods to unambiguously test validity are thus needed if there is to be any certainty when using edge models to improve integrated simulations of fusion reactors.

\begin{figure}[ht]
\centering
\includegraphics[width=0.625\linewidth]{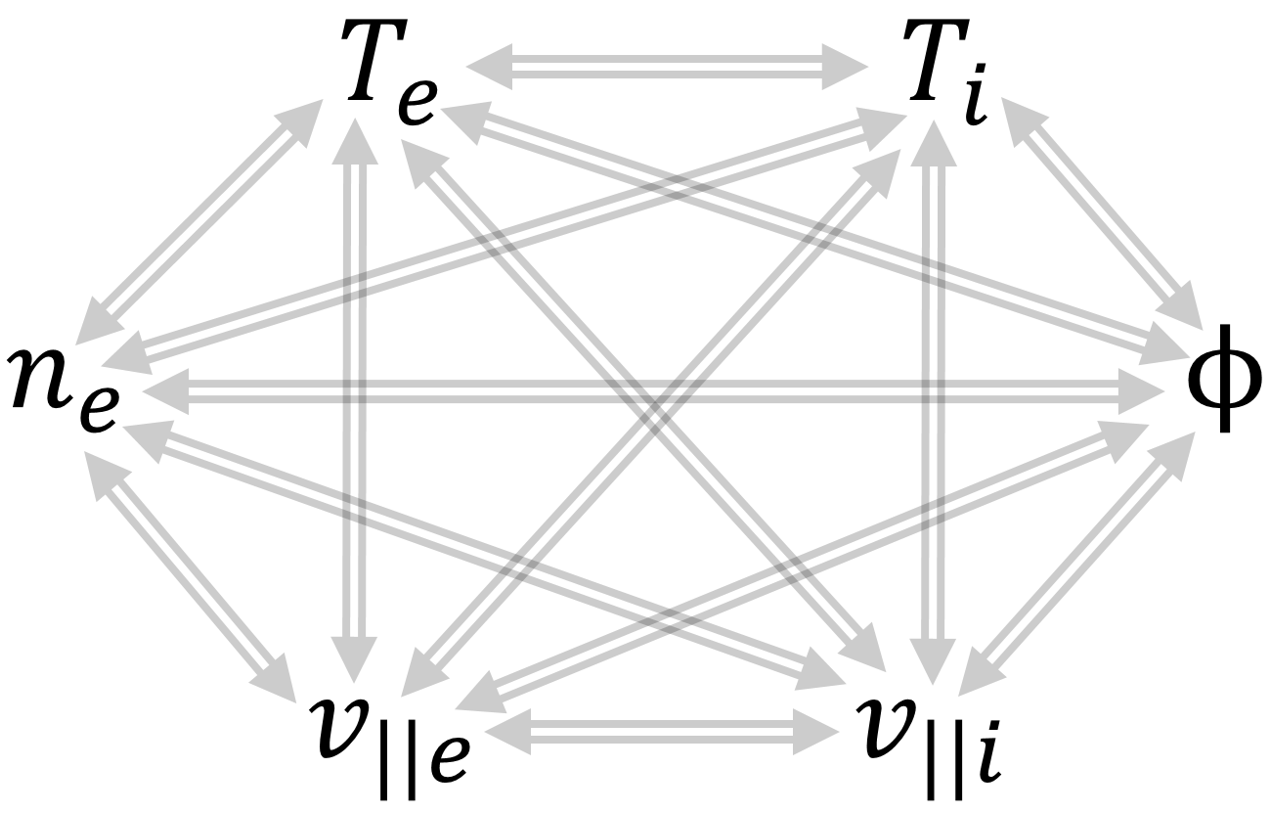}
\caption{\label{simple_network_graph}A visual representation of connections that exist and define multi-field turbulence models. The nonlinear relationship between $n_e$ and $T_e$ with $\phi$ will be examined in this thesis for plasma conditions relevant to nuclear fusion.}
\end{figure}

One hallmark of a turbulence theory is the nonlinear connection that it sets between dynamical variables as visualized in Figure \ref{simple_network_graph}. For ionized gases, perturbations in the plasma are intrinsically linked with perturbations of the electromagnetic field. Accordingly, the electric potential response concomitant with electron pressure fluctuations constitutes one of the key relationships demarcating a plasma turbulence theory and presents a significant consistency test en route to its full validation. The electric field perpendicular to magnetic field lines is a particularly important quantity as it is a strong drive of cross-field fluxes and thus edge pressure profiles on turbulent scales. These ${\bf E \times B}$ flows are observed to strongly influence edge plasma stability \cite{osti_1630104,osti_1762124} and the motion of coherent structures which can account for significant particle losses during standard operation of tokamaks \cite{blobs_review,Carralero_2017,Kube_2018}. Resultant turbulence-induced fluxes impacting walls can cause sputtering, erosion, and impurity injection which can further adversely affect safe operation and confinement of fusion plasmas \cite{blobs_review,kuang_SPARC,kuang_ARC}. Therefore, it is imperative that transport models for boundary plasmas are capable of accurately predicting the turbulent electric field response to electron pressure fluctuations. The aim of this thesis is to build a framework to examine this fundamental relationship and quantitative impacts from common approximations on turbulent fields as expected in drift-reduced Braginskii fluid theory. The machinery used for this objective are physics-informed neural networks (PINNs). 

\section{Deep learning of physics}

At its core, this thesis seeks to utilize machine learning---computation algorithms training to complete objectives by the use of experience and data---to model turbulent physical systems. It is instructive to translate the language of machine learning into physics for building familiarity and understanding why such efforts could potentially help advance turbulence modelling. To start, it is useful to consider a generic goal shared by physics and machine learning: predict physical variable(s) ${\bf y}$ based upon observation(s) of ${\bf x}$. Experimentally, one is interested in the distribution $p({\bf y} \lvert {\bf x})$ denoting the conditional probability of ${\bf y}$ given ${\bf x}$. Following the exposition in \cite{tegmark}, this statement can be expressed via Bayes’ theorem as 

\begin{equation}\label{eq:Bayes1}
    p({\bf y} \lvert {\bf x}) = p({\bf x} \lvert {\bf y})p({\bf y})/p({\bf x}) = p({\bf x} \lvert {\bf y})p({\bf y})/\sum\limits_{{\bf y}'} p({\bf x}' \lvert {\bf y}')p({\bf y}').
\end{equation}

The negative logarithm of these probabilities can be defined according to

\begin{equation}
 H_{\bf y}({\bf x}) \equiv -\ln p({\bf x} \lvert {\bf y}),
\end{equation}
\begin{equation}
 \mu_{\bf y} \equiv -\ln p({\bf y}),
\end{equation}

\noindent where $H_{\bf y}({\bf x})$ represents the Hamiltonian of ${\bf x}$ given ${\bf y}$ up to an arbitrary additive constant. Eq. \eqref{eq:Bayes1} can accordingly be recast in the usual Boltzmann form as

\begin{equation}
    p({\bf y} \lvert {\bf x}) = \frac{1}{N({\bf x})}e^{-[ H_{\bf y}({\bf x}) + \mu_{\bf y}]}
\end{equation}
with 
\begin{equation}
    N({\bf x}) = \sum_{{\bf y}'} e^{-[ H_{{\bf y}'}({\bf x}) + \mu_{{\bf y}'}]}.
\end{equation}



To understand how a neural network can map this vector-valued function given by Bayes' theorem, one recognizes that a simple $n$-layer feedforward network is just a series of linear and nonlinear transformations in succession equal to

\begin{equation}{\label{simple_network_eqn}}
    {\bf f(x)} = {\bf \sigma}_n {\bf A}_n ... {\bf \sigma}_2 {\bf A}_2 {\bf \sigma}_1 {\bf A}_1 {\bf x},
\end{equation}

\noindent where ${\bf \sigma}_i$ is a nonlinear operator, and  ${\bf A}_i$ are affine transformations of the form ${\bf A}_i {\bf x} = {\bf W}_i {\bf x} + {\bf b}_i$ with weight matrices ${\bf W}_i$ and bias vectors ${\bf b}_i$ that need to be tuned. The more layers present, the ``deeper'' the network. These weights and biases are analogous to the set of coefficients optimized in ordinary polynomial regression, but the functional form of the solution is not necessarily fixed {\it a priori}. This is motivated by multilayer feedforward neural networks being universal function approximators in theory capable of approximating any measurable function to arbitrary degrees of accuracy \cite{cybenko_approximation_1989,HORNIK}. These neurons (universal analog computing modules) constituting graph networks are akin to NAND gates (universal digital computing modules) in logic circuits: any computable function can be accurately described by a sufficiently large network of them. And just as NAND gates are not unique (e.g. NOR gates are also universal), nor is any particular activation function \cite{tegmark}. Regularly chosen nonlinear operators for neurons include local functions such as the hyperbolic tangent,

\begin{equation}
\text{tanh}(x) = (e^x - e^{-x})/(e^x + e^{-x}),
\end{equation}

\noindent or collective operations such as {\it max-pooling} (i.e. maximum of all vector elements). One function known as {\it softmax} exponentiates and normalizes all terms according to

\begin{equation}
    \sigma(x_i) \equiv e^{x_i}/\sum\limits_{i} e^{x_i}.
\end{equation}

For any Hamiltonian approximated by an $n$-layer feedforward neural network, one can therefore fully represent the sought physical variable's distribution by simply applying the computational graph's outputs through a softmax layer,

\begin{equation}
   {\bf  p({\bf y} \lvert x)} = \sigma[-{\bf H_y (x)} - {\bf \mu_y}].
\end{equation}

But this expression for the sought probability distribution is an accurate representation only if the graphs are adequately trained to learn the original Hamiltonian function. Whether the networks are practically capable of learning such dynamics, e.g. ${\bf H_y (x)}$, is not always clear. In practice, there are numerous deeply interconnected influences jointly acting on networks during learning including construction of the overall graph architecture, initialization of ${\bf W}_i$ and ${\bf b}_i$, complexity of the objective function (e.g. Hamiltonian), selection of nonlinear activation in neurons, potential lack of a deterministic relationship between the given inputs and targets, and the optimization procedure employed for training ${\bf W}_i$ and ${\bf b}_i$ \cite{GlorotAISTATS2010,tegmark,wang2020understanding,HORNIK}. While advancements in all categories are essential in the generalized learning of complex systems with artificial intelligence, for the task of representing edge plasma turbulence, physics-informed constraints (e.g. partial differential equations, algebraic expressions, correlations) are embedded into the training of networks to uncover unobserved dynamics ($\bf y$) given limited information  ($\bf x$). The ability to exactly differentiate graphs significantly enables physically useful and computationally efficient mathematical constructions that can serve as regularization mechanisms \cite{raissi2017physics}. If a network is trying to represent the Hamiltonian, for example, then its self-differentiation with respect to phase space coordinates would yield the classical equations of motion, which can act to physically inform and numerically guide ${\bf W}_i$ and ${\bf b}_i$ to help us uncover what we can. This ethos will be at the heart of this analysis of fusion plasmas in both simulation and experiment with networks representing turbulent fields in these systems.

\section{Outline of chapters}
The primary results composing this thesis include the development of a novel physics-informed deep learning technique in Chapter 2 to uncover the turbulent electric field consistent with drift-reduced Braginskii theory from just 2-dimensional electron pressure measurements aligned with the magnetic field. In Chapter 3, this computational technique enables the first direct comparisons of instantaneous turbulent fields between two distinct full-$f$ global plasma turbulence models: electrostatic drift-reduced Braginskii theory and electromagnetic long-wavelength gyrokinetics. Good agreement is found between the two models in magnetized helical plasmas at low normalized pressure while being quantitatively inconsistent at high-$\beta$. To transfer the above turbulent field analysis framework to boundary plasmas in experiment, Chapter 4 develops and demonstrates an entirely new optimization technique to translate brightness measurements of HeI line radiation into local plasma and neutral fluctuations via a novel integrated framework of networks that combines neutral transport physics and collisional radiative theory for the $3^3 D - 2^3 P$ transition in atomic helium. This analysis for ionized gases is transferable to both magnetized and unmagnetized environments with arbitrary geometries and extends the gas puff imaging approach in fusion plasmas. Based upon fast camera data on the Alcator C-Mod tokamak, the first 2-dimensional time-dependent experimental measurements of the turbulent electron density, electron temperature, and neutral density are presented in a fusion plasma using a single spectral line. These experimental measurements are then utilized in Chapter 5 to estimate the 2-dimensional turbulent electric field consistent with drift-reduced Braginskii theory under the framework of an axisymmetric fusion plasma with purely toroidal field. To further examine the effects of approximations in neutral dynamics on turbulent field predictions, calculations are performed with and without atomic helium effects on particle and energy sources within the reduced plasma turbulence model. The inclusion of neutral helium (which is locally puffed in the experiment) is found to strengthen correlations between the electric field and plasma pressure. The neutral dynamics are also associated with an observed broadening of turbulent field fluctuation amplitudes, underlining their quantitative importance on turbulent ${\bf E \times B}$ flows and edge shearing of coherent plasma structures.

\chapter{Uncovering turbulent fields from electron pressure observations}

\setlength{\epigraphwidth}{1.0\textwidth}
\epigraph{There is a physical problem that is common to many fields, that is very old, and that has not been solved. It is not the problem of finding new fundamental particles, but something left over from a long time ago---over a hundred years. Nobody in physics has really been able to analyze it mathematically satisfactorily in spite of its importance to the sister sciences. It is the analysis of circulating or turbulent fluids.}{Richard P. Feynman}


\setlength{\epigraphwidth}{0.5\textwidth}
\epigraph{Shall I refuse my dinner because I do not fully understand the process of digestion?}{Oliver Heaviside}


The boundary region is critical in determining a fusion reactor's overall viability since edge plasma conditions strongly influence a myriad of reactor operations \cite{Federici_2001,Li2020,Chang2021}. Validating edge turbulence models is accordingly a crucially important endeavour since gathering sufficient information to effectively test reduced turbulent transport models is vital towards developing predictive capability for future devices. These machines will access novel burning plasma regimes and operate with some of the largest temperature gradients in the universe, but existing models may be inaccurate and standard diagnostics incapable of surviving such harsh thermonuclear environments \cite{biel_diagnostics_2019}. Yet edge modelling continues to need improvement---comprehensive gyrokinetic codes suitable for the boundary of tokamaks are still under development and fluid simulations commonly lack essential physics necessary to study collisionless systems. On this point, this chapter introduces the transport model known as drift-reduced Braginskii theory \cite{Braginskii1965RvPP,Kim-1993,francisquez_thesis} which is relevant to boundary plasmas and widely applied to analyze edge turbulence. Various adaptations of these equations have been recently taken to investigate several important edge phenomena including pedestal physics \cite{Francisquez_2017}, blob dynamics \cite{Ricci_2012}, neutral effects \cite{Thrysoe_2018}, and heat fluxes impinging plasma-facing components \cite{nespoli_non-linear_2017}. While experimental trends are at times reproduced in these works \cite{Zholobenko_2021_validation}, direct quantitative agreement between the two-fluid turbulence theory and observations is generally lacking on a wide scale due to difficulty in aligning global simulations with plasma experiments where relevant measurements may be sparse or missing altogether. Fusion plasma diagnostic measurements are inherently noisy and limited in their spatiotemporal scope (e.g. 1- or 2-dimensional profiles of electron density and temperature \cite{griener_continuous_2020,Bowman_2020,Mathews2020,mathews2022deep}) and resultantly require suitable analysis techniques. To this end, Chapter 2 demonstrates a physics-informed deep learning framework to diagnose unknown turbulent field fluctuations consistent with drift-reduced Braginskii theory from limited electron pressure observations. Namely, the drift-reduced Braginskii model is represented via PINNs \cite{Lagaris_1998,raissi2017physics,SIRIGNANO20181339}---highly expressive function approximators trained to solve supervised learning tasks while respecting nonlinear partial differential equations---to infer unobserved field dynamics from partial measurements of a synthetic plasma. As will be illustrated through a readily adaptable multi-network machine learning framework, this paradigm is transferable to the broad study of quasineutral plasmas in magnetized collisional environments and presents novel pathways for the AI-assisted interpretation of plasma diagnostics. In ways previously inaccessible with classical analytic methods, this approach has the ability to improve the direct testing of reduced turbulence models in both experiment and simulation to inform physicists of the equations necessary to model the edge. The overall computational technique introduces a systematic pathway towards quantitatively testing plasma turbulence theories, and is to date among the most complex systems applied in physics-informed deep learning codes. To demonstrate this framework, Chapter 2 proceeds with a description of a synthetic plasma modelled computationally via drift-reduced Braginskii theory in Section \ref{sec:level2.1}, outlines a machine learning architecture for multi-field turbulence analysis in Section \ref{sec:level2.2}, presents results in the robust learning of unknown turbulent fields in Section \ref{sec:level2.3}, and concludes with a summary in Section \ref{sec:level2.4}.

\section{\label{sec:level2.1}Drift-reduced Braginskii modelling}

\begin{figure}[ht]
\centering
\includegraphics[width=0.48\linewidth]{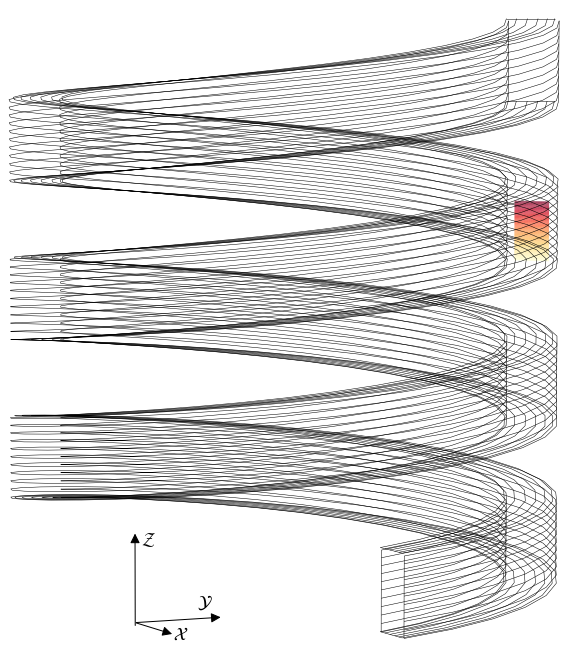}
\caption{\label{chapter2_geometry}A visualization of the geometry employed in \texttt{GDB} where the black contours represent the helical shaping of magnetic field lines with constant pitch angle. The shaded region indicates an example of a 2-dimensional field-aligned rectangular cross-section analyzed in the larger 3-dimensional domain of the full plasma.}
\end{figure}

The synthetic plasma analyzed is numerically simulated by the global drift-ballooning (\texttt{GDB}) finite difference code \cite{GDB, francisquez2020fluid} which solves the two-fluid drift-reduced Braginskii equations in the electrostatic limit relevant to low-$\beta$ conditions. This is a full-$f$ \cite{Belli_2008,Full-F,Held_2020,deltaf_DRB} fluid model in the sense that the evolution of the equilibrium and fluctuating components of the solution are not separated and relative perturbation amplitudes can be of order unity as found in experiments \cite{Garcia_SOL_fluctuations}. A 3-dimensional simulation domain is implemented with a shearless field-aligned coordinate system where ${\bf \hat{x}}$ is the unit vector along the radial direction (i.e. ${\bf{\hat{R}}}$), the helical magnetic field is oriented along ${\bf \hat{z}}$, and ${\bf \hat{y}}$ is perpendicular to both ${\bf \hat{x}}$ and ${\bf \hat{z}}$. Figure \ref{chapter2_geometry} displays a visualization of this geometry in Cartesian coordinates $(\mathcal{X},\mathcal{Y},\mathcal{Z})$ applied in \texttt{GDB}. 

The two-fluid theory encoded in the numerical simulation assumes the plasma is magnetized ($\Omega_i \gg \frac{\partial}{\partial t}$), collisional ($\nu_{ei} \gg \frac{\partial}{\partial t}$), and quasineutral ($\nabla \cdot \v{j} \approx 0$) with the reduced perpendicular fluid velocity given by ${\bf E \times B}$, diamagnetic, and ion polarization drifts. After neglecting collisional drifts and terms of order $m_e/m_i$, one arrives at the following equations (in Gaussian units) governing the plasma's density ($n \approx n_e \approx n_i$), vorticity ($\gvort$), parallel electron velocity ($\vpe$), parallel ion velocity ($\vpi$), electron temperature ($T_e$), and ion temperature ($T_i$) \cite{francisquez2020fluid,francisquez_thesis,Braginskii1965RvPP}

\begin{eqnal}\label{eq:nDotGDBH}
\d{^e n}{t} = -\frac{2c}{B}\left[n\curv{\phi}-\frac{1}{e}\curv{p_e}\right] -n\delpar{\vpe} +\nSrcN+\mathcal{D}_{n}
\end{eqnal}
\begin{eqnal}\label{eq:wDotGDBH}
\pd{\gvort}{t} &= \frac{2c}{eB}\left[\curv{p_e}+\curv{p_i}\right]-\frac{1}{em_i \Omega_i}\curv{G_i}+\frac{1}{e}\delpar{\cur}+\mathcal{D}_{\gvort}\\
&\quad-\div{\left\lbrace\frac{nc^2}{\Omega_i B^2}\left[\phi,\gradperp{\phi}+\frac{\gradperp{p_i}}{en}\right]+\frac{nc}{\Omega_i B}\vpi\delpar{\left(\gradperp{\phi}+\frac{\gradperp{ p_i}}{en}\right)}\right\rbrace}
\end{eqnal}
\begin{eqnal}\label{eq:vpeDotGDBH}
\d{^e\vpe}{t} &= \frac{1}{m_e}\left(e\delpar{\phi}-\frac{\delpar{p_e}}{n}-0.71\delpar{T_e} + e\eta_\parallel\cur \right) \\
&\quad + \frac{2}{3} \frac{\delpar{G_e}}{n} + \frac{2cT_e}{eB}\curv{\vpe}+\momSrce+\mathcal{D}_{\vpe}
\end{eqnal}
\begin{eqnal}\label{eq:vpiDotGDBH}
\d{^i\vpi}{t} &= \frac{1}{m_i}\left(-e\delpar{\phi}-\frac{\delpar{p_i}}{n}+0.71\delpar{T_e} - e\eta_\parallel\cur \right)\\
&+\frac{2 T_e}{3n}\frac{\delpar{G_i}}{n}-\frac{2cT_i}{eB}\curv{\vpi}+\momSrci+\mathcal{D}_{\vpi}
\end{eqnal}
\begin{eqnal}\label{eq:TeDotGDBH}
\d{^e T_e}{t} = \frac{2T_e}{3n}\left[\d{^e n}{t} + \frac{1}{T_e}\delpar \kappa^e_\parallel \delpar T_e + \frac{5n}{m_e \Omega_e} \curv{T_e} \right.\\ \left. + \eta_\parallel \frac{\cur^2}{T_e} + \frac{0.71}{e}(\delpar{\cur} - \frac{\cur}{T_e}\delpar{T_e}) + \frac{1}{T_e} \enerSrceN \right] + \mathcal{D}_{T_e}
\end{eqnal}
\begin{eqnal}\label{eq:TiDotGDBH}
\d{^i T_i}{t} &= \frac{2T_i}{3n}\left[\d{^i n}{t} + \frac{1}{T_i}\delpar \kappa^i_\parallel \delpar T_i  - \frac{5n}{m_i \Omega_i} \curv{T_i} + \frac{1}{T_i} \enerSrciN \right] + \mathcal{D}_{T_i}
\end{eqnal}

\noindent whereby the field-aligned electric current density is $\cur = en\left(\vpi - \vpe\right)$, the stress tensor's gyroviscous terms contain $G_s = \eta^s_0 \left\lbrace 2\delpar{v_{\parallel s}}+c\left[\curv{\phi} + \curv{p_s}/(q_s n)\right]\right\rbrace$, and $\eta^s_0$, $\Omega_s$, and $q_s$ are the species ($s = \{e,i\}$) viscosity, cyclotron frequency, and electric charge, respectively. The convective derivatives are $d^s f/dt = \partial_t f + (c/B)\left[\phi,f\right] + v_{\parallel s}\delpar{f}$ with $\left[F,G\right] = \bhatZ \times \nabla F \cdot \nabla G$ and $\bhatZ$ representing the unit vector parallel to the magnetic field. The field's magnitude, $B$, decreases over the major radius of the torus ($B\propto1/R$), and its curvature is $\gv{\kappa} = -{\bf{\hat{R}}}/R$. The curvature operator, $\curv{f} = \bhatZ \times \gv{\kappa} \cdot \grad{f}$, $\nabla_\parallel = -\partial / \partial z$, and $\bhatZ = -{\bf \hat{z}}$ follow past convention \cite{francisquez2020fluid}. The coefficients $\kappa^s_\parallel$
and $\eta^s_\parallel$ correspond to parallel thermal conductivity and electrical resistivity, respectively. Time-independent Gaussian-shaped density ($\nSrcN$) and energy sources ($S_{E,s}$) are placed at the left wall while zero external momentum ($S_{\mathcal{M}\parallel s}$) is explicitly forced upon the system. Explicit hyperdiffusion consisting of both fourth-order cross-field and second-order parallel diffusion is applied for numerical stability in the form of $\mathcal{D}_f = \chi_x \frac{\partial f} {\partial x^4} + \chi_y \frac{\partial f} {\partial y^4} + \chi_z \frac{\partial f} {\partial z^2}$. Under quasineutrality, electric fields arise not by local imbalance of charged particles but by the requirement that the electric current density is divergence free \cite{DRB_consistency3,Zholobenko_2021}. Accordingly, the electrostatic potential, $\phi$, is numerically solved for the synthetic plasma via the following boundary value problem: \begin{equation}\label{BVP_gvort_phi}
\div{ \frac{nc}{\Omega_i B}\left(\gradperp{\phi}+\frac{\gradperp{p_i}}{en}\right) } = \gvort.
\end{equation}

The synthetic plasma consists of deuterium ions and electrons with real masses (i.e. $m_i = 3.34 \times 10^{-27} \text{ kg}$ and $m_e = 9.11\times 10^{-31} \text{ kg}$) and on-axis magnetic field of $B_{axis} = 5.0 \text{ T}$ with minor and major radius of $a_0 = 0.22 \text{ m}$ and $R_0 = 0.68 \text{ m}$, respectively, consistent with characteristics of discharges in the Alcator C-Mod tokamak \cite{Hutch_CMod,Marmar_CMod,Alcator_Greenwald} for which there is evidence of fluid drift turbulence controlling edge profiles \cite{labombard_evidence_2005}. Moreover, drift-reduced models, where the ion gyration frequency is assumed to be faster than the plasma fluctuations (i.e. $\Omega_i \gg \frac{\partial}{\partial t}$), are generally good approximations to full velocity models when studying edge turbulence \cite{Leddy_full_velocity}. 

This discretized toroidal geometry is a flux-tube-like domain corresponding to Figure \ref{chapter2_geometry} on the outboard side (i.e. strictly bad curvature) of the tokamak scrape-off layer (SOL) with field lines of constant helicity wrapping around the torus and terminating on walls producing both resistive interchange and toroidal drift-wave turbulence. Transport is primarily along blobby field-aligned structures with increased pressure propagating due to perpendicular drifts which polarize the blob and yield outward ${\bf E \times B}$ drift of the filament. This is related to the Poynting vector representing the directional energy flux density of the electromagnetic field \cite{Thrysoe_2020,DRB_consistency3}. The physical dimensions of the entire simulation domain are $[L_x = 7.7\text{ cm}, L_y = 5.5 \text{ cm}, L_z = 1800.0 \text{ cm}]$ with spatiotemporal resolution of $[\Delta x = 0.03 \text{ cm}, \Delta y = 0.04 \text{ cm}, \Delta z = 56.25 \text{ cm}, \Delta t = 4.55 \times 10^{-11} \text{ s}]$. Periodic boundary conditions are employed in the binormal direction for all quantities. Homogeneous Neumann conditions (i.e. fixing the function's derivative normal to the boundary) are set on the radial surfaces for $n$, $\vpe$, $\vpi$, $T_e$, and $T_i$ while homogeneous Dirichlet conditions (i.e. fixing the function on the boundary) are used for $\gvort$ and $\phi$. By constraining $\phi = 0$ along the walls, this in principle enforces radial $\bf {E \times B}$ flows to go to zero on the simulation boundaries for the synthetic plasma being analyzed. The lower limit of the Bohm criterion, a necessary condition for the formation of a stationary Debye sheath \cite{Riemann_1991}---the transition from a plasma to a solid surface---is imposed as a parallel boundary condition,
\begin{equation}
\vpi(z=\pm \frac{L_z}{2}) = \mp c_{s} = -\sqrt{\frac{T_i + T_e}{m_i}},
\end{equation}
\begin{equation}
\vpe(z=\frac{\pm L_z}{2}) = \begin{cases} \mp c_{s}\exp(\Lambda - \frac{e\phi}{T_e}) &\mbox{if } \phi > 0 \\
\mp c_{s}\exp(\Lambda) &\mbox{if } \phi \leq 0 \end{cases},
\end{equation}
where $\Lambda = \log\sqrt{m_i/[2\pi m_e(1 + T_i/T_e)]}$. Since the direction of the flows at the sheath entrance are known, ghost cells in the $z$-direction are filled such that an upwind stencil ensues to evolve $n$, $\gvort$, $T_e$, and $T_i$ \cite{francisquez2020fluid}. For $T_e$ and $T_i$ specifically, finite conductive heat fluxes entering the sheaths are applied according to $q_{{\parallel},s} = -\kappa^s_{\parallel} \delpar T_s = \pm \gamma_s n v_{{\parallel},s} T_s$,
where the upper (lower) sign corresponds to the top (bottom) sheath and $\gamma_s$ is the sheath transmission coefficient. Its value for ions and electrons is taken to be $\gamma_i =5T_i/2T_e$ and $\gamma_e = 2 + \lvert e \phi \rvert/T_e$, respectively \cite{francisquez2020fluid}. Collisional coefficients and diffusivities are kept constant in the direct numerical simulation as they can be unphysically large at high temperatures due to the lack of kinetic effects and generally require closures going beyond Chapman-Enskog to account for plasma where $\nu_{ei} \not\gg \partial/ \partial t$ \cite{Chapman_1970}. 

To start the numerical simulation, electrons and ions are initialized with zero parallel velocity and vorticity fields along with truncated Gaussian density and temperature profiles. A second-order trapezoidal leap-frog time-stepping scheme evolves the system of equations forward with subcycling of parabolic terms (e.g. $\delpar \kappa^s_\parallel \delpar T_s$) \cite{computational_methods,francisquez_thesis} due to the low frequency turbulence structure changing slowly over the thermal diffusion timescale. The commonly applied Boussinesq approximation \cite{GRILLIX_2018} in Braginskii solvers is also used when evolving the generalized vorticity, $\gvort$. The normalizations applied to solve these partial differential equations in both the finite difference code and deep learning framework are sketched in the Appendix. A complete treatment of the numerical solver and greater specificity regarding the turbulence simulations can be found in \cite{GDB,francisquez2020fluid}.

\section{\label{sec:level2.2}Machine learning fluid theory}

Neural networks are operationally computational programs composed of elementary arithmetic operations (e.g. addition, multiplication) and functions (e.g. $\exp$, $\sin$, $\log$) which can be differentiated to arbitrary order up to machine precision via application of chain rule \cite{Raissi_JMLR,AD_2020}. While biases are presently inevitable \cite{wang2020eigenvector}, these regression models are in theory constructed without necessarily committing to a designated class of basis functions (e.g. polynomial, trigonometric). Automatic differentiation in conjunction with this adaptive capacity of neural networks permits them to effectively address nonlinear optimization problems in physics and engineering by training upon both partial differential equations and observational data via multi-task learning \cite{raissi2017physics}. Constraining classically underdetermined systems by physical laws and experimental measurements in this way presents an emerging technique in computational mechanics which this thesis extends to the deduction of unknown turbulent plasma dynamics. In this physics-informed deep learning framework, every dynamical variable in equations \eqref{eq:nDotGDBH}--\eqref{eq:TiDotGDBH} is approximated by its own fully-connected neural network, which is commonly known as a data-efficient universal function approximator \cite{cybenko_approximation_1989}, which can be molded to learn unknown turbulent fields given sufficient training. 

\begin{figure}[ht]
\centering
\includegraphics[width=0.85\linewidth]{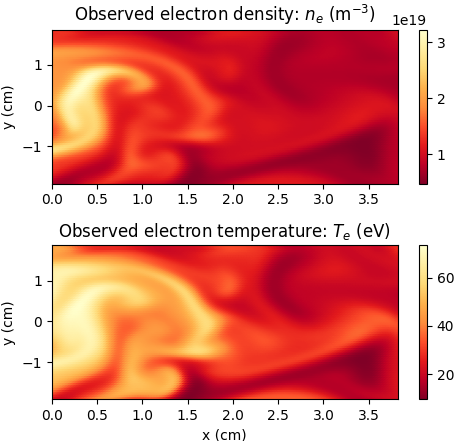}
\caption{\label{observed_dens_and_Te}These 2-dimensional slices of the simulated turbulent electron density and temperature over a short temporal window are the only observed variables used as inputs to the deep learning framework. The full 3D synthetic plasma exhibits field-aligned filamentary structures (i.e. blobs).}
\end{figure}

For analysis in the multi-network framework, partial measurements of $n_e$ and $T_e$ over time only come from a smaller 2-dimensional field-aligned domain in the interior of the synthetic plasma described in Section \ref{sec:level2.1} to emulate experiment with dimensions of $[L^*_x = 3.8 \text{ cm}, L^*_y = 3.8 \text{ cm}]$ and spatiotemporal resolution of $[\Delta^* x = 0.03 \text{ cm}, \Delta^* y = 0.04 \text{ cm}, \Delta^* t = 7.27 \times 10^{-7} \text{ s}]$ as depicted by a snapshot in Figure \ref{observed_dens_and_Te}. Each network consequently takes the local spatiotemporal points $(x,y,t)$ from the reduced domain for measurements as the only inputs to the initial layer while the dynamical variable being represented (e.g. $\phi$) is the sole output. In the middle of the architecture, every network consists of 5 hidden layers with 50 neurons per hidden layer and hyperbolic tangent activation functions ($\sigma$) using Xavier initialization \cite{GlorotAISTATS2010}. 

\begin{figure}[ht]
\centering
\includegraphics[width=0.8\linewidth]{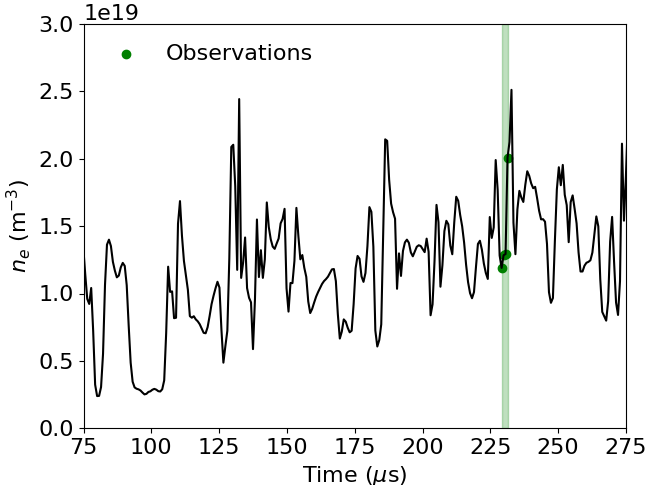}
\caption{\label{observed_1D_dens_time} A local time trace of the turbulent $n_e$ over 200 $\mu$s from the simulated plasma at $[x = 1.0 \text{ cm}, y = 0.0 \text{ cm}, z = -28.1 \text{ cm}]$. The observed synthetic data analyzed in the machine learning framework only comes from the small temporal window (green) which corresponds to just 4 points in time.}
\end{figure}

\begin{figure*}
\centering
\includegraphics[width=1.0\linewidth]{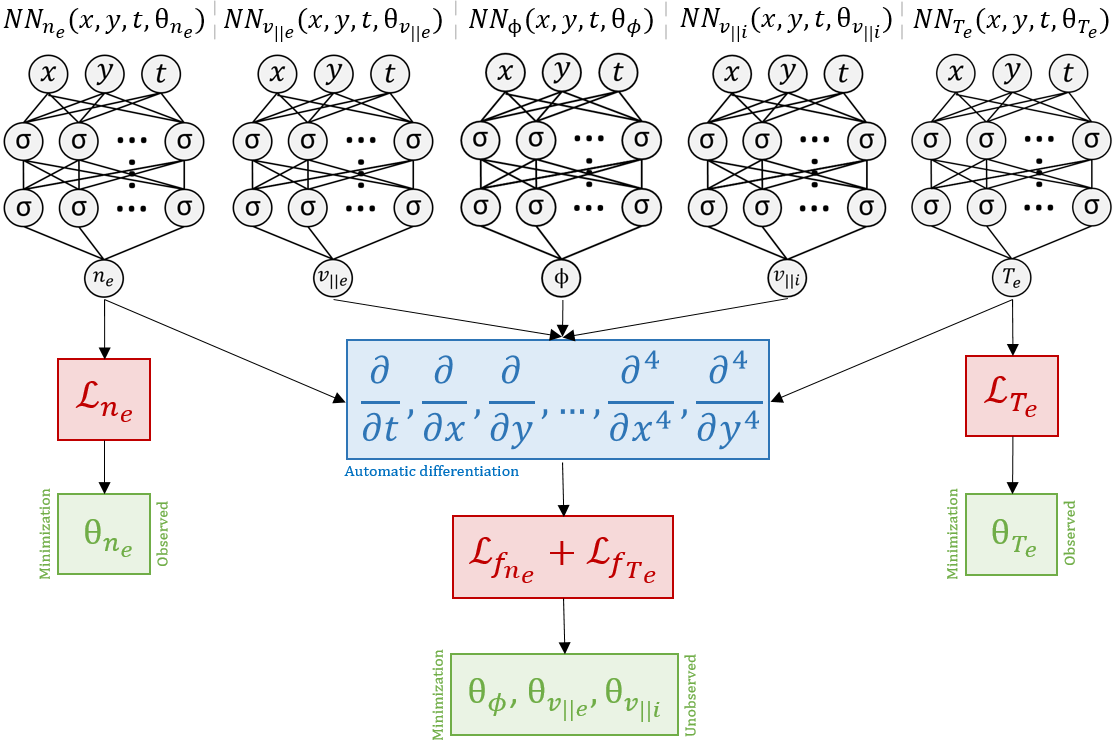}
\caption{\label{PlasmaPINN}Visualization of the physics-informed framework with individual networks---where $\theta_{n_e}$ represents the weights and biases of the $n_e$ network, for example---being selectively trained against loss functions comprising both partial observations, $\mathcal{L}_{n_e}$ and $\mathcal{L}_{T_e}$, and reduced theory, $\mathcal{L}_{f_{n_e}}$ and $\mathcal{L}_{f_{T_e}}$, to infer unobserved turbulent dynamics. All spatial gradients and time derivatives in $f_{n_e}$ and $f_{T_e}$ are represented using automatic differentiation of each individual variable's network which in practice extends the size of the computation graph being evaluated. To handle reduced 2-dimensional data from the 3-dimensional synthetic plasma, the z-coordinate is removed from the networks for simplicity and as a test for determining the minimal information necessary to learn $\phi$. If noisy data are observed, then $\theta_{n_e}$ (and $\theta_{T_e}$, if $T_e$ measurements are available) should be additionally trained against $\mathcal{L}_{f_{n_e}}$ (and $\mathcal{L}_{f_{T_e}}$).}
\end{figure*}

In actuality, the repeated differentiation and summation of networks to construct every single term's representation in the partial differential equations subsequently constructs a far larger resultant computation graph. The cumulative network is therefore a truly deep approximation---at least compared to the 5 hidden layers in each dynamical variable's individual network---of the plasma turbulence theory. Partial observations of the simulated plasma consist of only $n_e$ and $T_e$ measurements of 2-dimensional spatial extent, which is similar to experimental fluctuation diagnostics like gas puff imaging \cite{Zweben_2017,mathews2022deep}, as visualized in Figure \ref{observed_dens_and_Te} over just 4 separate time slices (i.e. 2.9 $\mu$s). For reference, the synthetic plasma's fluctuations have an approximate autocorrelation time of 1.5 $\mu$s and radial autocorrelation length of 0.4 cm. The narrow temporal extent of the strongly fluctuating $n_e$ observations at a local spatial point is further visualized in Figure \ref{observed_1D_dens_time}. Properties of all other dynamical variables in the 6-field turbulence theory are taken to be unknown, and the networks are simultaneously optimized against the drift-reduced Braginskii equations and observed data to better approximate the unmeasured quantities. Physical constraints are learned by the networks via minimization of ascribed loss functions encompassing both limited measurements of the plasma and two-fluid turbulence model. To be precise, partial synthetic measurements are learned by training the $n_e$ and $T_e$ networks against the average $\mathcal{L}_2$-norm of their respective relative errors
\begin{equation}\label{eq:L_n_DotGDBH}
    \mathcal{L}_{n_e} = \frac{1}{N_0}\sum_{i=1}^{N_0} \lvert n^*_{e}(x^i_0,y^i_0,z^i_0,t^i_0) - n^i_{e,0} \rvert^2,
\end{equation}
\begin{equation}\label{eq:L_Te_DotGDBH}
    \mathcal{L}_{T_e} = \frac{1}{N_0}\sum_{i=1}^{N_0} \lvert T^*_{e}(x^i_0,y^i_0,z^i_0,t^i_0) - T^i_{e,0} \rvert^2,
\end{equation}
where $\lbrace x_0^i,y_0^i,z_0^i,t_0^i,n_{e,0}^i,T_{e,0}^i\rbrace^{N_0}_{i=1}$ correspond to the set of observed data and the variables $n^*_e$ and $T^*_e$ symbolize predicted electron density and temperature, respectively, by the networks. The theory enforcing physical constraints in the deep learning framework is expressed by evaluating the individual terms in the model by differentiating the graphs with respect to input spatiotemporal coordinates via application of chain rule through automatic differentiation \cite{tensorflow2015-whitepaper}. Correspondingly, model loss functions are embedded during training by recasting \eqref{eq:nDotGDBH} and \eqref{eq:TeDotGDBH} in the following implicit form

\begin{eqnal}\label{eq:f_n_DotGDBH}
f_{n_e} &\coloneqq -\d{^e n}{t} -\frac{2c}{B}\left[n\curv{\phi}-\frac{1}{e}\curv{p_e}\right]-n\delpar{\vpe} +\nSrcN+\mathcal{D}_{n},
\end{eqnal}

\begin{eqnal}\label{eq:f_Te_DotGDBH}
f_{T_e} &\coloneqq -\d{^e T_e}{t} + \frac{2T_e}{3n}\left[\d{^e n}{t} + \frac{1}{T_e}\delpar \kappa^e_\parallel \delpar T_e + \eta_\parallel \right.\\& \left. + \frac{5n}{m_e \Omega_e} \curv{T_e} \frac{\cur^2}{T_e} + \frac{0.71}{e}(\delpar{\cur} - \frac{\cur}{T_e}\delpar{T_e}) + \frac{1}{T_e} \enerSrceN \right] + \mathcal{D}_{T_e},
\end{eqnal}
and then further normalized into dimensionless form matching the numerical code as in \eqref{eq:normlnDotGDBH} and \eqref{eq:normlTeDotGDBH} \cite{francisquez2020fluid}. This normalized implicit formulation is vital to learning via optimization since all physical terms collectively sum to zero when the equations are ideally satisfied. These physical constraints provided by the unitless evolution equations of $n_e$ and $T_e$ are jointly optimized using loss functions defined by
\begin{equation}\label{eq:L_f_n_DotGDBH}
    \mathcal{L}_{f_{n_e}} = \frac{1}{N_f}\sum_{i=1}^{N_f} \lvert f^{*}_{n_e}(x^i_f,y^i_f,z^i_f,t^i_f) \rvert^2,
\end{equation}
\begin{equation}\label{eq:L_f_Te_DotGDBH}
    \mathcal{L}_{f_{T_e}} = \frac{1}{N_f}\sum_{i=1}^{N_f} \lvert f^{*}_{T_e}(x^i_f,y^i_f,z^i_f,t^i_f) \rvert^2,
\end{equation}
where $\lbrace x_f^i,y_f^i,z_f^i,t_f^i\rbrace^{N_f}_{i=1}$ denote the set of collocation points, and $f^{*}_{n_e}$ and $f^{*}_{T_e}$ are the null partial differential equations prescribed by \eqref{eq:f_n_DotGDBH} and \eqref{eq:f_Te_DotGDBH} in normalized form directly evaluated by the neural networks. Optimization against the applied plasma theory is central to the methodology and enforces physical constraints in the deep learning framework by ensuring each sub-network respects the multi-field turbulence model's constraints as visualized in Figure \ref{PlasmaPINN}. This enables fine tuning of each neural networks' weights and biases by adjusting them in this generalized regression model to satisfy the physical laws governing the nonlinear connection sought between the sub-networks. The set of collocation points over which the partial differential equations are evaluated can be arbitrarily large and span any extent over the physical domain, but are taken in this example to correspond to the positions of the synthetic measurements being trained upon, i.e. $\lbrace x_0^i,y_0^i,z_0^i,t_0^i\rbrace^{N_0}_{i=1} = \lbrace x_f^i,y_f^i,z_f^i,t_f^i\rbrace^{N_f}_{i=1}$.

It should be once again noted that the only observed dynamical quantities in these equations are 2-dimensional views of $n_e$ and $T_e$ without any explicit information about boundary conditions nor initializations. All analytic terms encoded in these equations including high-order operators are computed exactly by the neural networks without any approximation (e.g. linearization) nor discretization. This machine learning framework with a collocation grid of arbitrarily high resolution uses a continuous spatiotemporal domain without time-stepping nor finite difference schema in contrast with the numerical \texttt{GDB} code. To handle 2-dimensional data, slow variation of dynamics is assumed in the $z$-coordinate and effectively set all parallel derivatives to zero (i.e. $\frac{\partial}{\partial z}\rightarrow 0$). Notwithstanding, parallel flows and Ohmic heating terms in the model are still kept. If measurements in the $z$-direction are available or more collocation points utilized during training with observational data of reduced dimensionality, this procedure may be relaxed---it is partly a trade-off between computational fidelity and stability. It is noteworthy that the temporal resolution of the data observed by the neural networks is several orders of magnitude lower than the time steps taken by the finite difference solver in \texttt{GDB} as required for numerical stability, i.e. $\Delta^* t \gg \Delta t$. Also, if sought, training on data sets viewed at oblique angles in 3-dimensional space over long macroscopic timescales can be easily performed via segmentation of the domain and parallelization, but a limited spatial view with reduced dimensionality is taken to emulate experimental conditions for field-aligned observations \cite{Zweben_2017} and theoretically test what information is indispensable to learn unobserved turbulent dynamics. 

Loss functions are optimized with mini-batch sampling where $N_0 = N_f = 500$ using stochastic gradient descent via Adam \cite{kingma2014adam} and the L-BFGS algorithm---a quasi-Newton optimization algorithm \cite{10.5555/3112655.3112866}---for 20 hours over 32 cores on Intel Haswell-EP processors which corresponds to approximately 8694 full iterations over both optimizers. If observing noisy data, it is found that expanding to larger sample sizes with $N_0 = N_f = 2500$ and training solely via L-BFGS is optimal for learning. Removing $\mathcal{L}_{f_{n_e}}$ and $\mathcal{L}_{f_{T_e}}$ from the optimization process (i.e. setting $N_f = 0$) would correspond to training of classical neural networks without any knowledge of the underlying governing equations which would then be incapable of learning turbulent field fluctuations. Setting $N_0 = 0$ instead while providing initial and boundary conditions for all dynamical variables would alternatively correspond to regularly solving the equations directly via neural networks. Overall, priming networks by firstly training in stages on observed data or prior constraints, i.e. priming, is useful to enhance stability and convergence in this multi-objective task. Additionally, encoding domain expertise such as subsonic bounds on parallel flows or regularizing temperature to be strictly positive via suitable output activation functions can assist training by constraining the admissible solution landscape. Networks constructed in this way can intrinsically abide by physical laws which is especially useful to uncover unknowns like $\vpi$ and $T_i$.

A fundamental goal in computational plasma modelling is determining the minimum complexity necessary (and no less) to develop sufficiently predictive tokamak simulations. With sparse availability of measurements in fusion experiments, designing diagnostic techniques for uncovering such information is crucial. On this point, it is emphasized that training is from scratch over just a single synthetic plasma discharge with no extraneous validation nor testing sets required since overfitting is technically not encountered in this physics-informed paradigm. The multi-network deep learning framework simply utilizes a single set of $n_e$ and $T_e$ measurements over a period of microseconds which corresponds to the small data regime of machine learning. Merging partial observational data of $n_e$ and $T_e$ along with physical laws in the form of partial differential equations governing the time-dependent evolution of $n_e$ and $T_e$ sufficiently constrains the set of admissible solutions for the previously unknown nonlinear mappings the neural networks ultimately learn. It is also quite general: due to quasineutrality, no significant adjustments are necessary to generalize the technique when multiple ions and impurities may be present in boundary plasmas beyond the inclusion of appropriate collisional drifts and sources in multi-species plasmas \cite{multi_species,DRB_consistency1}. This deep learning technique for diagnosing turbulent fields is hence easily transferable which permits its systematic application across magnetic confinement fusion experiments whereby the underlying physical model fundamental to the turbulent transport is consistent. The framework sketched can also be extended to different settings in the interdisciplinary study (both numerical and experimental) of magnetized collisional plasmas in propulsion engines and astrophysical environments.

\section{\label{sec:level2.3}Numerical Experiments}

\subsection{\label{sec:level2.3.1}Recovery of the unknown turbulent electric field}

\begin{figure}[ht]
\centering
\includegraphics[width=0.9\linewidth]{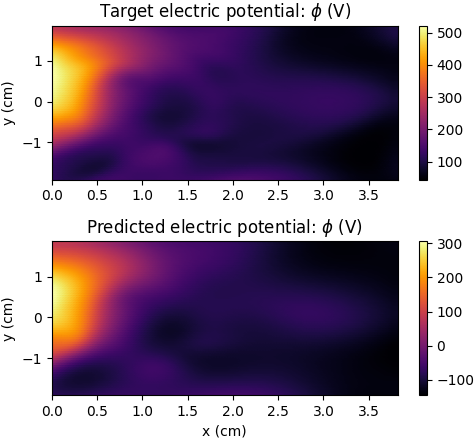}
\caption{\label{learned_phi}The synthetic plasma's unobserved electric potential (top) at $t = 229.9 \ \mu\text{s}$ is learned approximately up to an additive constant by the neural network (bottom).}
\end{figure}

\begin{figure}[ht]
\centering
\includegraphics[width=0.85\linewidth]{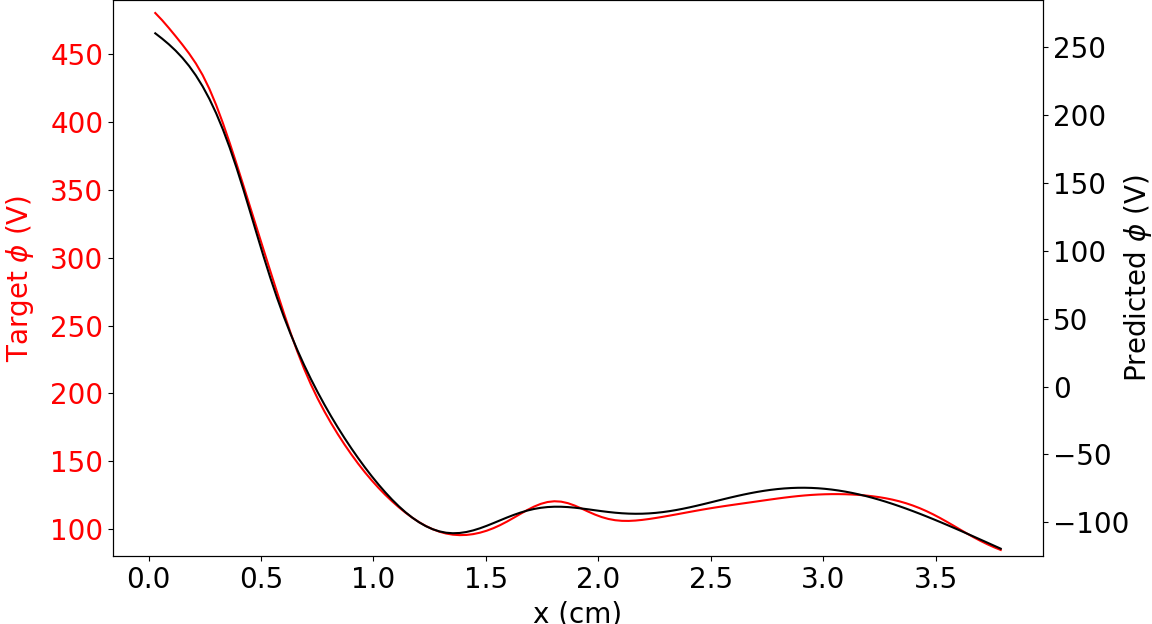}
\caption{\label{1d_phi}A 1-dimensional radial profile of the true and predicted $\phi$ at $[y = 0.0 \text{ cm}, z = -28.1 \text{ cm}, t = 229.9 \ \mu\text{s}]$, i.e. a slice of Figure \ref{learned_phi}. The ordinates are offset differently but both exactly span 410 V with equivalent spacing.}
\end{figure}

Accurate turbulent edge electric field fluctuation characterization is particularly significant to magnetic confinement fusion devices. By constraining the deep learning framework with the two-fluid turbulence theory and modest amounts of empirical information in the form of partial 2-dimensional observations of $n_e$ and $T_e$, it is found that physics-informed neural networks can accurately learn the plasma's electric potential on turbulent scales without the machine learning framework ever having observed this quantity, as displayed in Figures \ref{learned_phi} and \ref{1d_phi}. 

\begin{figure}[ht]
\centering
\includegraphics[width=0.9\linewidth]{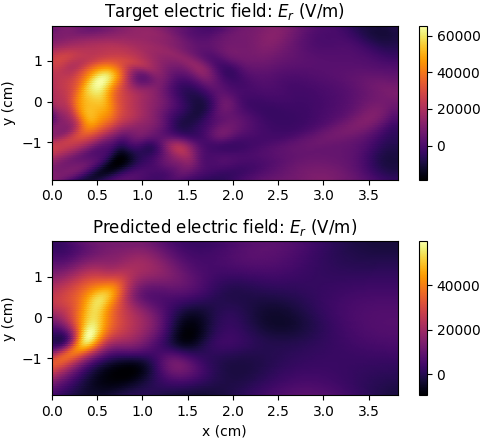}
\caption{\label{learned_E_r}The learned turbulent $E_r$ (bottom) closely agrees with the magnitude and structure of the true $E_r$ (top) despite $\gvort$, $\vpe$, $\vpi$, and $T_i$ being initially unknown.}
\end{figure}

It is notable that despite there being no knowledge of $\gvort, \vpe, \vpi,$ and $T_i$ (i.e. multiple unknowns existing in the partial differential equations and \eqref{BVP_gvort_phi} never even being directly invoked), the electric field is nonetheless learned consistently with the physical theory encoded  by the plasma turbulence model. Since $\phi$ is a gauge-invariant quantity exact up to an additive constant, it is accordingly uncovered up to a scalar offset which varies randomly due to the stochastic nature of the optimization employed in the machine learning framework. This difference physically arises because no direct boundary information was enforced upon the neural network when learning $\phi$, although it could be technically implemented. By contrast, the \texttt{GDB} code imposed zero potential on the outer walls of the simulation domain. General agreement in both magnitude and structure in the learned radial electric field is evident in Figure \ref{learned_E_r} with an average absolute error of 2619 V/m. 

To better interpret the learning process, normalized loss functions being trained upon are tabulated after $M$ full iterations by the optimizers in Table \ref{loss_tabulate}. After one iteration, \eqref{eq:L_f_n_DotGDBH} and \eqref{eq:L_f_Te_DotGDBH} are relatively small in magnitude, and this would correspond to a trivial result satisfying the partial differential equations given the nonuniqueness of its solutions. As training progresses, observational data is better captured in the deep learning framework and the neural networks proceed to converge toward the sought solution as opposed to trivial ones. A difference in the rates of learning for $n_e$, $T_e$, and $\phi$ also exist since the electric field is learned implicitly via the model instead of being trained upon directly. Each individual loss function being optimized therefore does not necessarily decrease perfectly monotonically, but it is instead the collective training against partial differential equations in conjunction with observational data that is key. Namely, while there are many potential solutions to \eqref{eq:f_n_DotGDBH} and \eqref{eq:f_Te_DotGDBH}---and while they may be more easily satisfied by trivial solutions---the limited $n_e$ and $T_e$ measurements compel the optimizer towards the physical solution of the partially observed plasma. In scenarios where inconsistencies in the true and learned model $E_r$ exist, one might repurpose this machine learning framework to iteratively test and thereby discover the {\it correct} partial differential equations altogether by quantitatively examining the proposed model's consistency with observations as in Table \ref{loss_tabulate}. For example, the analytic form of reduced source models in fluid theories \cite{Thrysoe_2018,Thrysoe_2020} can be inserted in the physics-informed deep learning framework to account for local turbulent ionization and inelastic collisions with kinetic neutrals by observing such measurements of $n_e$, $T_e$, and $\phi$ in global simulations \cite{Wersal_2015} and experiments \cite{mathews2022deep}.

\begin{table}[ht]
{\centering
\renewcommand{\arraystretch}{1.0}
\includegraphics[width=0.9\linewidth]{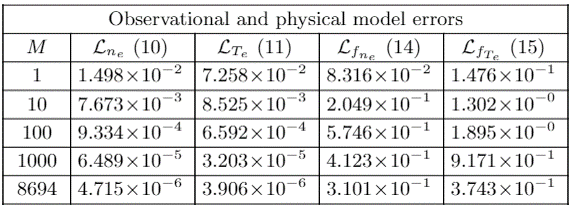}
\caption{\label{loss_tabulate}Each normalized loss function optimized in the machine learning framework is tabulated after $M$ full iterations, where $M = 8694$ corresponds to the final iteration after 20 hours of training against both the partial observations of $n_e$ and $T_e$ and their implicit evolution equations, i.e. \eqref{eq:L_n_DotGDBH}, \eqref{eq:L_Te_DotGDBH}, \eqref{eq:L_f_n_DotGDBH}, and \eqref{eq:L_f_Te_DotGDBH}.}
}
\end{table}

\begin{figure}[ht]
\centering
\includegraphics[width=0.9\linewidth]{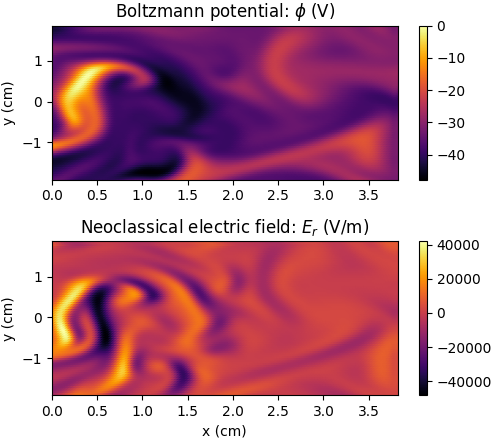}
\caption{\label{analytic_E_r}Estimates of the turbulent $\phi$ and $E_r$ as expected by the Boltzmann model or neoclassical estimates yield markedly errant predictions when compared to the true values displayed at the top of Figures \ref{learned_phi} and \ref{learned_E_r}.}
\end{figure}

\subsection{\label{sec:level2.3.2}Contrast with conventional calculations}

For comparison, classical and oft-employed models for calculating the electric potential with adiabatic electrons such as the Boltzmann relation, i.e. $n_e(\phi) = n_e(\phi_0) e^{q_e(\phi_0 - \phi)/T_e}$ \cite{Callen_adiabatic}, fail when computing perpendicular turbulent field fluctuations. Alternative approximations of $E_r$ from simple ion pressure balance as expected neoclassically, i.e. $\nabla \phi = -\nabla p_i/(Z n_i e)$ where $Z=1$ for deuterium ions \cite{Viezzer_2013}, would yield highly incorrect estimates of the turbulent electric field, too. Such methods ordinarily used in magnetic confinement fusion are only applicable to discerning equilibrium fields and dynamics parallel to the magnetic field in steady-state scenarios, but are erroneous in the analysis of microturbulence in nonquiescent plasmas. This is markedly observed when comparing Figure \ref{analytic_E_r} to the true $\phi$ and $E_r$ as plotted in Figures \ref{learned_phi} and \ref{learned_E_r}, respectively. This deep learning technique based upon drift-reduced Braginskii theory therefore provides a novel way to accurately measure the turbulent electric field in edge plasmas from just the electron density and temperature. As a further point of contrast compared to classical techniques, it is important to note that the inverse learning scenario has not been demonstrated thus far. In particular, given observations of $\phi$ and $T_e$, one cannot simply infer the turbulent $n_e$ fluctuations with the machine learning framework outlined. This one-way nature in learning indicates a division exists between the two pathways when attempting to constrain the admissible solutions of \eqref{eq:L_f_n_DotGDBH} and \eqref{eq:L_f_Te_DotGDBH} to uncover unknown nonequilibrium dynamics. Training is thus unidirectional and representative of asymmetries extant in the partial data and turbulence theory being learnt via optimization. 

\subsection{\label{sec:level2.3.3}Impacts of noise in physics-informed deep learning}

\begin{figure}
\centering
\includegraphics[width=1.0\linewidth]{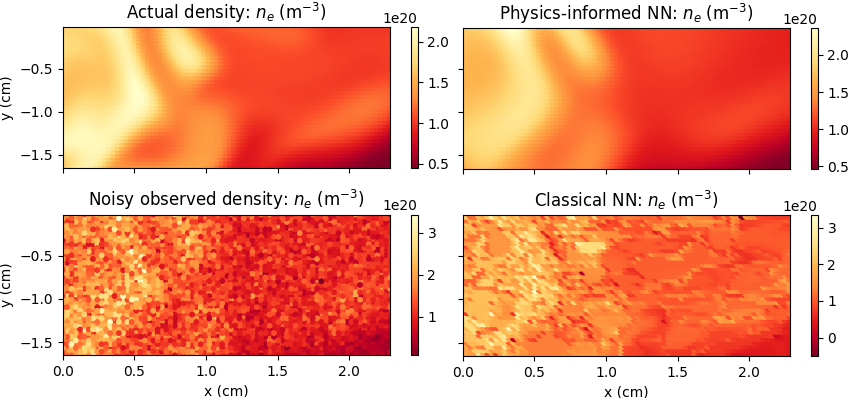}
\caption{\label{noisy_dens} The physics-informed deep learning framework is capable of recovering the true $n_e$ despite strong Gaussian noise, i.e. $\sigma = 0.25$, present. The classical solution corresponds to a standard feed-forward neural network where $N_f = 0$.}
\end{figure}

It is also important to examine the practicality of these results when noise corrupts the observations as inevitably expected when translating this framework to experiment. As an example test, it is found that if only observing density fluctuations with normally distributed errors, one can still largely recover the unmeasured perpendicular electric fields and even resolve the partially observed variables. Namely, given just $n_e$ measurements with Gaussian noise of 25\% as in Figure \ref{noisy_dens}, the deep learning framework is robust enough to learn the true turbulent density in this physics-informed paradigm, which can subsequently be used to infer the unmeasured electric field. If $E_r$ was already known, this technique could then precisely check the validity of the reduced turbulence theory against observations from experiment or kinetic simulations \cite{Mathews2021PoP}. But, if using a standard feed-forward neural network, one must be careful with convergence since the objective of simply minimizing $\mathcal{L}_{n_e}$ without sufficient regularization, as innately provided by $\mathcal{L}_{f_{n_e}}$, can result in overfitting of noisy data.

\section{\label{sec:level2.4}Conclusion}

These results illustrate a custom physics-informed deep learning framework with the novel capacity to learn unknown nonequilibrium dynamics in a multi-field turbulent transport model broadly relevant to magnetized collisional plasmas. This chapter specifically demonstrates the ability to determine unobserved turbulent electric fields consistent with the drift-reduced Braginskii equations from partial electron pressure observations, in contrast with with standard analytic techniques. This technique can be applied to infer field fluctuations that may be difficult to measure or when plasma diagnostics provide only partial information. On the other hand, if experimental electric field measurements exist, then the quantitative validity of the plasma turbulence model embedded in the neural networks can be directly assessed. This technique is also quite robust since, due to quasineutrality, it can be used to study ionized gases in magnetized environments with multiple ions and impurities present as commonly found in astrophysical settings and fusion energy and space propulsion systems. From a mathematical physics standpoint, it is remarkable that nonlinear dynamics can be accurately recovered from partial data and theory in a 6-field model. Inferring completely unknown turbulent fields from just 2-dimensional measurements and representations of the evolution equations given by \eqref{eq:f_n_DotGDBH} and \eqref{eq:f_Te_DotGDBH} demonstrates a massive reduction in the original 3-dimensional turbulence model indicating redundancy and the existence of reduced theory characterizations. Going forward, this framework has the potential capability to be generalized (e.g. to learn $T_e$, $T_i$, $\vpe$, and $\vpi$ in addition to $\phi$ using just 1-dimensional $n_e$ measurements) and transform how turbulence theories are systematically and quantitatively validated in both plasma simulations and experiments. The interpretable physics-informed machine learning methodology outlined should also be transferable across models (e.g. collisionless fluids, gyrokinetic, electromagnetic, atomic physics) and complex geometries. Furthermore, known limitations and {\it unknown} corrections to Braginskii's theory exist \cite{Catto-paper}, which can be introduced in the deep learning framework to automate testing and discovery of reduced plasma turbulence models when high fidelity data is observed. These extensions in theory and computing are left for future works.

 

\chapter{Turbulent field fluctuations in gyrokinetic and fluid plasmas}
\setlength{\epigraphwidth}{0.48\textwidth}
\epigraph{All models are wrong, but some are useful.}{George E. P. Box}

Validating the accuracy of reduced turbulence theories is amongst the greatest challenges to the advancement of plasma modelling. But meaningfully evaluating dynamical connections between turbulent fields across theories or experiment has been an intricate task for conventional numerical methods. Namely, when comparing global plasma simulations to other numerical models or diagnostics, one must precisely align initial states, boundary conditions, and sources of particles, momentum, and energy (e.g. plasma heating, sheath effects, atomic and molecular processes) \cite{francisquez2020fluid,Mijin_2020}. These descriptions of physics can be increasingly difficult to reconcile when applying different representations---for example, fluid or gyrokinetic---or when measurements characterizing the plasma are sparse as commonly encountered in thermonuclear fusion devices. Additionally, imperfect conservation properties (both in theory and numerics due to discretization) and the limited accuracy of time-integration schema can introduce errors further misaligning comparative modelling efforts. Such difficulties exist even when examining the same nominal turbulence theory across numerical codes \cite{nonlinear_GK_comparison_new,Tronko_ORB5vGENE,nonlinear_GK_comparison1,nonlinear_GK_comparison2} and become magnified for cross-model comparisons \cite{francisquez2020fluid}. To overcome these classical limitations, Chapter 3 applies the custom physics-informed deep learning framework developed in Chapter 2 to demonstrate the first direct comparisons of instantaneous turbulent fields between electrostatic drift-reduced Braginskii theory and electromagnetic long-wavelength gyrokinetic simulations. Good overall quantitative agreement is found between the two turbulence models in magnetized helical plasmas at low normalized pressure.

In the fusion community, both comprehensive full-$f$ global gyrokinetic \cite{mandell_hakim_hammett_francisquez_2020,XGC,GENE,COGENT,ELMFIRE,ORB5,GYSELA,PICLS} and two-fluid \cite{BOUT++,TOKAM3X,GBS,Stegmeir2018,GDB} direct numerical simulations are under active development to improve predictive modelling capabilities and the design of future reactors, but no single code can currently capture all the dynamics present in the edge with sufficient numerical realism. It is therefore of paramount importance to recognize when certain numerical or analytical simplifications make no discernible difference, or when a given theoretical assumption is too risky, to clearly identify the limits of present simulations \cite{francisquez2020fluid}. Adaptations of the drift-reduced Braginskii equations have recently investigated several important edge phenomena \cite{Francisquez_2017,Ricci_2012,Thrysoe_2018,nespoli_non-linear_2017}, but precise quantitative agreement with experiment is generally lacking on a wide scale. The predictive accuracy of these codes in fusion reactors is uncertain, especially since kinetic effects may not be negligible as edge plasma dynamics approach the ion Larmor radius and collision frequencies are comparable to turbulent timescales \cite{Braginskii1965RvPP,blobs_review}. Yet these fluid codes remain in widespread usage for their reduced computational cost, which offers the ability to perform parameter scans with direct numerical simulations that can help uncover underlying physics and optimize reactor engineering designs. Nevertheless, the Vlasov-Maxwell system of equations is the gold standard in plasma theory, but such modelling is prohibitively expensive when wholly simulating the extreme spatiotemporal scales extant in fusion reactors. For the purposes of more tractable yet high fidelity numerical simulations, this system is averaged over the fast gyro-motion of particles to formulate 5D gyrokinetics. Fluid-gyrokinetic comparisons therefore present a way to analyze the robustness of reduced modelling while also illuminating when collisional fluid theories may be insufficient. Quantifying potential shortcomings is thus vital for the testing and development of improved reduced models \cite{waltz2019kinetic}. 

While explicit comparisons of global nonlinear dynamics across turbulence models were previously impractical using standard numerical simulations, the physics-informed deep learning \cite{raissi2017physics} technique from Chapter 2 enables this direct quantitative cross-examination of plasma turbulence theories by uncovering the turbulent electric field expected in two-fluid and gyrokinetic models based upon an electron pressure fluctuation. This turbulence model validation framework is potentially transferable across magnetic confinement fusion experiments where the underlying physical model governing turbulent transport is consistent, which permits its systematic application. The framework sketched can thus be extended to different settings especially in the interdisciplinary study (both numerical and experimental) of magnetized collisional plasmas in space propulsion systems and astrophysical environments if sufficient observations exist. Overall, these physics-informed neural networks present a new paradigm in the testing of reduced plasma turbulence models.


To demonstrate these results, this chapter describes a gyrokinetic model based upon \cite{mandell_hakim_hammett_francisquez_2020} in Section \ref{sec:level3.1}, outlines pertinent aspects of the physics-informed machine learning architecture from Chapter 2 now being applied anew on gyrokinetic simulation data in Section \ref{sec:level3.2}, presents the first direct comparisons of instantaneous turbulent fields in Section \ref{sec:level3.3}, and concludes with a summary in Section \ref{sec:level3.4}.

\section{\label{sec:level3.1}Gyrokinetic modelling}

Following the full treatment outlined in \cite{mandell_thesis} and \cite{mandell_hakim_hammett_francisquez_2020}, the long-wavelength gyrokinetic modelling analyzed in this chapter constitutes nonlinear global electromagnetic turbulence simulations on open field lines using the \texttt{Gkeyll} computational plasma framework. The full-$f$ electromagnetic gyrokinetic equation is solved in the symplectic formulation \cite{brizard2007foundations}, which describes the evolution of the gyrocenter distribution function $f_s(\v{Z},t)=f_s(\v{R},v_\parallel,\mu,t)$ for each species ($s = \{e,i\}$), where $\v{Z}$ is a phase-space coordinate composed of the guiding center position $\v{R}=(x,y,z)$, parallel velocity $v_\parallel$, and magnetic moment $\mu=m_s v_\perp^2/2B$. In terms of the gyrocentre Hamiltonian and the Poisson bracket in gyrocentre coordinates with collisions $C[f_s]$ and sources $S_s$, the gyrokinetic equation is
\begin{align}
    \pderiv{f_s}{t} + \{f_s, H_s\} - \frac{q_s}{m_s}\pderiv{A_\parallel}{t}\pderiv{f_s}{v_\parallel} = C[f_s] + S_s, \label{liouville}
\end{align}
or, equivalently,
\begin{align}
    \pderiv{f_s}{t} + \dot{\v{R}}\v{\cdot}\nabla f_s + \dot{v}^H_\parallel \pderiv{f_s}{v_\parallel}- \frac{q_s}{m_s}\pderiv{A_\parallel}{t}\pderiv{f_s}{v_\parallel} = C[f_s] + S_s,
\end{align}
where the gyrokinetic Poisson bracket is defined as
\begin{equation}
  \{F,G\} = \frac{\v{B^*}}{m_s B_\parallel^*} \v{\cdot} \left(\nabla F \frac{\partial G}{\partial v_\parallel} - \frac{\partial F}{\partial v_\parallel}\nabla G\right) - \frac{\uv{b}}{q_s B_\parallel^*}\times \nabla F \v{\cdot} \nabla G, 
\end{equation}
and the gyrocentre Hamiltonian is
\begin{equation}
    H_s = \frac{1}{2}m_sv_\parallel^2 + \mu B + q_s  \phi.
\end{equation}
The nonlinear phase-space characteristics are given by
\begin{equation}
    \dot{\v{R}} = \{\v{R},H_s\} = \frac{\v{B^*}}{B_\parallel^*}v_\parallel + \frac{\uv{b}}{q_s B_\parallel^*}\times\left(\mu\nabla B + q_s \nabla \phi\right),
\end{equation}
\begin{eqnal}\label{vpardot}
    \dot{v}_\parallel &= \dot{v}^H_\parallel -\frac{q_s}{m_s}\pderiv{A_\parallel}{t} = \{v_\parallel,H_s\}-\frac{q_s}{m_s}\pderiv{A_\parallel}{t} \\ \quad &= -\frac{\v{B^*}}{m_s B_\parallel^*}\v{\cdot}\left(\mu\nabla B + q_s \nabla \phi\right)-\frac{q_s}{m_s}\pderiv{A_\parallel}{t}.
\end{eqnal}

\noindent Here, $B_\parallel^*=\uv{b}\v{\cdot} \v{B^*}$ is the parallel component of the effective magnetic field $\v{B^*}=\v{B}+(m_s v_\parallel/q_s)\nabla\times\uv{b} + \delta \v{B}$, where $\v{B} = B \uv{b}$ is the equilibrium magnetic field and $\delta \v{B} = \nabla\times(A_\parallel \uv{b})\approx \nabla A_\parallel \times \uv{b}$ is the perturbed magnetic field {(assuming that the equilibrium magnetic field varies on spatial scales larger than perturbations so that $A_\parallel\nabla\times\uv{b}$ can be neglected)}. Higher-order parallel compressional fluctuations of the magnetic field are neglected so that $\delta\v{B}=\delta\v{B}_\perp$ and the electric field is given by ${\mathbf E} = -\nabla\phi - (\partial{A_\parallel}/\partial t) \uv{b}$. The species charge and mass are $q_s$ and $m_s$, respectively. In \eqref{vpardot}, $\dot{v}_\parallel$ has been separated into a term coming from the Hamiltonian, $\dot{v}^H_\parallel = \{v_\parallel,H_s\}$, and another term proportional to the inductive component of the parallel electric field, $(q_s/m_s)\partial {A_\parallel}/\partial{t}$. In the absence of collisions and sources, \eqref{liouville} can be identified as a Liouville equation demonstrating that the distribution function is conserved along the nonlinear phase space characteristics. Accordingly, the gyrokinetic equation can be recast in the following conservative form,
\begin{eqnal}
\pderiv{(\mathcal{J}f_s)}{t} &+ \nabla\v{\cdot}( \mathcal{J} \dot{\v{R}} f_s) + \pderiv{}{v_\parallel}\left(\mathcal{J}\dot{v}^H_\parallel f_s\right) - \pderiv{}{v_\parallel}\left(\mathcal{J}\frac{q_s}{m_s}\pderiv{A_\parallel}{t} f_s \right) = \mathcal{J} C[f_s] + \mathcal{J} S_s, \label{emgk} 
\end{eqnal}
where $\mathcal{J} = B_\parallel^*$ is the Jacobian of the gyrocentre coordinates and $\uv{b}\v{\cdot}\nabla\times\uv{b}\approx0$ so that $B_\parallel^*\approx B$. The symplectic formulation of electromagnetic gyrokinetics is utilized with the parallel velocity as an independent variable instead of the parallel canonical momentum $p_\parallel$ in the Hamiltonian formulation \cite{brizard2007foundations,hahm1988nonlinear}. This notation is used for convenience to explicitly display the time derivative of $A_\parallel$, which is characteristic of the symplectic formulation of electromagnetic gyrokinetics. The electrostatic potential, $\phi$, is determined by quasi-neutrality,
\begin{equation}
    \sigma_g + \sigma_\text{pol} = \sigma_g - \nabla\v{\cdot}\v{P} = 0,
\end{equation}
with the guiding centre charge density (neglecting gyroaveraging in the long-wavelength limit) given by
\begin{equation}
    \sigma_g = \sum_s q_s \int d\v{w}\ \mathcal{J} f_s.
\end{equation}
Here, $d\v{w}= 2\pi\, m_s^{-1} dv_\parallel\, d\mu = m_s^{-1} dv_\parallel\, d\mu \int d\alpha$ is the gyrocentre velocity-space volume element
$(d{\bf v}=m_s^{-1}dv_\parallel\, d\mu\, d\alpha\, \mathcal{J})$ with the gyroangle $\alpha$ integrated away and the Jacobian factored out (formally, $\mathcal{J}$ should also be included in $d\v{w}$). The polarization vector is then
\begin{eqnal}
    \v{P} &= -\sum_s \int d\v{w}\ \frac{m_s}{B^2}\mathcal{J}f_s \nabla_\perp \phi \approx -\sum_s \frac{m_s n_{0s}}{B^2} \nabla_\perp \phi \equiv - \epsilon_\perp \nabla_\perp \phi,
\end{eqnal}
where $\nabla_\perp=\nabla-\uv{b}(\uv{b}\v{\cdot}\nabla)$ is the gradient orthogonal to the background magnetic field. A linearized time-independent polarization density, $n_0$, is assumed which is consistent with neglecting a second-order ${\bf E \times B}$ energy term in the Hamiltonian. Such an approximation in the SOL is questionable due to the presence of large density fluctuations, although a linearized polarization density is commonly used in full-$f$ gyrokinetic simulations for computational efficiency and reflective of common numerical modelling practices \cite{ku2018fast,shi2019full,mandell_hakim_hammett_francisquez_2020}. Adding the nonlinear polarization density along with the second-order ${\bf E \times B}$ energy term in the Hamiltonian are improvements kept for future work. Consequently, the quasi-neutrality condition can be rewritten as the long-wavelength gyrokinetic Poisson equation,
\begin{equation}
    -\nabla \v{\cdot} \sum_s \frac{m_s n_{0s}}{B^2} \nabla_\perp \phi = \sum_s q_s \int d\v{w}\ \mathcal{J}f_s, \label{poisson}
\end{equation}
where, even in the long-wavelength limit with no gyroaveraging, the first-order polarization charge density on the left-hand side of \eqref{poisson} incorporates some finite Larmor radius (FLR) or $k_\perp$ effects in its calculation. It is worth emphasizing that this “long-wavelength” limit is a valid limit of the full-$f$ gyrokinetic derivation since care was taken to include the guiding-center components of the field perturbations at $O(1)$. Further, although one may think of this as a drift-kinetic limit, the presence of the linearized ion polarization term in the quasineutrality equation distinguishes the long-wavelength gyrokinetic model from versions of drift-kinetics that, for example, include the polarization drift in the equations of motion.
The parallel magnetic vector potential, $A_\parallel$, is determined by the parallel Amp\`ere equation,
\begin{equation}
    -\nabla_\perp^2 A_\parallel = \mu_0 \sum_s q_s m_s \int  v_\parallel \mathcal{J} f_s\,d\v{w}. \label{ampere1}
\end{equation}
Note that one can also take the time derivative of this equation to get a generalized Ohm's law which can be solved directly for $\pderivInline{A_\parallel}{t}$, the inductive component of the parallel electric field $E_\parallel$ \cite{reynders1993gyrokinetic, cummings1994gyrokinetic, chen2001gyrokinetic}:
\begin{equation}
    -\nabla_\perp^2 \pderiv{A_\parallel}{t} = \mu_0 \sum_s q_s m_s \int v_\parallel \pderiv{(\mathcal{J} f_s)}{t}\, d\v{w}.
\end{equation}
Writing the gyrokinetic equation as
\begin{equation}
    \pderiv{(\mathcal{J}f_s)}{t} = 
    \pderiv{(\mathcal{J}f_s)}{t}^\star + \pderiv{}{v_\parallel}\left(\mathcal{J} \frac{q_s}{m_s}\pderiv{A_\parallel}{t} f_s\right), \label{fstar}
\end{equation}
where $\partial{(\mathcal{J}f_s)^\star}/\partial{t}$ denotes all terms in the gyrokinetic equation (including sources and collisions) except $\pderivInline{A_\parallel}{t}$, Ohm's law can be, after integration by parts, rewritten 
\begin{equation}
    \left(-\nabla_\perp^2 + \sum_s \mu_0 q_s^2 \int\mathcal{J} f_s\, d\v{w}\right) \pderiv{A_\parallel}{t} 
    \notag
    = \mu_0 \sum_s q_s m_s \int v_\parallel \pderiv{(\mathcal{J}f_s)}{t}^\star\,d\v{w}. \label{ohmstar}
\end{equation}

To model collisions, the code uses a conservative Lenard--Bernstein (or Dougherty) operator \cite{Lenard1958plasma,Dougherty1964model,francisquez2020conservative},

\begin{eqnal} \label{eq:GkLBOEq}
\mathcal{J}C[f_s] &= \nu\left\lbrace\pderiv{}{v_\parallel}\left[\left(v_\parallel - u_\parallel\right)\mathcal{J} f_s+v_{th,s}^2\pderiv{(\mathcal{J} f_s)}{v_\parallel}\right]\right.\\&\left.\quad +\pderiv{}{\mu}\left[2\mu \mathcal{J} f_s+2\mu\frac{m_s}{B}v_{th,s}^2\pderiv{(\mathcal{J} f_s)}{\mu}\right]\right\rbrace,
\end{eqnal}
where $n_s u_\parallel^2 = \int d\v{w} \mathcal{J}v_\parallel^2f_s$, $3 n_s v_{th,s}^2 = 2\int d\v{w} \mathcal{J}\mu Bf_s/m_s$, $n_s u_\parallel = \int d\v{w} \mathcal{J} v_\parallel f_s$, $n_s = \int d\v{w} \mathcal{J} f_s$, and $T_s = m_s v_{th,s}^2$. This collision operator contains the effects of drag and pitch-angle scattering, and it conserves number, momentum, and energy density. Consistent with the present long-wavelength treatment of the gyrokinetic system, FLR effects are ignored. In this work, both like-species and cross-species collisions among electrons and ions are included. The collision frequency $\nu$ is kept velocity-independent, i.e. $\nu\neq\nu(v)$. Further details about this collision operator, including its conservation properties and discretization, can be found in \cite{francisquez2020conservative}.

To clarify the approximations undertaken in deriving the gyrokinetic model formulated above and its consequent effects on turbulent fields, the key assumptions are reviewed: The orderings in gyrokinetic theory that effectively reduce the full phase space's dimensionality are ${\omega}/{\Omega_s} \ll 1$ and $k_\perp/k_\parallel \gg 1$. These orderings express that the charged particle gyrofrequency ($\Omega_s = q_s B/m_s$) in a magnetic field is far greater than the characteristic frequencies of interest ($\omega$) with perpendicular wavenumbers ($k_\perp$) of Fourier modes being far larger than parallel wavenumbers ($k_\parallel$). Such properties are generally expected and observed for drift-wave turbulence in magnetically confined fusion plasmas where $\omega \ll \Omega_s$ \cite{Zweben_2007}. An additional “weak-flow” ordering \cite{Dimits_1992,Parra_Catto_2008,Dimits_2012} is applied where $v_{\bf E \times B}/v_{th,s} \approx k_\perp \rho_s q_s \phi/T_s \ll 1$ which allows for large amplitude perturbations. This approximation is also generally valid for electrons and deuterium ions in edge tokamak plasmas \cite{Belli_2018} as it assumes ${\bf E \times B}$ flows are far smaller than the thermal velocity, $v_{th,s}$, as observed in experiment \cite{blobs_review,Zweben_2015}. By constraining gradients of $\phi$ instead of $\phi$ itself, this weak-flow ordering simultaneously allows perturbations of order unity ($q_s \phi/T_s \sim 1$) at long wavelengths ($k_\perp \rho_s \ll 1$) and small perturbations ($q_s \phi/T_s \ll 1$) at short wavelengths ($k_\perp \rho_s \sim 1$) along with perturbations at intermediate scales. Alternatively, this approximation can be intuitively viewed as the potential energy variation around a gyro-orbit being small compared to the kinetic energy, $q_s\phi(\v{R} + {\bm \rho_s}) - q_s\phi(\v{R}) \approx q_s {\bm \rho_s} \cdot \nabla_\perp \phi \ll T_s$ \cite{mandell_thesis}. Here ${\bm \rho_s}$ is the gyroradius vector which points from the center of the gyro-orbit $\v{R}$ to the particle location $\v{x}$. To ensure consistency in the gyrokinetic model at higher order (although the guiding-centre limit is eventually taken in the continuum simulations, i.e. $k_\perp \rho_s \ll 1$), a strong-gradient ordering is also employed which assumes the background magnetic field varies slowly relative to edge profiles \cite{Zweben_2007}. As noted above, the long-wavelength limit is taken and variations of $\phi$ on the scale of the gyroradius is neglected. This yields guiding-center equations of motion which do not contain gyroaveraging operations. While extensions to a more complete gyrokinetic model are in progress, these contemporary modelling limitations are worth noting for the scope of the present results. In accordance with \cite{mandell_hakim_hammett_francisquez_2020}, the gyrokinetic turbulence is simulated on helical, open field lines as a rough model of the tokamak SOL at NSTX-like parameters. 

A field-aligned geometry \cite{beer1995field} is employed for numerical modelling whereby $x$ is the radial coordinate, $z$ is the coordinate parallel to the field lines, and $y$ is the binormal coordinate which labels field lines at constant $x$ and $z$. These coordinates map to physical cylindrical coordinates ($R,\varphi,Z)$ via $R=x$, $\varphi=(y/\sin\theta+z\cos\theta)/R_c$, $Z=z\sin\theta$. The field-line pitch $\sin\theta=B_v/B$ is taken to be constant, with $B_v$ the vertical component of the magnetic field (analogous to the poloidal field in tokamaks), and $B$ the total magnitude of the background magnetic field. The open field lines strike material walls at the top and bottom of the domain consistent with the simple magnetized torus configuration studied experimentally via devices such as the Helimak \cite{gentle2008} and TORPEX \cite{fasoli2006}. This system without magnetic shear contains unfavorable magnetic curvature producing the interchange instability that drives edge turbulence. There is no good curvature region to produce conventional ballooning-mode structure in the current setup. Further, $R_c=R_0+a$ is the radius of curvature at the centre of the simulation domain, with $R_0$ the major radius and $a$ the minor radius. This geometry is equivalent to the one in Chapter 2, with curvature operator

\begin{equation}
    (\nabla\times\uv{b})\v{\cdot}\nabla f(x,y,z)\approx \left[(\nabla\times\uv{b})\v{\cdot} \nabla y\right]\pderiv{f}{y} =-\frac{1}{x}\pderiv{f}{y} \label{eq:GkCurv},
\end{equation}
where $\v{B}=B_\text{axis}(R_0/R)\uv{e}_z$ in the last step and $B_\text{axis}$ is the magnetic field strength at the magnetic axis.
This geometry represents a {flux-tube-like domain} on the outboard strictly bad curvature side that wraps helically around the torus and terminates on conducting plates at each end in $z$. {Note that although the simulation is on a flux-tube-like domain, it is not performed in the local limit commonly applied in $\delta f$ gyrokinetic codes; instead, the simulations are effectively global as they include radial variation of the magnetic field and kinetic profiles.} The simulation box is centred at $(x,y,z)=(R_c,0,0)$ with dimensions $L_x=56\rho_{i0}\approx 16.6$ cm, $L_y=100\rho_{i0}\approx 29.1$ cm, and $L_z=L_p/\sin\theta=8.0$ m, where $\rho_{i0} = \sqrt{m_i T_{i0}}/q_i B_0$ and $L_p=2.4$ m approximates the vertical height of the SOL. The velocity-space grid has extents $-4v_{th,s0}\leq v_\parallel \leq 4 v_{th,s0}$ and $0\leq\mu\leq6T_{s0}/B_0$, where $v_{th,s0}=\sqrt{T_{s0}/m_s}$ and $B_0=B_\text{axis}R_0/R_c$. The low-$\beta$ simulations presented here use piecewise-linear ($p=1$) basis functions, with $(N_x,N_y,N_z,N_{v_\parallel},N_\mu)=(32,64,16,10,5)$ the number of cells in each dimension. At high-$\beta$, due to the increased computational cost, the resolution is lowered to $(N_x,N_y,N_z,N_{v_\parallel},N_\mu)=(16,32,14,10,5)$. For $p=1$, one should double these numbers to obtain the equivalent number of grid points for comparison with standard grid-based gyrokinetic codes.

The radial boundary conditions model conducting walls at the radial ends of the domain, given by the Dirichlet boundary condition $\phi=A_\parallel=0$. The condition $\phi=0$ effectively prevents ${\bf E \times B}$ flows into walls, while $A_\parallel=0$ makes it so that (perturbed) field lines never intersect the walls. For the latter, one can think of image currents in the conducting wall that mirror currents in the domain, resulting in exact cancellation of the perpendicular magnetic fluctuations at the wall. Also, the magnetic curvature and $\nabla B$ drifts do not have a radial component in this helical magnetic geometry. These boundary conditions on the fields are thus sufficient to guarantee that there is no flux of the distribution function into the radial walls. Conducting-sheath boundary conditions are applied in the $z$-direction \cite{shi2017gyrokinetic,shi2019full} to model the Debye sheath (the dynamics of which is beyond the gyrokinetic ordering), which partially reflects one species (typically electrons) and fully absorbs the other species depending on the sign of the sheath potential. This involves solving the gyrokinetic Poisson equation for $\phi$ at the $z$-boundary (i.e. sheath entrance), and using the resulting sheath potential to determine a cutoff velocity below which particles (typically low energy electrons) are reflected by the sheath. {Notably, this boundary condition allows current fluctuations in and out of the sheath. This differs from the standard logical sheath boundary condition \cite{parker1993suitable} which imposes zero net current to the sheath by assuming ion and electron currents at the sheath entrance are equal at all times.} The fields do not require a boundary condition in the $z$-direction since only perpendicular derivatives appear in the field equations. The simulations are carried out in a sheath-limited regime but there can be electrical disconnection from the plasma sheath if the Alfv\'en speed is slow enough. Periodic boundary conditions are used in the $y$-direction. 

The simulation parameters roughly approximate an H-mode deuterium plasma in the NSTX SOL: $B_\text{axis}=0.5$ T, $R_0=0.85$ m, $a=0.5$ m, $T_{e0}=T_{i0}=40$ eV. To model particles and heat from the core crossing the separatrix, a non-drifting Maxwellian source of ions and electrons is applied,
\begin{equation}
    S_{s} = \frac{n_S(x,z)}{(2\pi T_S/m_{s})^{3/2}}\exp\left(-\frac{m_{s} v^2}{2 T_S}\right),
\end{equation}
with source temperature $T_{S}=70$ eV for both species and $v^2 = v_\parallel^2 + 2 \mu B/m_s$. The source density is given by
\begin{equation}
    n_S(x,z) = \begin{cases}
    S_0\exp\left(\frac{-(x-x_S)^2}{(2\lambda_S)^2}\right)\qquad \qquad &|z|<L_z/4\\
    0 \qquad &\mathrm{otherwise}
    \end{cases}
\end{equation}
so that $x_S-3\lambda_S < x < x_S+3\lambda_S$ delimits the source region, and the code sets $x_S=1.3$ m and $\lambda_S=0.005$ m. This localized particle source structure in $z$-space results in plasma ballooning out largely in the centre of the magnetic field line. The source particle rate $S_0$ is chosen so that the total (ion plus electron) source power matches the desired power into the simulation domain, $P_\mathrm{src}$. Since \texttt{Gkeyll} simulates a flux-tube-like fraction of the whole SOL domain, $P_\mathrm{src}$ is related to the total SOL power, $P_\mathrm{SOL}$, by $P_\mathrm{src} = P_\mathrm{SOL}L_y L_z/(2\pi R_c L_\mathrm{pol})\approx0.115 P_\mathrm{SOL}$. Amplitudes are adjusted to approximate $P_\mathrm{SOL}=5.4$ MW crossing into these open flux surfaces at low-$\beta$ conditions relevant to NSTX \cite{Zweben_2015}. An artificially elevated density case with $P_\mathrm{SOL}=54.0$ MW is also tested to study edge turbulence at high-$\beta$. The collision frequency is comparable in magnitude to the inverse autocorrelation time of electron density fluctuations at low-$\beta$. For the high-$\beta$ case, $\nu$ is found to be about $10 \times$ larger, and the plasma thus sits in a strongly collisional regime. Simulations reach a quasi-steady state with the sources balanced by end losses to the sheath, although there is no neutral recycling included yet which is a focus of ongoing work \cite{bernard2020a}. Unlike in \cite{shi2019full}, no numerical heating nor source floors are applied in the algorithm to ensure positivity.

In summary, the full-$f$ nonlinear electromagnetic gyrokinetic turbulence simulations of the NSTX plasma boundary region employ the lowest-order, i.e. guiding-center or drift-kinetic limit, of the system. Implementing gyroaveraging effects given by the next order terms in advanced geometries is the focus of future work \cite{noah_private}. These present approximations in modern full-$f$ global gyrokinetic simulations should be kept in mind when attributing any similarities or differences to two-fluid theory in the next sections since the gyrokinetic formulation can itself be improved. The turbulence simulations presented are thus a reflection of the current forefront of numerical modelling. A full exposition of the derivation and benchmarking including the energy-conserving discontinuous Galerkin scheme applied in \texttt{Gkeyll} for the discretization of the gyrokinetic system in 5-dimensional phase space along with explicit time-stepping and avoidance of the Amp\`ere cancellation problem is found in \cite{mandell_thesis}. 

\section{\label{sec:level3.2}Machine learning fluid theory (again)}

A vital goal in computational plasma physics is determining the minimal complexity necessary in a theory to sufficiently represent observations of interest for predictive fusion reactor simulations. Convection-diffusion equations with effective mean transport coefficients are widely utilized \cite{Reiter_1991,SOLPS_DEKEYSER2019125} but insufficient in capturing edge plasma turbulence where scale separation between the equilibrium and fluctuations is not justified \cite{NAULIN2007,mandell_thesis}. Other reduced models such as classical magnetohydrodynamics are unable to resolve electron and ion temperature dynamics with differing energy transport mechanisms in the edge \cite{TeTi_2008,TeTi_2009,TeTi_2016}. Following the framework set forth in Chapter 2 (with few small differences noted here), Chapter 3 considers the widely used first-principles-based two-fluid drift-reduced Braginskii equations \cite{Braginskii1965RvPP,francisquez_thesis} in the electrostatic limit relevant to low-$\beta$ conditions in the edge of fusion experiments \cite{Zweben_2015} for comparison to gyrokinetic modelling \cite{mandell_hakim_hammett_francisquez_2020}. Drift-reduced Braginskii theory is also a full-$f$ \cite{Belli_2008,Full-F,Held_2020} nonlinear model but evaluated in the fluid limit. To align coordinates with the gyrokinetic plasma, one adjustment here is that $\bhatZ = +{\bf \hat{z}}$ instead of $\bhatZ = -{\bf \hat{z}}$ as in Chapter 2. The 3-dimensional shearless field-aligned coordinate system over which the fluid equations are formulated in the physics-informed machine learning framework thus exactly matches the gyrokinetic code's geometry. 

To consistently calculate the electric field response and integrate \eqref{eq:nDotGDBH}--\eqref{eq:TiDotGDBH} in time, classically one would compute the turbulent $\phi$ in drift-reduced Braginskii theory by directly invoking quasineutrality and numerically solving the boundary value problem given by the divergence-free condition of the electric current density \cite{DRB_consistency3,Zholobenko_2021}. For the purposes of comparison with gyrokinetic modelling, this novel technique can compute the turbulent electric field consistent with the drift-reduced Braginskii equations without explicitly evaluating $\nabla \cdot \v{j} = 0$ nor directly applying the Boussinesq approximation. Namely, this work applies the validated physics-informed deep learning framework from Chapter 2 to infer the gauge-invariant $\phi$ directly from \eqref{eq:nDotGDBH} and \eqref{eq:TeDotGDBH} for direct analysis of electron pressure and electric field fluctuations in nonlinear global electromagnetic gyrokinetic simulations and electrostatic two-fluid theory. The only existing requirement on the observational data in this deep learning framework is that the measurements' spatial and temporal resolutions should be be finer than the autocorrelation length and time, respectively, of the turbulence structures being analyzed. This scale condition is well-satisfied for the low-$\beta$ case and marginally-satisfied for the high-$\beta$ data analyzed. The set of collocation points over which the partial differential equations are evaluated correspond to the positions of the observed electron pressure data, i.e. $\lbrace x_0^i,y_0^i,z_0^i,t_0^i\rbrace^{N_0}_{i=1} = \lbrace x_f^i,y_f^i,z_f^i,t_f^i\rbrace^{N_f}_{i=1}$ once again. One difference is that loss functions are optimized with mini-batch sampling using $N_0 = N_f = 1000$ using just L-BFGS \cite{10.5555/3112655.3112866} for 20 hours over 32 cores on Intel Haswell-EP processors. The only locally observed dynamical quantities in these equations are 2-dimensional views of $n_e$ and $T_e$ (to emulate gas puff imaging \cite{mathews2022deep}) without any explicit information about boundary conditions nor initializations nor ion temperature dynamics nor parallel flows. The collocation grid consists of a continuous spatiotemporal domain without time-stepping nor finite difference schema in contrast with standard numerical codes. All analytic terms encoded in these equations including high-order operators are computed exactly by the neural networks without discretization. In that sense, it is a potentially higher fidelity continuous representation of the continuum equations. While the linearized polarization density---analogous to the Boussinesq approximation---is employed in the gyrokinetic simulations, no such approximations are explicitly applied by the neural networks.

With the availability of measurements often sparse in fusion experiments, designing diagnostic techniques for validating turbulence theories with limited information is important. On this point, it is noteworthy that this framework can potentially be adapted to experimental measurements of electron density and temperature \cite{griener_continuous_2020,Bowman_2020,Mathews2020,mathews2022deep}. To handle the particular case of 2-dimensional turbulence data, one essentially assumes slow variation of dynamics in the $z$-coordinate and effectively set all parallel derivatives to zero. In computational theory comparisons where no such limitations exist, training on $n_e$ and $T_e$ in 3-dimensional space over long macroscopic timescales can be easily performed via segmentation of the domain and parallelization, but a limited spatial view away from numerical boundaries with reduced dimensionality is taken to mirror experimental conditions for field-aligned observations \cite{Zweben_2017,mathews2022deep} and fundamentally test what information is indispensable to compare kinetic and fluid turbulence. To compare the gyrokinetic and two-fluid theories as directly as possible, the toroidal simulations are analyzed at $z = L_z/3$ where no applied sources are present. Further, beyond the inclusion of appropriate collisional drifts and sources, this technique is generalizable to boundary plasmas with multiple ions and impurities present due to the quasi-neutrality assumptions underlying the two-fluid theory \cite{multi_species}. 
 
\section{\label{sec:level3.3}Quantifying plasma turbulence consistency}

\begin{figure*}[ht]
\includegraphics[width=1.0\linewidth]{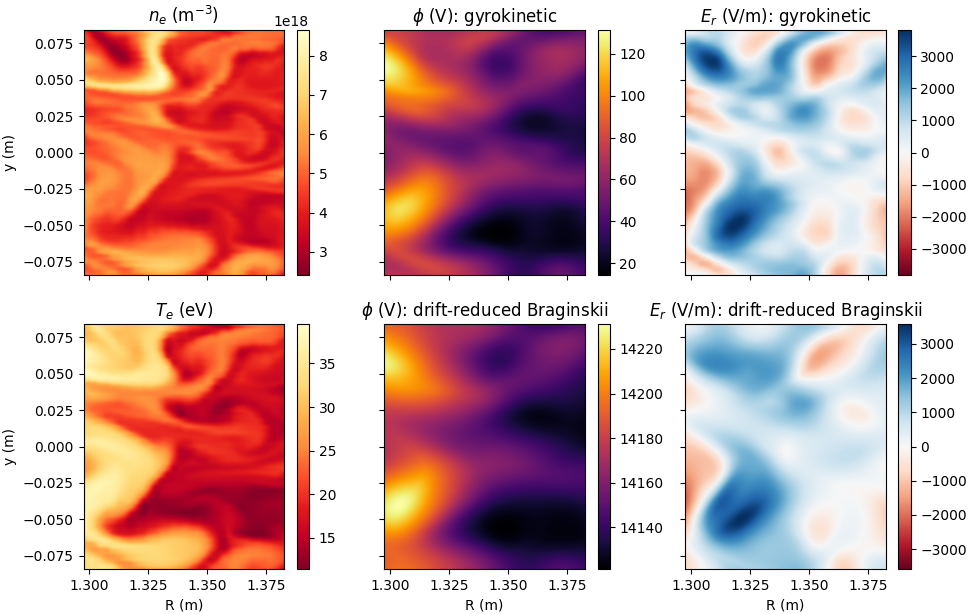}
\caption{\label{PlasmaPINN_lowbeta}The turbulent electric potential, $\phi$ (a gauge-invariant quantity which is equivalent up to a scalar constant offset), and radial electric field, $E_r$, concomitant with electron pressure fluctuations as predicted by electrostatic drift-reduced Braginskii theory and electromagnetic gyrokinetic modelling in low-$\beta$ conditions are in good quantitative agreement. The two-fluid theory's $\phi$ and $E_r$ are based upon the training the physics-informed deep learning framework while the gyrokinetic results are from the discontinuous Galerkin numerical solver.}
\end{figure*}

\begin{figure}[ht]
\includegraphics[width=0.9\linewidth]{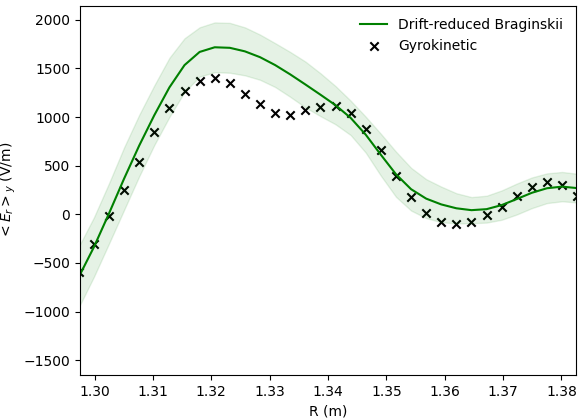}
\caption{\label{PlasmaPINN_lowbeta_avg}The $y$-averaged turbulent radial electric field, $\langle E_r\rangle_y$, as predicted by electrostatic drift-reduced Braginskii theory and electromagnetic gyrokinetic modelling at low-$\beta$. The results plotted for the drift-reduced Braginskii output here are based upon collecting 100 independently-trained physics-informed neural networks.}
\end{figure}

A defining characteristic of a nonlinear theory is how it mathematically connects dynamical variables. The focus of this work is quantitatively examining the nonlinear relationship extant in plasma turbulence models between electron pressure and electric field fluctuations. As outlined in Section \ref{sec:level3.2}, using a custom physics-informed deep learning framework whereby the drift-reduced Braginskii equations are embedded in the neural networks via constraints in the form of implicit partial differential equations, this thesis demonstrates the first ever direct comparisons of instantaneous turbulent fields between electrostatic two-fluid theory and electromagnetic long-wavelength gyrokinetic modelling in low-$\beta$ helical plasmas with results visualized in Figure \ref{PlasmaPINN_lowbeta}. This multi-network physics-informed deep learning framework enables direct comparison of drift-reduced Braginskii theory with gyrokinetics and bypasses demanding limitations in standard forward modelling numerical codes which must precisely align initial conditions, physical sources, numerical diffusion, and boundary constraints (e.g. particle and heat fluxes into walls) in both gyrokinetic and fluid representations when classically attempting comparisons of turbulence simulations and statistics. Further, the theoretical and numerical conservation properties of these simulations ordinarily need to be evaluated which can be a significant challenge especially when employing disparate numerical methods with differing discretization schema and integration accuracy altogether \cite{francisquez2020fluid}. This framework overcomes these hurdles by training on partial electron pressure observations simultaneously with the plasma theory sought for comparison. Figure \ref{PlasmaPINN_lowbeta} specifically shows that the turbulent electric field predicted by drift-reduced Braginskii two-fluid theory is largely consistent with long-wavelength gyrokinetic modelling in low-$\beta$ helical plasmas. The consistency is also evident if analyzing $y$-averaged radial electric field fluctuations and accounting for the inherent scatter ($\sigma_{PINN}$) from the stochastic optimization employed as displayed in Figure \ref{PlasmaPINN_lowbeta_avg}. To clarify the origins of $\sigma_{PINN}$, every time this physics-informed deep learning framework is trained from scratch against the electron pressure observations, the learned turbulent electric field will be slightly different due to the random initialization of weights and biases for the networks along with mini-batch sampling during training. To account for this stochasticity, the framework is trained anew 100 times while collecting the predicted turbulent $E_r$ after each optimization. the quantity $\sigma_{PINN}$ corresponds to the standard deviation from this collection. The two-fluid model's results displayed in Figure \ref{PlasmaPINN_lowbeta_avg} are thus based upon computing $\langle E_r\rangle_y$ from 100 independently-trained physics-informed machine learning frameworks. Their turbulent outputs are averaged together to produce the mean, while the scatter inherently associated with the 100 realizations---which approximately follows a normal distribution---comprises the shaded uncertainty interval spanning 2 standard deviations. As visualized, the $\langle E_r\rangle_y$ profiles predicted by the the electromagnetic gyrokinetic simulation and electrostatic drift-reduced Braginskii model are generally in agreement within error bounds at low-$\beta$.

\begin{figure*}[ht]
\includegraphics[width=1.0\linewidth]{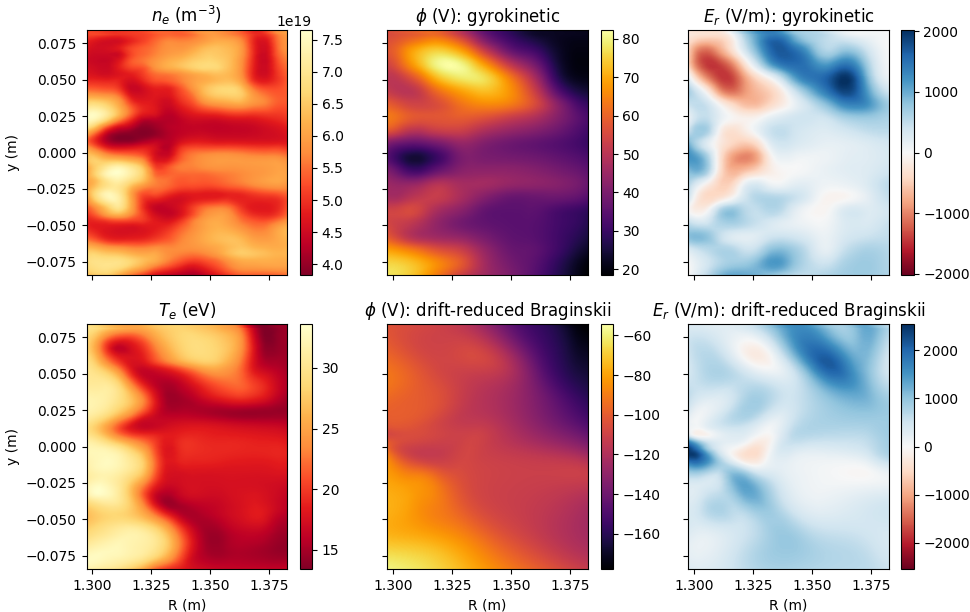}
\caption{\label{PlasmaPINN_highbeta}The turbulent electric potential, $\phi$ (a gauge-invariant quantity which is equivalent up to a scalar constant offset), and radial electric field, $E_r$, concomitant with electron pressure fluctuations as predicted by electrostatic drift-reduced Braginskii theory and electromagnetic gyrokinetic modelling in high-$\beta$ conditions are quantitatively inconsistent. The two-fluid theory's $\phi$ and $E_r$ are based upon the training of the physics-informed deep learning framework while the gyrokinetic results are from the discontinuous Galerkin numerical solver.}
\end{figure*}
\begin{figure}[ht]
\includegraphics[width=0.9\linewidth]{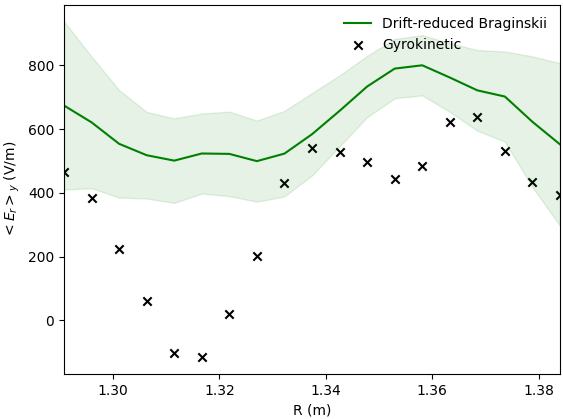}
\caption{\label{PlasmaPINN_highbeta_avg}The $y$-averaged turbulent radial electric field, $\langle E_r\rangle_y$, as predicted by electrostatic drift-reduced Braginskii theory and electromagnetic gyrokinetic modelling at high-$\beta$. The results plotted for the drift-reduced Braginskii output here are based upon collecting 100 independently-trained physics-informed neural networks.}
\end{figure}

In high-$\beta$ conditions where the particle source is artificially increased by 10$\times$, electromagnetic effects become important and the electrostatic two-fluid theory is inconsistent with electromagnetic gyrokinetic simulations as displayed by Figure \ref{PlasmaPINN_highbeta}. As remarked above, multiple realizations are conducted to analyze the sample statistics of the learned turbulent fields consistent with drift-reduced Braginskii two-fluid theory based solely upon the intrinsic scatter during training to account for the stochastic nature of the optimization. By collecting 100 independently-trained realizations, the uncertainty linked to this intrinsic scatter can be evaluated as demonstrated in Figure \ref{PlasmaPINN_highbeta_avg}. These discrepancies indicate that fluctuations in $\bf B_\perp$, which are evaluated by solving the parallel Amp\`ere equation (or, equivalently, generalized Ohm's law), cannot be neglected when considering plasma transport across the inhomogeneous background magnetic field as in electrostatic theory. 

\begin{figure}
\includegraphics[width=1.0\linewidth]{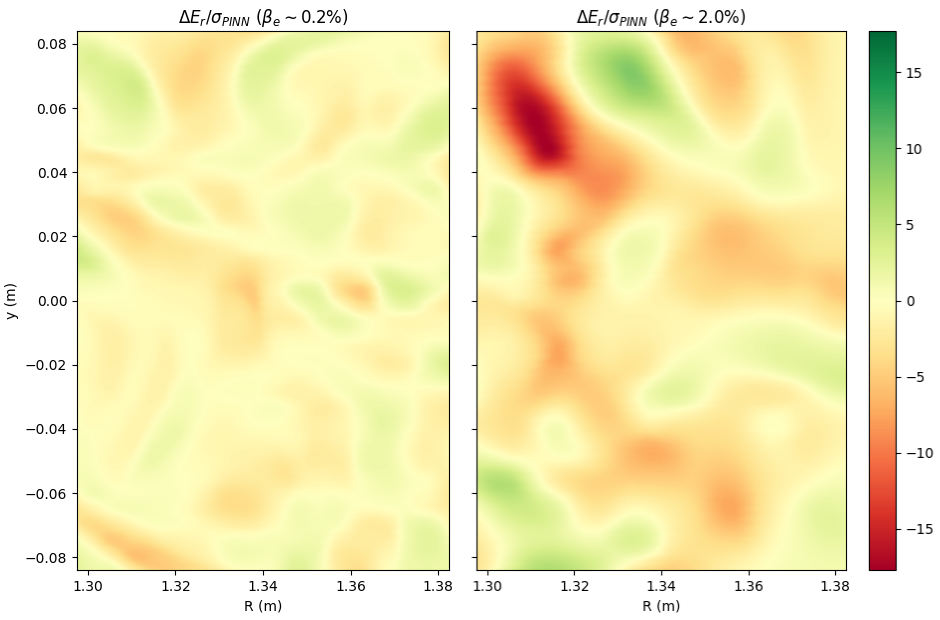}
\caption{\label{PlasmaPINN_inst_err}The relative error in the instantaneous radial electric field fluctuations ($\Delta E_r / \sigma_{PINN}$) between electrostatic drift-reduced Braginskii theory and electromagnetic long-wavelength gyrokinetic modelling is displayed in low-$\beta$ (top) and high-$\beta$ (bottom) conditions. While all errors in the low-$\beta$ scenario are generally within approximately 3--4 standard deviations and representative of mostly good quantitative agreement, one must go to over $15\sigma_{PINN}$ to fully account for the turbulent fields using an electrostatic two-fluid theory at $\beta_e \sim 2\%$. The results are based upon collecting 100 independently-trained physics-informed neural networks to compute the turbulent $E_r$ and the intrinsic scatter in these predictions.}
\end{figure}

While the fluid approximation is generally expected to be increasingly accurate at high density due to strong coupling between electrons and ions, these results underline the importance of electromagnetic effects even in shear-free high-$\beta$ plasmas as found in planetary magnetospheres and fusion experiments (e.g. dipole confinement \cite{LDX-experiment}), and for the first time enables the degree of error between instantaneous fluctuations to be precisely quantified across turbulence models. Uncertainty estimates stemming from the stochastic framework in both regimes are reflected in Figure \ref{PlasmaPINN_inst_err}. 
One should note that there are novel and different levels of errors to be mindful of in this evaluation. For example, poor convergence arising from nonuniqueness of the turbulent $E_r$ found during optimization against the drift-reduced Braginskii equations, or $\mathcal{L}_{f_{n_e}}$ and $\mathcal{L}_{f_{T_e}}$ remaining non-zero (and not below machine precision) even after training \cite{Mathews2021PRE}. These potential errors exist on top of standard approximations in the discontinuous Galerkin numerical scheme representing the underlying gyrokinetic theory such as the implemented Dougherty collision operator. Notwithstanding, when comparing the electrostatic drift-reduced Braginskii theory to electromagnetic long-wavelength gyrokinetic simulations at low-$\beta$, the results represent good consistency in the turbulent electric field and all observed discrepancies are mostly within the stochastic optimization's expected underlying scatter. Alternatively, when analyzing high-$\beta$ conditions, it is observed that the electrostatic two-fluid model cannot explain the turbulent electric field in the electromagnetic gyrokinetic simulations. In particular, $\Delta E_r \gtrsim 15\sigma_{PINN}$ in the bottom plot of Figure \ref{PlasmaPINN_inst_err}, where $\Delta E_r$ is the difference in the instantaneous $E_r$ predicted by two-fluid theory and gyrokinetics. This signals that the two sets of turbulent $E_r$ fluctuations are incompatible at high-$\beta$ conditions and precisely quantifies the separation in the models' predictions. 

Our multi-network physics-informed deep learning framework demonstrates the suitability of electrostatic two-fluid theory as a good approximation of turbulent electric fields in modern gyrokinetic simulations for low-$\beta$ helical plasmas with sufficient initial and boundary conditions. Conversely, the electrostatic turbulence model is demonstrably insufficient for high-$\beta$ plasmas. This finding is indicative of the importance of including electromagnetic effects such as magnetic flutter in determining cross-field transport even at $\beta_e \sim 2\%$. But field line perturbations and the reduced numerical representation of gyrokinetic theory are not the only effects at play causing mismatch at high-$\beta$: due to the nature of the strong localized particle source in $z$-space to produce high-$\beta$ conditions, parallel dynamics including electron flows along field lines and Ohmic heating effects become increasingly important. These variables should now be observed or learnt first from the 2-dimensional electron pressure measurements to accurately reconstruct the turbulent electric field which signifies a departure from the expected low-$\beta$ edge of tokamaks. Going forward, consideration of advanced magnetic geometries with squeezing and shearing of fusion plasmas near null-points, which may even couple-decouple upstream turbulence in tokamaks, will be important in the global validation of reduced turbulence models with realistic shaping \cite{Ryutov_geometry,Kuang2018APS,kuang_thesis,Nespoli_2020,Myra_2020}. Also, while there is good convergence in the low-$\beta$ gyrokinetic simulations at parameters relevant to the edge of NSTX, numerical convergence in the artificially elevated high-$\beta$ case is currently questionable and a potential source of discrepancy. Running the high-$\beta$ gyrokinetic simulation with proper collision frequency at an improved resolution of $(N_x,N_y,N_z,N_{v_\parallel},N_\mu)=(48,96,18,10,5)$ would cost $\sim$4 million CPU-hours to check and such investigations are left for the future.

As for distinctiveness, the techniques used for computing the displayed turbulent electric fields in the two cases are markedly different. In particular, the long-wavelength gyrokinetic Poisson equation, which is originally derived from the divergence-free condition of the electric current density, is employed in the gyrokinetic simulations. In contrast, simply the electron fluid evolution equations are used to infer the unknown turbulent field fluctuations consistent with drift-reduced Braginskii theory \cite{Mathews2021PRE}. A principle underlying these models is quasineutrality, but this condition is not sufficient on its own. If one were to apply equilibrium models such as the Boltzmann relation or simple ion pressure balance as expected neoclassically, the turbulent electric field estimates for these nonequilibrium plasmas with nontrivial cross-field transport would be highly inaccurate as in Chapter 2. Further, no external knowledge of boundary conditions such as sheath effects are explicitly provided to the physics-informed deep learning framework, but this information implicitly propagates from the walls into the observations of electron pressure. This novel approach resultantly permits using limited 2D measurements to compare a global 3D fluid turbulence model directly against 5D gyrokinetics beyond statistical considerations for characterizing non-diffusive intermittent edge transport \cite{NAULIN2007}. All in all, the agreement between drift-reduced Braginskii theory and gyrokinetics supports its usage for predicting turbulent transport in low-$\beta$ shearless toroidal plasmas, but an analysis of all dynamical variables is required for the full validation of drift-reduced Braginskii theory. Alternatively, when 2-dimensional experimental electron density and temperature measurements are available \cite{Furno2008,mathews2022deep}, this technique can be used to infer $E_r$ and the resulting structure of turbulent fluxes heading downstream. 


\section{\label{sec:level3.4}Conclusion}

To probe the fundamental question of how similar two distinct turbulence models truly are, using a novel technique to analyze electron pressure fluctuations, this chapter has directly demonstrated that there is good agreement in the turbulent electric fields predicted by electrostatic drift-reduced Braginskii theory and electromagnetic long-wavelength gyrokinetic simulations in low-$\beta$ helical plasmas. At $\beta_e \sim 2\%$, the 2-dimensional electrostatic nature of the utilized fluid theory becomes insufficient to explain the microinstability-induced particle and heat transport. Overall, by analyzing the interconnection between dynamical variables in these global full-$f$ models, physics-informed deep learning can quantitatively examine this defining nonlinear characteristic of turbulent physics. In particular, one can now unambiguously discern when agreement exists between multi-field turbulence theories and identify disagreement when incompatibilities exist with just 2-dimensional electron pressure measurements. This machine learning tool can therefore act as a necessary condition to diagnose when reduced turbulence models are unsuitable, or, conversely, iteratively construct and test theories till agreement is found with observations.

While this work focuses on the electric field response to electron pressure, extending the analysis to all dynamical variables (e.g. $T_i, \vpe, \vpi$) for full validation of reduced multi-field turbulence models in a variety of regimes (e.g. collisionality, $\beta$, closed flux surfaces with sheared magnetic field lines) using electromagnetic fluid theory is the subject of future work. Also, since plasma fluctuations can now be directly compared across models, as gyrokinetic codes begin including fully kinetic neutrals, this optimization technique can help validate reduced source models to accurately account for atomic and molecular interactions with plasma turbulence in the edge of fusion reactors \cite{Thrysoe_2018,Thrysoe_2020} since these processes (e.g. ionization, recombination) affect the local electric field. Further progress in the gyrokinetic simulations such as the improved treatment of gyro-averaging effects \cite{brizard2007foundations}, collision operators \cite{francisquez2020conservative}, and advanced geometries \cite{mandell_thesis} will enable better testing and discovery of hybrid reduced theories as well \cite{zhu2021drift}. For example, in diverted reactor configurations, electromagnetic effects become increasingly important for transport near X-points where $\beta_{p} \rightarrow \infty$. A breakdown of Alfv\'{e}n's theorem in these regions can also arise due to the impact of Coulomb collisions and magnetic shear contributing to an enhanced perpendicular resistivity \cite{Myra-X-point} which presents an important test case of non-ideal effects within reduced turbulence models. While this work supports the usage of electrostatic two-fluid modelling, with adequate initial and boundary conditions, over long-wavelength gyrokinetics for low-$\beta$ magnetized plasmas without magnetic shear, a comparison of all dynamical variables beyond the turbulent electric field is required for a full validation of the reduced model. Further investigations into reactor conditions may suggest the modularization of individually validated fluid-kinetic turbulence models across different regions in integrated global device modelling efforts \cite{hakim2019discontinuous,Merlo2021_XGC_GENE}. This task can now be efficiently tackled through pathways in deep learning as demonstrated by this new class of validation techniques. In addition, precisely understanding the fundamental factors--both physical and numerical--determining the prediction interval, $\sigma_{PINN}$, is the subject of ongoing research in analyzing the nature (e.g. uniqueness, smoothness) of chaotic solutions to fluid turbulence equations and the chaotic convergence properties of physics-informed neural networks. 
\chapter{Plasma and neutral fluctuations from gas puff turbulence imaging in the SOL of Alcator C-Mod}

\setlength{\epigraphwidth}{0.85\textwidth}
\epigraph{Melody and harmony are like lines and colors in pictures. A simple linear picture may be completely beautiful; the introduction of color may make it vague and insignificant. Yet color may, by combination with lines, create great pictures, so long as it does not smother and destroy their value.}{Rabindranath Tagore, interviewed by Albert Einstein}



Diagnosing edge plasmas is an essential task towards testing turbulence models and better understanding plasma fueling and confinement in fusion devices. Gas puff imaging (GPI) of turbulence is a widely applied experimental diagnostic that captures line emission based upon the interaction of neutrals with the hot plasma. As a technique with decades of application in a range of settings \cite{Zweben_2017}, optical imaging of fluctuations provides a view of plasma turbulence. This transport is critical to the operation and energy gain of nuclear fusion reactors, but interpretation (e.g. velocimetry \cite{velocimetry_GPI}) of these fluctuations to directly test reduced physics models is not always straightforward. By tracing the atomic theory underlying the nonlinear dynamics of observed HeI line emission, this chapter outlines a novel spectroscopic method for quantifying the 2-dimensional electron pressure and neutral dynamics on turbulent scales based upon high-resolution visible imaging. This framework is independent of the previous chapters, but will begin enabling the physics-informed deep learning techniques of Chapters 2 and 3 to be translated into experiment.

The plasma edge in magnetic fusion devices is characterized by neighbouring regions: confined plasmas where temperatures can exceed 10$^{6}\text{ K}$, and the colder SOL where gaseous particles may not be completely ionized. These regions exist tightly coupled to one another and inseparable in many respects. Consequently, accounting for neutral transport in conjunction with ion and electron turbulence is essential in wholly analyzing boundary plasma fluctuations. Edge turbulence is characterized by a broadband spectrum with perturbation amplitudes of order unity and frequencies ranging up to 1 MHz. Edge localized modes and intermittent coherent structures convecting across open field lines can be responsible for significant particle losses and plasma-wall interactions that strongly influence operations. To model the vast dynamical scales present in fusion plasmas accordingly requires sufficiently complex modelling techniques. This chapter introduces custom neural architectures within a multi-network deep learning framework that bounds outputs to abide by collisional radiative theory \cite{Fujimoto1979ACM,GOTO_2003} and searches for solutions consistent with continuity constraints on neutral transport \cite{Wersal_2017}. Learning nonlinear physics via optimization in this way provides a new way to examine edge turbulence using 587.6 nm line emission observed by GPI. While the methodology is not fixed to any device, this chapter focuses on 2-dimensional experimental brightness measurements from open flux surfaces on the Alcator C-Mod tokamak \cite{Hutch_CMod, Alcator_Greenwald}, where a good signal-to-noise ratio is found. The validation techniques outlined in Chapters 2 and 3 emphasize the importance of comprehensive diagnostic coverage of electron dynamics on turbulent spatial and temporal scales. To this end, this chapter describes the first calculations of the 2-dimensional turbulent electron density, electron temperature, and neutral density that self-consistently include fluctuation-induced ionization using only observations of the 587.6 nm line via fast camera imaging. This novel turbulence diagnostic analysis paves new ways in systematically measuring edge plasma and neutral fluctuations.

To demonstrate this framework, the present chapter evaluates the validity of collisional radiative theory in conditions relevant to fusion plasmas for atomic helium line emission in Section \ref{sec:level4.1}, overviews the experimental setup for GPI on the Alcator C-Mod tokamak in \ref{sec:level4.2}, outlines a custom physics-informed machine learning optimization technique designed for turbulence imaging in Section \ref{sec:level4.3}, presents results from the analysis applied to experimental fast camera data in section \ref{sec:level4.4}, and concludes with a summary and future outlook in Section \ref{sec:level4.5}.

\section{\label{sec:level4.1}Time-dependent analysis of quantum states in atomic helium}

The electronic transition from $3^3 D$ to $2^3 P$ quantum states in atomic helium results in photon emission with a rest frame wavelength of 587.6 nm. Atomic physics modelling of this line radiation in a plasma correspondingly requires tracking all relevant electron transition pathways that can populate or depopulate $3^3 D$ \cite{Fujimoto1979ACM,GOTO_2003}. The starting point in this analysis is to consider the full rate equations where a quantum state $p$ follows

\begin{figure}[ht]
\centering
\includegraphics[width=0.65\linewidth]{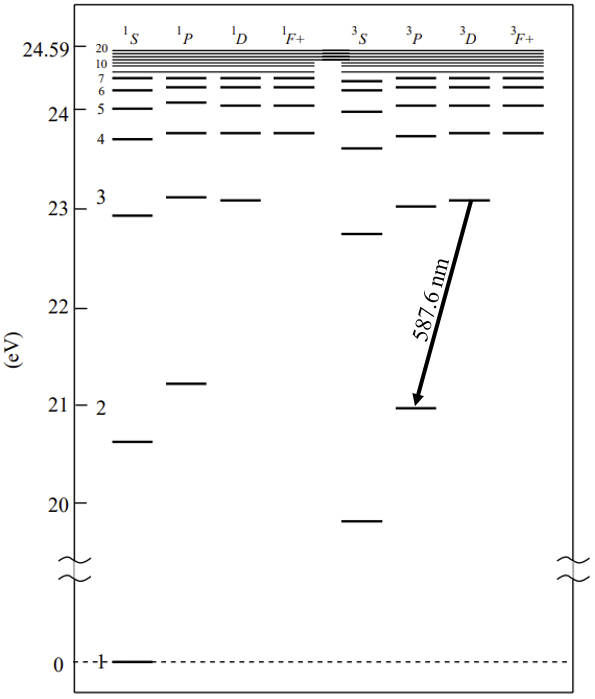}
\caption{\label{HeI_states} Energy level diagram for atomic helium considered in the calculations. An arrow connects $3^3D \rightarrow 2^3P$, which is the origin of the 587.6 nm photon emission. The labels $^{1,3}F+$ denote the quantum states representing all levels with $L \geq 3$. Figure reprinted from \cite{GOTO_2003} with permission from Elsevier.}
\end{figure}

\begin{eqnal}\label{eq:full_rate_eqn}
\frac{dn(p)}{dt} &= \sum\limits_{q \neq p} \lbrace {C(q,p)n_e + A(q,p)} \rbrace n(q) \\&- \lbrace {\sum\limits_{q \neq p} C(p,q)n_e + \sum\limits_{q < p} A(p,q)} + {S(p)n_e }\rbrace n(p) \\&+ \lbrace {\alpha(p)n_e + \beta(p) + \beta_d(p)} \rbrace n_i n_e,
\end{eqnal}

\noindent where $n(p)$ is the population density of the $p = n^{2S + 1} L$ state, in which $n$ is the principal quantum number, $S$ is the spin, and $L$ is the orbital angular momentum quantum number. The notation $q < p$ indicates that the quantum state $q$ lies energetically below $p$. Eq. \eqref{eq:full_rate_eqn} includes the spontaneous transition probability from $p$ to $q$ given by the Einstein A coefficient $A(p,q)$, electron impact transitions $C(p,q)$, electron impact ionization $S(p)$, three-body recombination $\alpha(p)$, radiative recombination $\beta(p)$, and dielectronic recombination $\beta_q(p)$, with $n_e$ and $n_i$ denoting the electron density and hydrogen-like He$^+$ density, respectively. All aforementioned rate coefficients except $A(p,q)$ have a dependence on the electron temperature ($T_e$) that arises from averaging cross-sections over a Maxwellian velocity distribution for electrons, which are based upon calculations with the convergent close-coupling (CCC) \cite{CCC1,CCC2,CCC3} and R-matrix with pseudostates (RMPS) \cite{RMPS} methods using high precision calculations of helium wavefunctions \cite{Drake1,Drake2}. The numerical framework applied follows \cite{GOTO_2003,Zholobenko_thesis} to model atomic helium with a corresponding energy level diagram visualized in Figure \ref{HeI_states}. Quantum states with $L \leq 2$ are resolved for $n < 8$ while states with $L \geq 3$ are bundled together into a single level denoted as ``$F+$''. For $n \geq 8$, $L$ is not resolved, while those with $n \geq 11$ are approximated as hydrogenic levels with statistical weights twice those of hydrogen. All energy levels up to $n=26$ are included with $n \geq 21$ being given by the Saha-Boltzmann equilibrium \cite{McWhirter_1963,Fujimoto1979ACM,Zholobenko_thesis}. 

For application in magnetized plasmas (e.g. tokamaks), where rate coefficients vary with magnetic field strength due to wavefunction mixing, spin-orbit interactions are included to account for mixing between the singlet and triplet fine structure levels \cite{GOTO_2003,Zholobenko_thesis}. Finite magnetic fields largely influence the modelling of metastable species and higher orbital quantum numbers \cite{Zholobenko_2021_validation}. To quantify radiation trapping effects, the dimensionless optical depth for a Doppler-broadened line transition between states $j \rightarrow k$ can be expressed as \cite{Huba2013}

\begin{eqnal}\label{eq:optical_depth}
\tau_{j \rightarrow k} = 5.4 \times 10^{-3} f_{j \rightarrow k} \lambda_{j \rightarrow k} n_j (\mu_j/T_j)^{\frac{1}{2}} L
\end{eqnal}

\noindent where $f_{j \rightarrow k}$ is the absorption oscillator strength, $\lambda_{jk} \ [\text{nm}]$ is the line center wavelength, $\mu_j$ is the mass ratio of the emitting species relative to a proton, $L \ [\text{cm}]$ is the physical depth of the gas along the viewing chord, and $n_j \ [10^{13} \ \text{cm}^{-3}]$ and $T_j \ [\text{eV}]$ are the density and temperature, respectively, of particles in state $j$. For 587.6 nm HeI line emission in conditions relevant to magnetic confinement fusion devices, where $f_{2^3P \rightarrow 3^3D} \sim 0.6$ \cite{HeI_coeff}, $\tau_{2^3P \rightarrow 3^3D} \ll 1$. This.results in the edge being optically thin for spectroscopic analysis of a localized gas puff \cite{Zholobenko_thesis,Zweben_2017}.
The rate equations \eqref{eq:full_rate_eqn} for an optically thin plasma can be equivalently expressed in matrix form as \cite{STOTLER_2007,Zholobenko_thesis}

\begin{eqnal}\label{eq:matrix_full_rate_eqn}
\frac{d{\bf{n}}}{dt} = {\bf{M}}(n_e,T_e) {\bf{n}} + {\bf{\Gamma}}(n_e,T_e,n_i)
\end{eqnal}

\noindent where ${\bf{n}}$ is a vector of the $N$ atomic states, ${\bf{M}}$ represents the $N \times N$ matrix of rates for collisional ionization, excitation, de-excitation, radiative decay, and recombination as above, and ${\bf{\Gamma}}$ symbolizes sources. Since time-evolving every state in atomic helium is computationally expensive, effective atomic physics models known as collisional radiative (CR) theories are often constructed. This involves separating the $N$ states into $P$ and $Q$ spaces of sizes $N_P$ and $N_Q$, respectively, such that \eqref{eq:matrix_full_rate_eqn} becomes

\begin{eqnal}\label{eq:matrix_P_Q}
\frac{d}{dt}
\begin{bmatrix}
   {\bf n_{P}} \\
   {\bf n_{Q}}
\end{bmatrix}
=
\begin{bmatrix}
   {\bf M_{P}} & {\bf M_{PQ}} \\
   {\bf M_{QP}} & {\bf M_{Q}}
\end{bmatrix}
\begin{bmatrix}
   {\bf n_{P}} \\
   {\bf n_{Q}}
\end{bmatrix}
+
\begin{bmatrix}
   {\bf \Gamma_{P}} \\
   {\bf \Gamma_{Q}}
\end{bmatrix}
= 
\begin{bmatrix}
   {\frac{d \bf n_{P}}{dt}} \\
   {0}
\end{bmatrix}
\end{eqnal}

Here the $Q$ space is assumed to be time-independent, under the expectation that they evolve on timescales faster than those of plasma turbulence fluctuations, this allows one to fold the dynamics of the $Q$ space into effective rates which depend upon $\bf n_P$. This can be written as

\begin{eqnal}\label{eq:Qspace}
{\bf n_{Q}} = - {\bf M_{Q}}^{-1}(\bf M_{QP} n_P + \bf \Gamma_{Q})
\end{eqnal}
\begin{eqnal}\label{eq:Pspace}
\frac{d}{dt}{\bf n_{P}} &= (\bf M_{P} - M_{PQ}M_Q^{-1}M_{QP})n_P - {\bf M_{PQ}M_Q^{-1}\Gamma_{Q}} + {\bf \Gamma_{P}} \\
&= {{\bf M}_{\text{eff}}}{\bf n_P} + {{\bf \Gamma}_{\text{eff}}}
\end{eqnal}

But the applicability of such a separation in dynamical space needs to be quantitatively tested. In particular, for the constructed CR model to be applicable, it should satisfy Greenland's criteria \cite{Greenland_CR,Greenland_full,STOTLER_2007}, which requires evaluating the normalized eigenvalues and eigenvectors of ${\bf M}(n_e,T_e)$. The $N$ eigenvectors are arranged as the columns of an $N \times N$ matrix $\bf T$, in order of increasing eigenvalue, $\lambda$, and can be partitioned into 4 submatrices:

\begin{eqnal}\label{eq:Tspace}
{\bf T} =
\begin{bmatrix}
   {\bf T_{P}} & {\bf T_{PQ}} \\
   {\bf T_{QP}} & {\bf T_{Q}}
\end{bmatrix}
\end{eqnal}

In terms of these quantities, Greenland's criteria require that (i) $\lvert \lvert {\bf T_{QP}} \rvert \rvert \ll 1$ and (ii) $\lvert \lvert {\bf T_{QP}} {\bf T_{P}^{-1}} \rvert \rvert \ll 1$. From this point onwards, an $N_P = 1$ CR model is adopted where the $P$ space consists of only the ground state for atomic helium being dynamically evolved. In this formulation, meta-stable species (e.g. $2^1S$, $2^3S$) are taken to be in steady state. Greenland's criteria for the $N_P = 1$ CR theory were previously examined in a range of conditions relevant to fusion plasmas and found to widely satisfy (i) and (ii) \cite{STOTLER_2007}, but there is an additional unresolved practical condition: (iii) the shortest timescales over which $P$ space states are evolved should be larger than the inverse of the smallest $Q$ space eigenvalue, i.e. $\frac{\partial}{\partial t} < \lvert \lambda_Q \rvert$. In more concrete terms, phenomena on timescales faster than $\tau_Q \equiv 1/\lvert \lambda_Q \rvert$ are not resolved. As a result, $\tau_Q$ represents the slowest timescale in $Q$ space, which is not tracked, and the ground state of atomic helium should be evolved on timescales slower than $\tau_Q$ for the separation of the two dynamical spaces to be consistent since all timescales faster than $\tau_Q$ are effectively instantaneous. For the CR formulation to be subsequently applicable in the spectroscopic analysis of plasma turbulence, the autocorrelation time of $n_e$ ($\tau_{n_e}$) and $T_e$ ($\tau_{T_e}$) must be larger than $\tau_Q$. Additionally, the exposure time of the experimental imaging diagnostic, $\tau_{GPI}$, should satisfy the timescale criterion of

\begin{figure}[ht]
\centering
\includegraphics[width=0.7\linewidth]{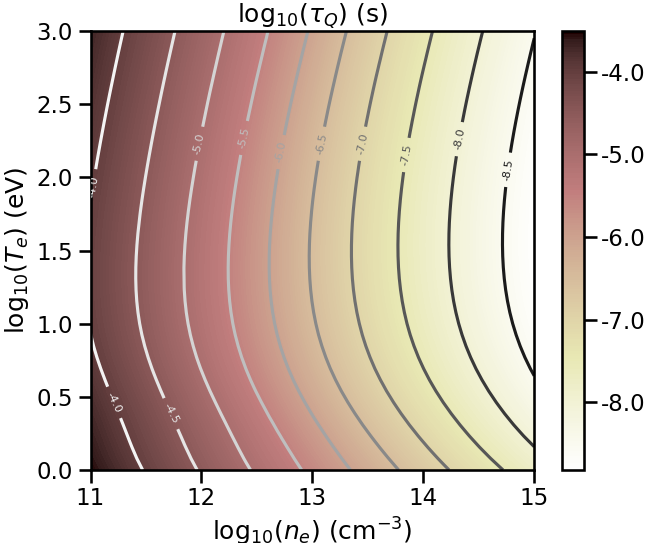}
\caption{\label{tauQ_5876} A contour plot of $\tau_Q$ for the $N_P = 1$ CR model scanned over a range of relevant electron densities and temperatures for magnetically-confined fusion plasmas. A logarithmic scale is applied on all axes including the colourbar.}
\end{figure}

\begin{eqnal}\label{eq:tauQ_validity}
\tau_Q < \tau_{GPI} < \tau_{n_e},\tau_{T_e}
\end{eqnal}

\noindent for consistency. This ensures the experimentally observed line emission in a single exposure time is based upon neutrals nominally excited by a unique $n_e$ and $T_e$ instead of a range of contributing magnitudes. Using revised cross-sections from \cite{RALCHENKO2008,Zholobenko_2018}, $\tau_Q$ is calculated under the $N_P = 1$ CR formulation in Figure \ref{tauQ_5876} at a range of $n_e$ and $T_e$ relevant to fusion plasmas. This quantity demarcates the temporal domain of validity. An important trend from the plot is that as $n_e$ increases, the limit on the temporal resolution of turbulent fluctuation measurements improves. For high plasma density fluctuations such as coherent filamentary structures, the resolution is roughly $\tau_Q \lesssim 1 \ \mu\text{s}$ for even $n_e \sim 10^{13} \ \text{cm}^{-3}$. As $n_e$ increases in higher field devices, the theoretical limit for resolving temporal scales improves. This aids the application of this GPI analysis for studying plasma turbulence in new regimes on upcoming tokamaks. A lower limit on spatial resolution for turbulence diagnostic imaging is set by $v_{HeI}/A(3^3D, 2^3P)$, provided that it is shorter than $v_{HeI} \tau_{n_e}$---or $v_{HeI} \tau_{T_e}$, if smaller---where $v_{HeI}$ is the drift velocity of the atomic helium. The validity criteria for the $N_P = 1$ CR formulation are generally satisfied in analyzing the $3^3D \rightarrow 2^3P$ transition for fusion plasmas of sufficient density, but one should take care when checking validity in specialized scenarios. For example, if applying CR theory to cameras imaging different electronic transitions (e.g. for analysis of line ratios \cite{HeI_line_ratio1,HeI_line_ratio2,Griener_2018}) with long exposure times where $\tau_{n_e}, \tau_{T_e} < \tau_{GPI}$, the formulated CR theory is technically invalid as the condition given by Eq. \eqref{eq:tauQ_validity} is no longer met. This could potentially cause errors in $n_e$ and $T_e$ profiles in existing experimental diagnostics towards closed flux surfaces, where plasma fluctuations are temporally faster than the observed autocorrelation time of far SOL turbulence \cite{labombard_evidence_2005}. Farther in the SOL as the plasma pressure drops, one should check that $\tau_Q < \tau_{n_e}, \tau_{T_e}$. Diagnosing edge fluctuations thus necessitates sufficiently high resolution for both the experimental diagnostic and applied CR theory.

The $N_P = 1$ CR formulation permits any excited state population density in $Q$ space to be written as

\begin{eqnal}\label{eq:nQ_CR}
{\bf n_Q}\lvert_q = n(q) = R_0(q)n_en_i + R_1(q)n_en(1^1S)
\end{eqnal}

\noindent where $R_0(q)$ and $R_1(q)$ are known as population coefficients associated with recombination and electron impact physics. The temporal evolution of the ground state, the only species in $P$ space for this CR model, follows 

\begin{eqnal}\label{eq:nP_CR}
\frac{d}{dt}{\bf n_P} = \frac{d}{dt}{n(1^1S)} = \alpha_{CR}n_en_i - S_{CR}n_en(1^1S)
\end{eqnal}

\noindent where $\alpha_{CR}$ and $S_{CR}$ are the recombination and ionization rate coefficients, respectively. To generate photon emissivity coefficients from this CR model, Eq. \eqref{eq:nQ_CR} is multiplied by the Einstein A coefficient for the given radiative transition. For the 587.6 nm line, $A(3^3D, 2^3P) = 2 \times 10^7 \ {\text{s}}^{-1}$. If $q = 3^3 D$, by multiplying Eq. \eqref{eq:nQ_CR} with the corresponding spontaneous decay rate, one can compute

\begin{eqnal}\label{eq:PECexc}
\text{PEC}^{exc} = R_1(3^3D) A(3^3D, 2^3P)
\end{eqnal}
\begin{eqnal}\label{eq:PECrec}
\text{PEC}^{rec} = R_0(3^3D) A(3^3D, 2^3P)
\end{eqnal}

\begin{figure*}[ht]
\centering
\includegraphics[width=1.0\linewidth]{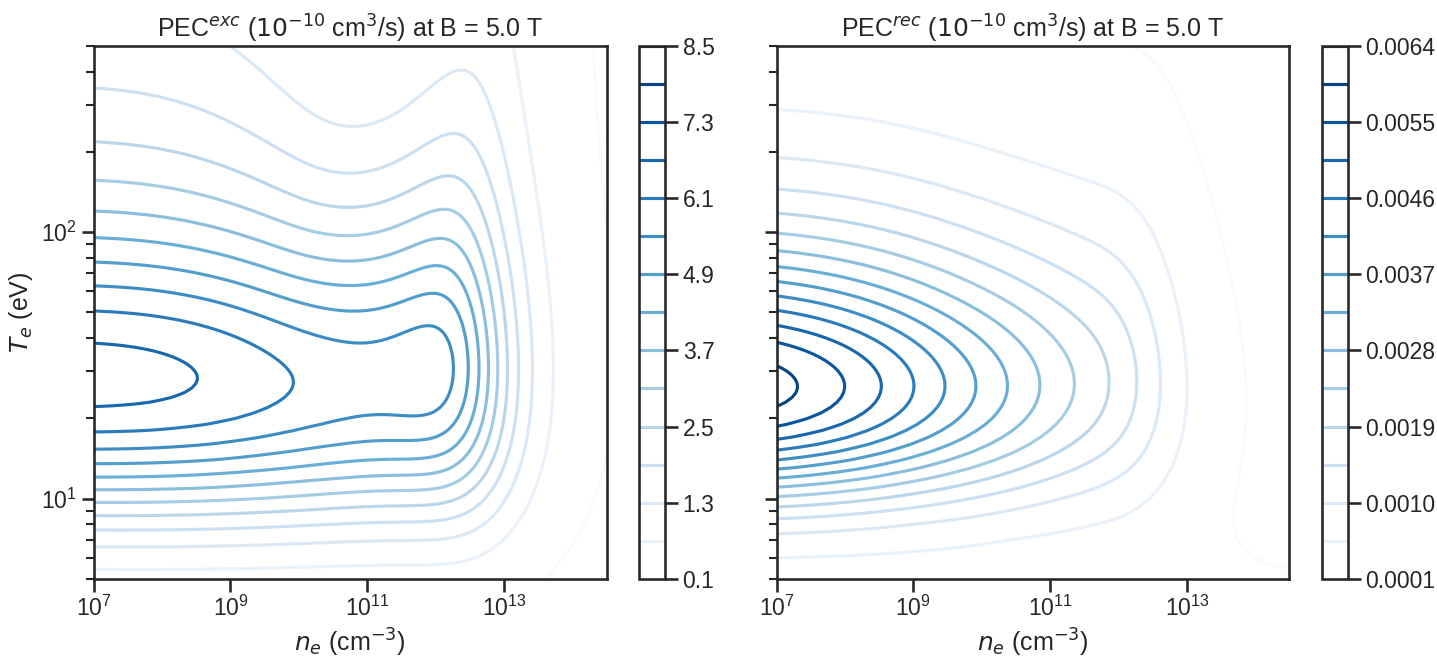}
\caption{\label{CR_rate3}Photon emissivity coefficients for the HeI 587.6 nm line based upon electron impact excitation (left) and recombination (right). These quantities follow from the $N_P = 1$ CR model's population coefficients, i.e. Eqs. \eqref{eq:PECexc} and \eqref{eq:PECrec}.}
\end{figure*}
\begin{figure*}[ht]
\centering
\includegraphics[width=1.0\linewidth]{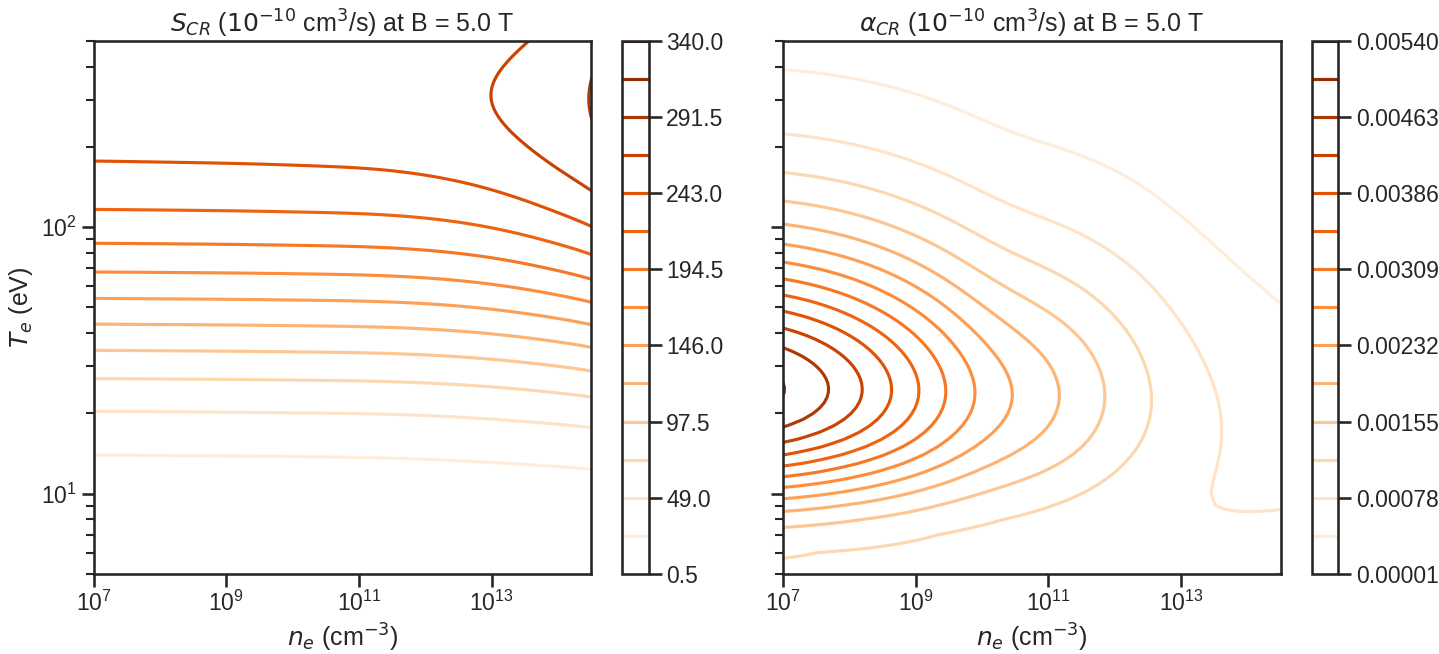}
\caption{\label{CR_rate5}Ionization (left) and recombination (right) rate coefficients derived from the $N_P = 1$ CR model. These quantities represent sinks and sources in Eq. \eqref{eq:nP_CR} for atomic helium when considering their transport in fusion plasmas.}
\end{figure*}

Contours of all coefficients along with their dependence on $n_e$ and $T_e$ are visualized at a magnetic field of $B = 5 \ \text{T}$ in Figures \ref{CR_rate3} and \ref{CR_rate5}. The plotted coefficients do not vary appreciably over $1 <$ B (T) $< 10$, which is relevant to Alcator C-Mod. Given these rates, one can further simplify the expressions for Eqs. \eqref{eq:nQ_CR} and \eqref{eq:nP_CR} when modelling 587.6 nm line emission in the presence of edge plasma turbulence by removing the effects of volumetric recombination, i.e. $\text{PEC}^{exc} \gg \text{PEC}^{rec}$ and $S_{CR} \gg \alpha_{CR}$, which are negligible for edge fusion plasmas unless $n_i \gg n_0 \equiv n(1^1S)$, i.e. only if the HeII density is far greater than the ground state neutral helium density. The effects of charge-exchange are also neglected as the reaction rate is small compared to electron impact ionization for atomic helium as long as $5 \ \text{eV} < T_e < 5 \ \text{keV}$ \cite{AMJUEL}. Note that this is not necessarily true for other atomic or molecular species, e.g. deuterium \cite{Helander_Catto_K1994}, but allows for an expression of 587.6 nm photon emissivity given by

\begin{eqnal}\label{eq:emissivity_GPI}
I = C n_0 n_e \text{PEC}^{exc}(n_e,T_e) = Cn_0 f(n_e,T_e)
\end{eqnal}

\noindent where $f(n_e,T_e)$ can be interpreted as the photon emission rate per neutral consistent with the $N_P = 1$ CR model. Using an oft-applied exponential model of $f(n_e,T_e) \propto n_e^{\alpha_n} T_e^{\alpha_T}$ and treating $\alpha_n$ and $\alpha_T$ as constants could yield erroneous emissivity predictions where fluctuations of order unity are beyond the perturbative regime. For example, $\alpha_T$ varies by a factor of 5 when $T_e$ goes from 4 eV to 10 eV \cite{Zweben_2017}. It is important to therefore retain the full range of dependency on $n_e$ and $T_e$. A constant factor $C$ is introduced in \eqref{eq:emissivity_GPI} to account for instrument calibration along with effects introduced by the finite thickness of the  observed emission cloud.

\section{\label{sec:level4.2}Experimental imaging of helium line emission on turbulent scales in the SOL of Alcator C-Mod}
Our experimental analysis technique is generic to any plasma discharge on Alcator C-Mod where good fast camera data exist for the 587.6 nm line. The plasma discharge chosen for this work is numbered 1120711021. This is a majority deuterium, lower single null diverted ohmic plasma with an on-axis toroidal magnetic field of 5.4 T and plasma current of 0.83 MA. The tokamak itself has a major radius of 0.68 m and minor radius of 0.22 m. The discharge has independent diagnostic measurements from a main chamber scanning probe equipped with a mirror Langmuir probe (MLP) biasing system run in a swept mode in the edge plasma \cite{MLP_origin,LaBombard_probevGPI}. Based on Thomson scattering \cite{JWHughes-TS-diagnostic} and electron cyclotron emission diagnostic measurements \cite{Basse_CMod}, the core electron density and temperature are $2.0 \times 10^{20} \ \text{m}^{-3}$ and $1.5$ keV, respectively.


\subsection{\label{sec:level4.2.1}Gas puff imaging on Alcator C-Mod}

For the present work, the GPI diagnostic on the Alcator C-Mod tokamak \cite{2002_Zweben,GPImanual} was configured to capture visible light at a wavelength of 587.6 nm arising from the interaction of the edge plasma with neutral helium puffed locally to the imaged region. This is a commonly used technique akin to other plasma diagnostics such as beam emission spectroscopy (BES) \cite{BES_McKee}. Helium is an ideal choice for 2-dimensional turbulence imaging for several reasons: its low atomic number results in radiative losses minimally perturbing the plasma state; its larger ionization energy allows for greater neutral penetration than thermal deuterium; its lack of molecular interactions reduces complexity in modelling; and its neutrality keeps its transport independent of external magnetic fields. The spatially localized HeI also provides a greater contrast to the background emissivity in fusion plasmas primarily fueled by hydrogen isotopes. 

Line emission from atomic helium was imaged onto a Phantom 710 fast camera, installed on Alcator C-Mod in 2009 to view the outboard midplane region \cite{GPImanual}. The camera has a maximum framing rate of 400,000 frames/s at 2.1 $\mu$s-exposure/frame when 64 $\times$ 64 pixels are being read out, and each pixel is approximately $20 \ \mu$m $\times \ 20 \ \mu$m. The diagnostic's resultant temporal resolution is 2.5 $\mu$s as it takes 0.4 $\mu$s to read values from the pixel array. The fast camera has a built-in positive offset of approximately 80 counts, which is subtracted from all GPI signals before analysis of the experimental data \cite{GPImanual}. Based upon the manufacturer’s specifications and sample bench tests, the fast camera measurements are expected to vary linearly with light level over the pixels analyzed. The brightness is thus offset in absolute magnitude by a constant scale factor and accounted for in the framework.

\begin{figure}[ht]
\centering
\includegraphics[width=1.0\linewidth]{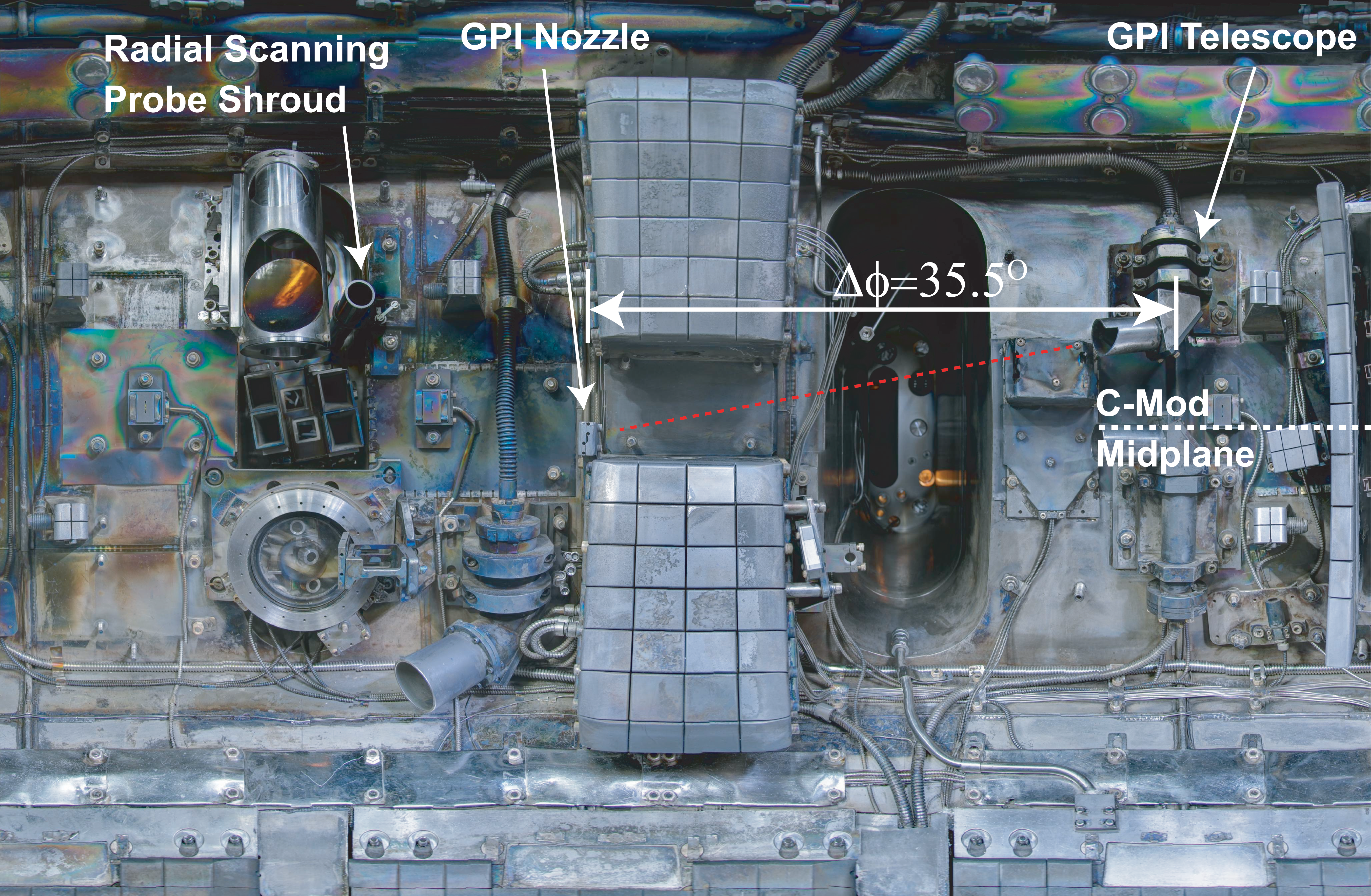}
\caption{\label{CMOD_Phantom}A section of a panoramic photo of the Alcator C-Mod outer wall, showing approximately one quarter of the device and centered on the split poloidal limiter next to the gas puff imaging measurement location. Labels indicate the position of the GPI nozzle, its imaging telescope, and an approximate line of sight (red dashed). The position of the radial scanning probe, which provides the mirror Langmuir probe measurements, is also exhibited.}
\end{figure}

\begin{figure}[ht]
\centering
\includegraphics[width=1.0\linewidth]{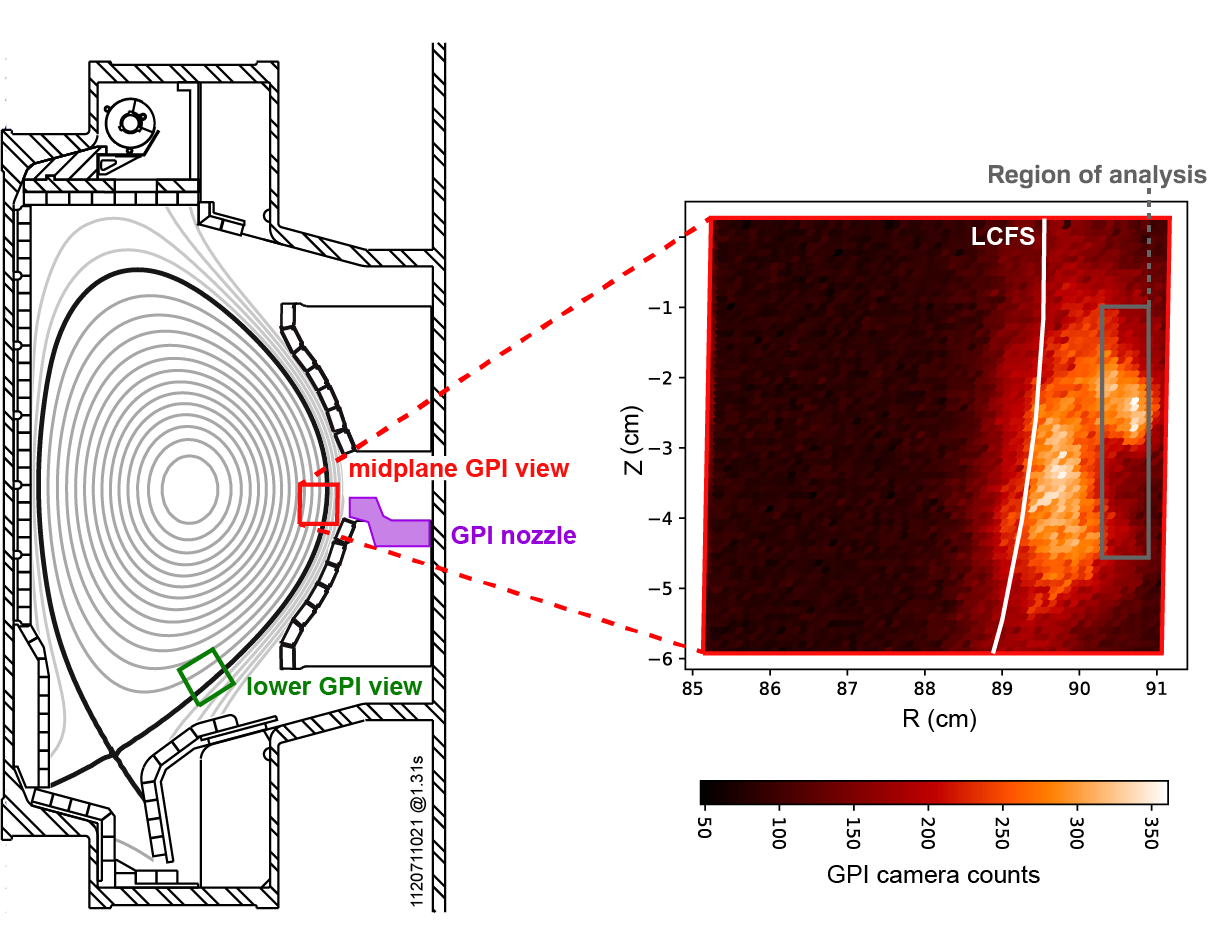}
\caption[Visualization of the experimental GPI setup on a poloidal cross section of a lower single null diverted plasma discharge (1120711021) on Alcator C-Mod. The expansion at right shows raw counts measured by the fast camera at $t = 1.312858$ s, and includes overlays of both the last closed flux surface and the approximate domain of the analysis described in Section \ref{sec:level4.4}.]{\label{CMOD_GPI}Visualization of the experimental GPI setup on a poloidal cross section of a lower single null diverted plasma discharge (1120711021) on Alcator C-Mod. In this chapter, measurements are used from the midplane fast camera with a 587.6 nm optical filter with full width at half maximum of 11.4 nm which has a largely field-aligned view of edge fluctuations. The expansion at right shows raw counts measured by the fast camera at $t = 1.312858$ s, and includes overlays of both the last closed flux surface and the approximate domain of the analysis described in Section \ref{sec:level4.4}.}
\end{figure}

A coherent fiber bundle/image guide was used to couple light from viewing optics mounted on the outer wall of the vacuum vessel to the Phantom camera detector array. The optics imaged a roughly 60 mm $\times$ 60 mm region in the $(R,Z)$-plane just in front of a gas puff nozzle through a vacuum window onto the image guide. The viewing chords pointed downwards at a fixed angle of $11.0^{\circ}$ below horizontal towards the vertically-stacked 4-hole gas nozzle displaced from the telescope by approximately $35.5^{\circ}$ in toroidal angle as displayed in Figure \ref{CMOD_Phantom}. The central ray of the imaged view thus pierced the gas puff plane approximately parallel with the local magnetic field line \cite{GPImanual}. This aligns the GPI optics with field-aligned fluctuations for typical operational parameters of an on-axis toroidal field of 5.4 T and plasma current of 1.0 MA. For discharge 1120711021 having a plasma current of 0.83 MA, the viewing chords are oriented at an angle of approximately 2$^{\circ}$ to the local field. Spatial blurring due to this angular misalignment, $\theta_B$, consequently limit resolution to $\Delta x = L_{||} \tan \theta_B$, where $L_{||}$ is the emission cloud's length parallel to the local magnetic field line. For $L_{||}$ between 5 -- 40 mm, the smearing will be 0.2 -- 1.4 mm in addition to the 1 mm pixel spot size in the image plane. Since the gas cloud expands after exiting the 4-hole nozzle and the local magnetic field's pitch angle varies, the smearing increases for those chords farther away from the nozzle depending upon the collimation of the gas cloud \cite{Zweben_2009,Zweben_2017}. With this setup and under these plasma conditions, the spatial resolution over the portion of the field-of-view analyzed is estimated to be approximately 1--2 mm. A visualization of the experimental setup is displayed in Figure \ref{CMOD_GPI}. Diagnostics injecting HeI with smaller angular half-width like thermal helium beams \cite{BES_WEST,BES_McKee} are thus helpful and this physics-informed deep learning technique can be directly transferred for their analysis as long as Greenland’s criteria are satisfied.

For background on the optics \cite{GPImanual}, the telescope contains no active shutter but a cylindrical shield on the front end, which is mounted on the outboard vessel wall at [$R = 102.5$ cm, $Z = 9.0$ cm]. A stainless steel mirror is located at the shield's back to direct light upward to several quartz lenses in vacuum. The image formed by the lenses is sent through a small vacuum window and mounted at the end of the bellows which carries the quartz fiber optics. This optics bundle is always in air, with one end connected to the camera, which uses a 5-m-long, $0.158” \times 0.158”$-sized coherent quartz fiber optic bundles to transmit the images. These have a better transmission than glass whilst not experiencing radiation browning which can darken glass bundles, but the quartz bundles have only $57 \times 57$ fibers (converse to $400 \times 400$ for glass in the past). These were hand-made by Fiberoptic Systems Inc. The quartz bundles were enclosed in custom vacuum bellows to transfer light from inside the vessel. The exterior end of the bellows was attached to a flange at the top of Alcator C-Mod, and the fiber optics came out of this end of the bellows and was connected in air to the camera. The square quartz optical bundle was imaged by a large 75 mm focal length commercial C-mount lens, focused to infinity for passage through the optical filter, and then $3 \times$ de-magnified and imaged onto the camera with a commercial 25 mm focal length C-mount lens. An Andover optical line transmission filter at 587.6 nm with full width at half maximum of 11.4 nm was screwed onto the larger lens. This optic was covered by a black cloth during operation. 


As noted above, helium gas is injected into the vessel via four vertically-displaced plasma-facing capillaries located at $Z = -4.2, -3.4, -2.6, \text{and} -1.9$ cm, which are mounted in a port on a shelf just below the outer midplane sitting in the shadow of two outboard limiters. The position $Z = 0$ corresponds to the vertical location of the machine midplane. The gas tubes' orifices are positioned at $R = 91.94$ cm with the channel exit diameter being 3 mm. The helium atoms are supplied by the Neutral gas INJection Array (NINJA) storage and delivery system \cite{NINJA_thesis} which has a pneumatically-controlled valve at the plenum which is connected to a 3.48-m-long, 1-mm-diameter capillary that feeds the 4 diverging gas tubes. Previous measurements indicate that the gas cloud exiting a single 1-mm-diameter capillary expands with angular half-width of 25$^{\circ}$ in both the poloidal and toroidal directions \cite{Terry_private}. This is the basis for estimating a spatial resolution of 1--2 mm given above. For discharge 1120711021, the plenum backing pressure was 434 torr and the total helium gas input was 6.27 torr$\cdot$L over two puffs with valve duration times of 0.08 s for each puff. The trigger time for the first NINJA valve opening was $t = 1.05$ s, while the second sustainment puff's trigger was applied at $t = 1.27$ s. The HeI flow rate at $t = 1.31$ s is estimated to be $1.21 \times 10^{20}$ s$^{-1}$ \cite{Terry_private}. Due to the tubes' spatial displacement, the helium gas puff is intended to be relatively uniform in the vertical direction. By definition, there is a shock at (or near) the vacuum-nozzle interface for this sonic flow since only particles moving downstream can escape and there is consequently no information being communicated to upstream particles \cite{shocks_in_tubes_Parks_and_Wu}. The neutral dynamics thus transition from a fluid regime in the gas tube to a kinetic regime upon entering the tokamak. The HeI exiting the diverging nozzles is approximately modelled by a drifting, cut-off Maxwellian distribution with a mean radial velocity of $-900$ m/s and a mean vertical velocity of $-20$ m/s since the direction of the non-choked flow in the capillaries is roughly 2.4$^{\circ}$ away from being oriented purely radially \cite{Terry_private}.


\subsection{\label{sec:level4.2.2}Experimental validity of $N_P = 1$ HeI CR theory}

\begin{figure*}[ht]
\centering
\includegraphics[width=1.0\linewidth]{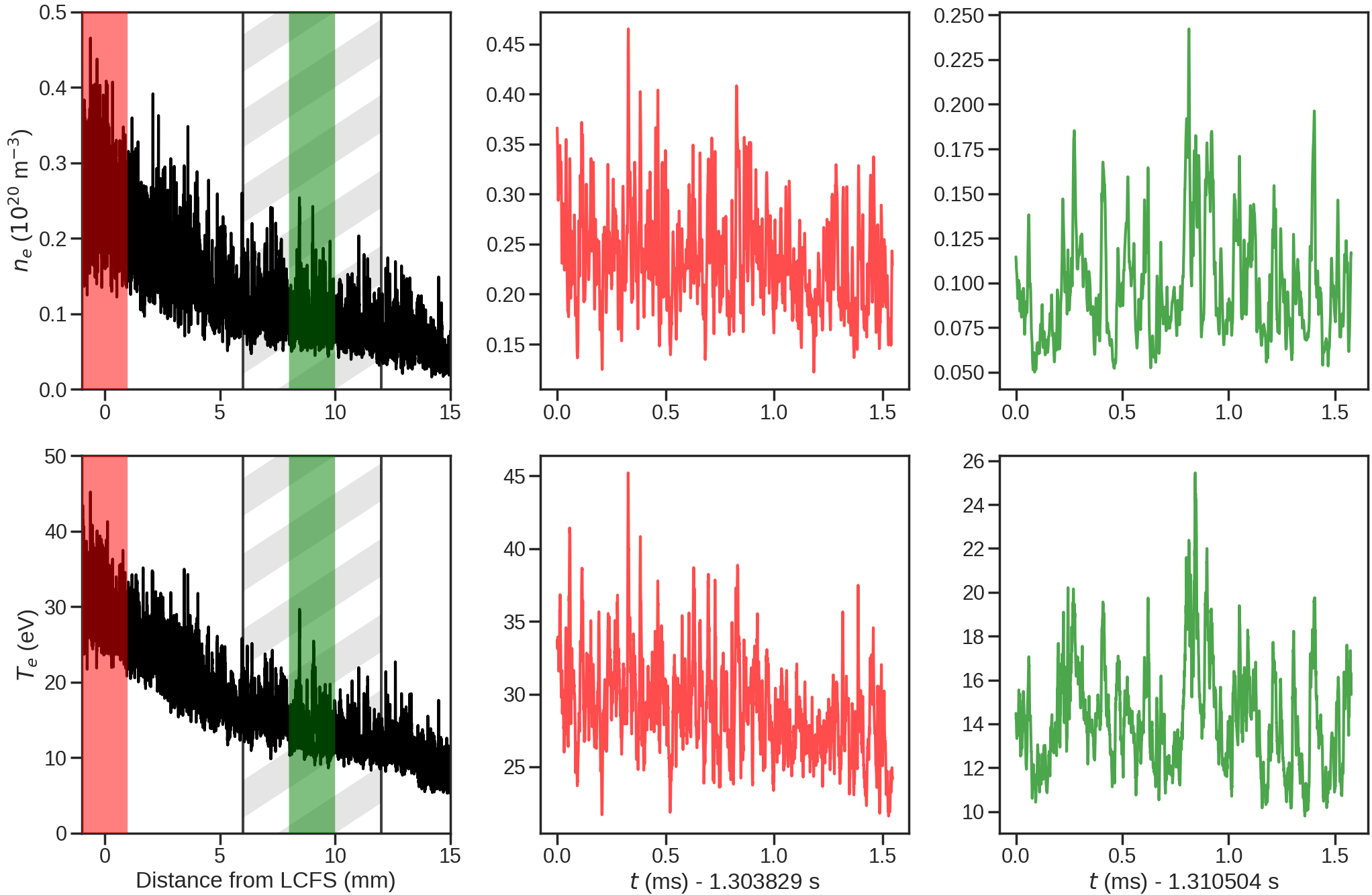}
\caption{\label{probe_1120711021} Experimental $n_e$ and $T_e$ measurements in discharge 1120711021 from the independent scanning mirror Langmuir probe. The full probe plunge duration is $1.288 < t \ \text{(s)} < 1.318$. As the probe is scanning back from closed flux surfaces, time series of $n_e$ and $T_e$ are plotted to compute autocorrelation times. Near the LCFS (red), $\tau_{n_e}$ and $\tau_{T_e}$ are approximately 4.2 $\mu$s and 6.1 $\mu$s, respectively. Farther into the SOL (green), $\tau_{n_e}$ and $\tau_{T_e}$ increase to 15.6 $\mu$s and 22.9 $\mu$s, respectively. For reference, the gray region roughly corresponds to the radial extent of the GPI data analyzed.}
\end{figure*}

To examine the experimental relevance of applying the $N_P = 1$ CR theory outlined in Section \ref{sec:level4.1} for analysis of edge plasma turbulence on Alcator C-Mod, a few key characteristic parameters of interest are reviewed based upon scanning MLP measurements of $n_e$ and $T_e$ in plasma discharge 1120711021. Magnetically disconnected from the GPI field of view, the scanning MLP in Figure \ref{CMOD_Phantom} is located at $Z = 11.1$ cm roughly $20^{\circ}$ in toroidal angle from the GPI view and radially traverses the tokamak plasma from the far edge to just inside the last closed flux surface (LCFS) with a temporal resolution of 0.3 $\mu$s. Measurements mapped to the midplane radius are visualized in Figure \ref{probe_1120711021} based upon a probe plunge nearly coincident temporally with the GPI analysis of this plasma discharge. While the probe bias is inherently perturbative due to the collection of charged particles, its effects on local plasma conditions are assumed to be negligible \cite{kuang_thesis}. From the MLP data, one can obtain autocorrelation times of fluctuations near the LCFS and approximately 8 - 10 mm radially outward into the SOL when mapped to the midplane radius. Towards closed flux surfaces, $\tau_{n_e}$ and $\tau_{T_e}$ are approximately 4.2 $\mu$s and 6.1 $\mu$s, respectively. In the far SOL, $\tau_{n_e}$ and $\tau_{T_e}$ increase to 15.6 $\mu$s and 22.9 $\mu$s, respectively. Since the probe has a finite velocity and the autocorrelation length of fluctuations is finite, these estimates of $\tau_{n_e}$ and $\tau_{T_e}$ act as conservative lower bounds as long as there is no aliasing nor phase-alignment between the probe's motion and turbulence structures. The fast camera exposure time of 2.1 $\mu$s is expected to be suitable for analysis of edge plasma fluctuations in this ohmic discharge, although faster cameras could be helpful in analyzing plasma conditions. Further, for turbulence near the LCFS where $n_e \gtrsim 10^{19} \ \text{m}^{-3}$  and $T_e \gtrsim 20$ eV, then $\tau_Q < 1 \ \mu$s, and the condition of $\tau_Q < \tau_{exp} < \tau_{n_e},\tau_{T_e}$ is well-satisfied. For fluctuations farther out into the SOL, this condition is still generally valid especially in the treatment of high pressure filaments, but one should be careful when $n_e$ drops below $2.5 \times 10^{18} \ \text{m}^{-3}$ in fusion plasmas. For spectroscopic techniques analyzing line intensities, each optical camera's exposure time needs to be suitably adjusted to satisfy the timescale condition. This is especially important towards closed flux surfaces where long camera exposure periods and shorter autocorrelation times would render examination of brightness ratios arising from turbulent fluctuations as inconsistent.

Our framework outlined in the next sections can be applied to regions with arbitrary geometries (e.g. X-point, divertor) if using sufficiently planar helium beams where the width of the collimated gas is smaller than the parallel autocorrelation length of the plasma fluctuations in the direction of the viewing chords. Since the viewing chords are roughly field-aligned over the pixels being analyzed, this parallel scale condition is expected to be satisfied. Finally, it is noted that the signal-to-noise ratio degrades in the inboard portion of the field-of-view, which includes plasma close to or on closed flux surfaces where the electron pressure and ionization rate increase sharply \cite{jwhughes1,jwhughes2}. Accordingly, fluctuations are analyzed a few millimetres away from the LCFS on a 2-dimensional $(R,Z)$-grid co-located at the nominal gas puff plane. In future work, if greater neutral penetration can be achieved such that high signal-to-noise can be attained on closed flux surfaces, the capability to then probe pedestal dynamics also exists where priming the optimization framework on available 1-dimensional data may help. Such training on background profiles was found to aid stability, and this will likely be especially helpful when approaching or on closed flux surfaces. Accordingly, Appendix B outlines a generalized regression technique developed to output edge background profiles in any confinement regime \cite{Mathews2020}. This opportunity may already be viable on devices with smaller  line-integrated $n_e$ and the methodology can extend to unmagnetized plasmas, too. 




\section{\label{sec:level4.3}Deep learning of time-dependent neutral transport physics and collisional radiative theory}

A novel multi-network deep learning framework custom-built for analysis of 587.6 nm helium line emission in fusion plasmas is outlined to uncover $n_e$, $T_e$, and $n_0$. Combining the theory governing atomic emission and neutral transport with experimental turbulence measurements via fast camera imaging into an integrated analysis framework requires sufficiently sophisticated modelling techniques. Neural networks only receiving experimental brightness measurements from GPI are thus used while being optimized against the $N_P = 1$ CR theory for photon emissivity along with the continuity equation for neutral transport which accounts for ionization of helium atoms on turbulent scales. In this way, one can combine training upon both mathematical laws and observational data. To begin, the unobserved quantities $n_e$, $T_e$, and $n_0$ each represented with their own neural network. The initial layer inputs correspond to the local spatiotemporal points $(x,y,t)$, with the $(x,y)$-coordinate being equivalent to $(R,Z)$, from the approximately 2-dimensional domain viewed by the fast camera in the poloidal plane of the gas puff nozzle. The only output of each network is the respective dynamical variable being represented. Every network's inner architecture consists of 5 hidden layers with 150 neurons per hidden layer and hyperbolic tangent activation functions ($\sigma$) using Xavier initialization \cite{GlorotAISTATS2010}. To provide reasonable intervals for the optimization bounds, the networks for $n_e$, $T_e$, and $n_0$ are constrained via output activation functions to be between $2.5 \times 10^{18} < n_e \ (\text{m}^{-3}) < 7.5 \times 10^{19}$, $2.5 < T_e \ (\text{eV}) < 150.0$, and $0.1 < n_0 \ (\text{arb. units}) < 10.0$, i.e. $n_0$ is assumed to not vary by more than two orders of magnitude in the region of the GPI analysis. While required for numerical stability, care must be taken since solutions existing near these limits may not be fully converged. The learnt constant calibration factor, $C$, is similarly represented by a network but does not vary spatially nor temporally. Physically, this results in $n_0$ being determined up to a constant scaling. The scalar constant also accounts for the 2-dimensional approximation of the localized gas puff, which has a finite toroidal width from the helium atoms exiting the capillaries. By assuming $n_e$, $T_e$, and $n_0$ to be roughly uniform along the camera's sightline, the effect of this finite volume is absorbed when learning the calibration factor. While the 2-dimensional approximation is reasonable for sufficiently planar gas injection, the deep framework can technically be generalized towards natively handling 3-dimensional space since it employs a continuous domain without any discretization. This is a future extension.

\begin{figure}[ht]
\centering
\includegraphics[width=0.7\linewidth]{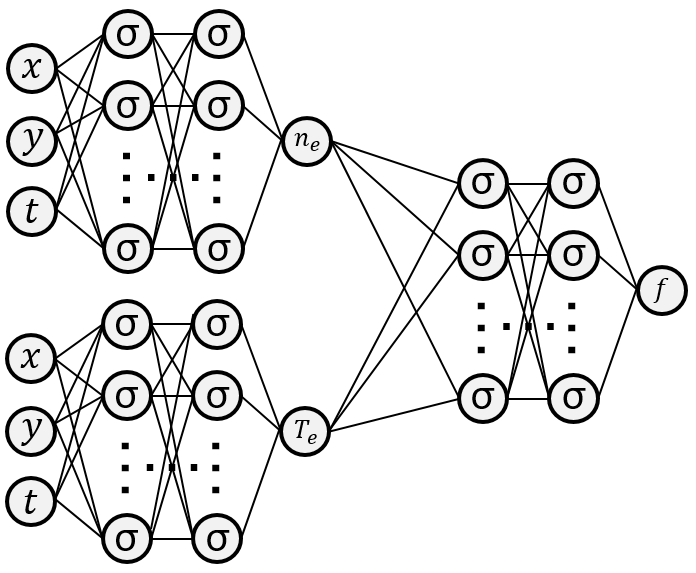}
\caption{\label{network_structure_f} Structure of networks to represent $f(n_e, T_e) = n_e \text{PEC}(n_e, T_e)$ which is one of the terms composing the total emissivity function, $I = C n_0 f(n_e, T_e)$. The ionization rate per neutral, $n_e S_{CR}(n_e, T_e)$, is similarly represented when applied in the transport equation and important to account for ``shadowing'' of neutrals \cite{2002_Zweben,STOTLER2003,Wersal_2017}. The left side of the overall network consists of the networks for the predicted $n_e$ and $T_e$, while the right side output represents the photon emissivity per neutral, where $\text{PEC}(n_e, T_e)$ is given by Figure \ref{CR_rate3}.}
\end{figure}

Our optimization is conducted in stages. To begin learning CR theory, novel neural network structures are constructed such that the outputs of the $n_e$ and $T_e$ networks serve as inputs to a new architecture representing the photon emissivity per neutral, $f \equiv f(n_e, T_e)$. The connectivity of the neurons conjoining $n_e$ and $T_e$ towards the network's output, $f$, is visualized in Figure \ref{network_structure_f}. These weights and biases are trained against $n_e \text{PEC}(n_e, T_e)$, which is derived from the $N_P = 1$ CR theory. The corresponding emissivity coefficient is plotted in Figure \ref{CR_rate3}. The ionization rate per neutral, $n_e S_{CR}(n_e, T_e)$, which is based upon the coefficient plotted in Figure \ref{CR_rate5}, is similarly represented by an architecture with $n_e$ and $T_e$ serving as inputs. All this training of the two architectures representing $f(n_e, T_e)$ and $n_e S_{CR}(n_e, T_e)$ is conducted in the first stage prior to any optimization against the fast camera data. This ensures the next stages involving training with embedded collisional radiative constraints take place under an integrated optimization framework with all quantities being represented by neural networks. For numerical purposes, $n_e$ are $T_e$ are normalized by $10^{19} \ \text{m}^{-3}$ and $50 \ \text{eV}$, respectively, and time is converted to units of microseconds during the optimization. For low temperature plasmas where $T_e < 2$ eV, training with the networks and output CR coefficients from \cite{GOTO_2003,Zholobenko_thesis} based upon fitted electron impact cross-sections should be carefully checked due to potential corrections to fits for collision strengths at such low energies. The $n_e$ and $T_e$ networks are trained only against constants of $10^{19} \ \text{m}^{-3}$ and 50 eV, respectively, for initialization. The priming, i.e. initial stage of overall training, of $n_e$ and $T_e$ and learning of CR coefficients by their respective networks takes place over the first 5 of 20 total hours of training. 

Next, the $n_e$, $T_e$, and $C$ networks are trained against Eq. \eqref{eq:emissivity_GPI} such that that the predicted brightness intensity consistent with CR theory matches experimental measurements from the fast camera. Additional constraints are thus placed in the optimizer such that solutions where $n_e$ and $T_e$ are correlated are favoured. This helps to avoid the learning of trivial solutions, e.g. purely $n_e$ fluctuations with zero $T_e$ fluctuations. Based upon past high-resolution MLP data, edge turbulent fluctuations were observed to exhibit strong correlations between the electron density and electron temperature which is a further motivation for applying this constraint \cite{KUBE2019_neTecorr}. Namely, the full loss function being collectively trained upon in this second stage is

\begin{eqnal}\label{eq:loss_GPI}
\mathcal{L}^{C,n_e,T_e} &= \frac{1}{N_0}\sum_{i=1}^{N_0} (\mathcal{L}_{GPI} + C_1 \mathcal{L}_{corr} + C_2 \mathcal{L}_{relcorr}),
\end{eqnal}

\noindent where

\begin{eqnal}\label{eq:loss_GPI1}
\mathcal{L}_{GPI} &= \lvert I^*(x^i_0,y^i_0,t^i_0) - I_0 \rvert^2 
\end{eqnal}

\begin{eqnal}\label{eq:loss_corr}
\mathcal{L}_{corr} &= - [n^*_e(x^i_0,y^i_0,t^i_0) - \langle n^*_e(x^i_0,y^i_0,t^i_0) \rangle] \times \\& [T^*_e(x^i_0,y^i_0,t^i_0) - \langle T^*_e(x^i_0,y^i_0,t^i_0) \rangle]
\end{eqnal}

\begin{eqnal}\label{eq:loss_relcorr}
\mathcal{L}_{relcorr} &= \frac{\lvert \langle n^*_e(x^i_0,y^i_0,t^i_0) - \langle n^*_e(x^i_0,y^i_0,t^i_0) \rangle \rangle \rvert^2}{\lvert \langle T^*_e(x^i_0,y^i_0,t^i_0) - \langle T^*_e(x^i_0,y^i_0,t^i_0) \rangle \rangle \rvert^2} \\&+ \frac{\lvert \langle T^*_e(x^i_0,y^i_0,t^i_0) - \langle T^*_e(x^i_0,y^i_0,t^i_0) \rangle \rangle \rvert^2}{\lvert \langle n^*_e(x^i_0,y^i_0,t^i_0) - \langle n^*_e(x^i_0,y^i_0,t^i_0) \rangle \rangle \rvert^2}
\end{eqnal}

\noindent with $I^*(x^i_0,y^i_0,t^i_0)$ following Eq. \eqref{eq:emissivity_GPI}, and the points $\lbrace x_0^i,y_0^i,t_0^i,I_{0}^i\rbrace^{N_0}_{i=1}$ corresponding to the set of observed data from GPI. The superscript notation on $\mathcal{L}$ identifies the multiple networks being simultaneously trained during optimization of the respective loss function, e.g. $\mathcal{L}^{C,n_e,T_e}$ indicates that the networks for $C$, $n_e$, and $T_e$ are being jointly optimized against this particular loss function. We note that the results from the converged solutions reported in Section \ref{sec:level4.4} are largely unchanged by removing \eqref{eq:loss_relcorr} in the optimization framework, although keeping it was found to enhance stability and thus the total number of realizations that converge. Better physics-informed optimization constraints may exist and should be investigated going forward to advance this turbulence analysis. While the coefficients $C_1$ and $C_2$ in Eq. \eqref{eq:loss_GPI} can be adaptively adjusted for optimal training in each iteration, they are set to constants of 1000 and 1, respectively, in the present implementation. The variables with asterisks symbolize predictions by their respective networks at the spatiotemporal points being evaluated during training. The notation $\langle X \rangle$ denotes the batch sample mean of $X$. This second training stage lasts for 100 minutes.

The next stage involves optimizing the $n_0$ network against both Eq. \eqref{eq:loss_GPI1} and its transport equation which accounts for drifts and fluctuation-induced ionization. Namely, in implicit form,

\begin{figure*}[ht]
\centering
\includegraphics[width=1.0\linewidth]{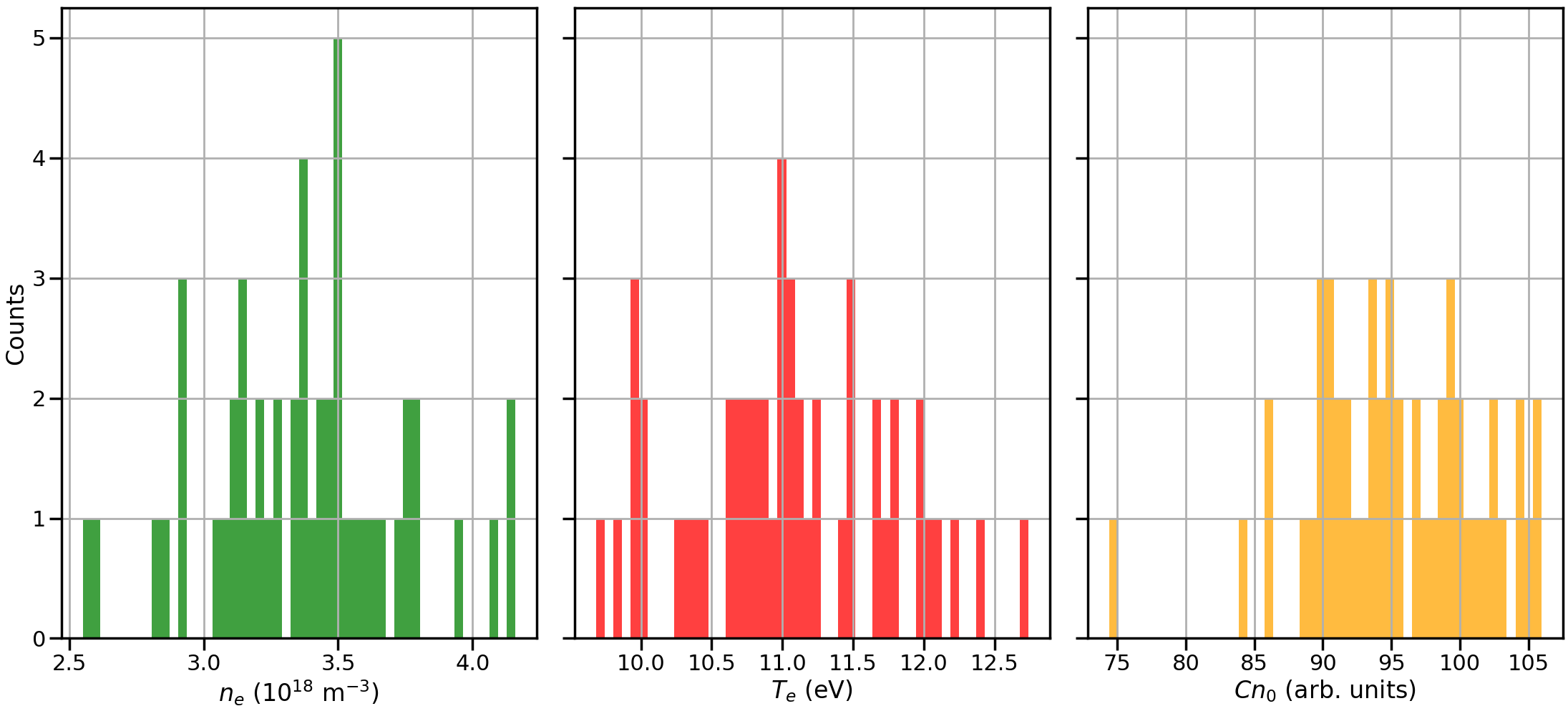}
\caption{\label{learned_1D_turbulence_histogram}Histogram of the $n_e$, $T_e$, and $Cn_0$ fluctuations at $[R = 90.7 \text{ cm}, Z = -4.2 \text{ cm}, t = 1.312850 \text{ s}]$, i.e. with corresponding normalized poloidal magnetic flux coordinate $\psi_n=1.07$, based upon 50 converged realizations when training against experimental GPI data from plasma discharge 1120711021 in the optimization framework. The ensemble mean and standard deviation of these realizations are used to construct the results presented in Section \ref{sec:level4.4}.}
\end{figure*}

\begin{eqnal}\label{eq:loss_fn0}
f_{n_0} = \frac{\partial n_0}{\partial t} + \frac{\partial (n_0 v_x)}{\partial x} +  \frac{\partial (n_0 v_y)}{\partial y} + n_0 n_e S_{CR}
\end{eqnal}

\noindent where, based upon 3-dimensional Monte Carlo neutral transport simulations of this region, closures of $v_x \sim -900$ m/s and $v_y \sim -20$ m/s are applied for modelling HeI as it exits the capillaries into the GPI frame of view \cite{SGBAEK_DEGAS2_GPI_CMOD,Terry_private}. This approximation of HeI with constant drift may be reasonable for a narrow radial region, but the true velocity distribution characterizing helium gas particles becomes increasingly skewed the farther one goes away from the gas nozzles. Modelling other atomic and molecular species (e.g. deuterium) in this way may be inadequate as charge-exchange and recombination effects on trajectories are increasingly important. Also, neutral-neutral collisions and their impacts on velocity closures are presently neglected in this treatment. This allows for a scaling constant to be factored out of Eq. \eqref{eq:loss_fn0}, i.e. permitted by its linearity in $n_0$. If using a sufficiently high spectral resolution spectrometer to view the emission cloud, the Doppler shift can potentially be experimentally measured. This task for further exploring momentum transport physics and potentially even learning the velocity closure directly from the GPI data within the optimization framework in an additional stage of training is left for future work. 

The null formulation following Eq. \eqref{eq:loss_fn0} is vital for training since all physical terms collectively sum to zero when the unknown dynamical variables in the equation are correctly solved to self-consistently account for neutral propagation and ionization. The physical theory is computationally expressed by differentiating the $n_0$ neural network with respect to its input spatiotemporal coordinates via application of chain rule through automatic differentiation \cite{tensorflow2015-whitepaper}. By then multiplying and adding the graph outputs to construct representations of the physical constraints, the network for $n_0$ can be trained against \eqref{eq:loss_GPI} and \eqref{eq:loss_fn0} to satisfy the physical theory constraining the nonlinear connection between networks. This accounting of Eq. \eqref{eq:loss_fn0} is particularly essential since the Kubo number ($Ku = V_0/\lambda \omega$ where $V_0$ is the unperturbed drift, $\lambda$ is the autocorrelation length scale \cite{Kubo_def}, and $\omega$ is the fluctuation frequency) which quantifies the strength of turbulent perturbations on neutral transport, is large ($\gtrsim 1$) for helium \cite{Kubo_original,Kubo_number,SGBAEK_DEGAS2_GPI_CMOD}. There are no explicit boundary conditions applied for $n_0$, but instead its network is trained against the fast camera's experimentally measured intensities to learn how $n_0$ should be treated around the boundaries of the analyzed camera image. Namely, the loss function in this third following stage is given by

\begin{eqnal}\label{eq:loss_n0_train}
\mathcal{L}^{n_0} &= \frac{1}{N_0}\sum_{i=1}^{N_0} \mathcal{L}_{GPI} + \frac{C_{f_{n_0}}}{N_f}\sum_{j=1}^{N_f} \mathcal{L}_{f_{n_0}}
\end{eqnal}

\noindent with

\begin{eqnal}\label{eq:loss_fn0_train}
\mathcal{L}_{f_{n_0}} &=  \lvert f^*_{n_0}(x^j_f,y^j_f,t^j_f) \rvert^2 ,
\end{eqnal}

\noindent where $\lbrace x_f^j,y_f^j,t_f^j\rbrace^{N_f}_{j=1}$ denote the set of collocation points which can span any arbitrary domain but taken to be equivalent to the ones encompassed by $\lbrace x_0^i,y_0^i,t_0^i,I_{0}^i\rbrace^{N_0}_{i=1}$, and $f^*_{n_0}$ is the null partial differential equation prescribed by Eq. \eqref{eq:loss_fn0} in normalized form directly evaluated by the neural networks. For the remainder of the training time, i.e. after the first stage of priming and two subsequent stages training with Eq. \eqref{eq:loss_GPI} and then Eq. \eqref{eq:loss_n0_train} , the networks are further optimized sequentially against Eqs. \eqref{eq:loss_GPI} and \eqref{eq:loss_n0_train} in repeating intervals of 100 minutes to iteratively find convergence in their respective networks. The only difference in these later stages is that $C$ is no longer a free parameter whilst training against Eq. \eqref{eq:loss_GPI}, and $C_{f_{n_0}}$ in Eq. \eqref{eq:loss_n0_train} is increased from $10^2$ to $10^6$ to improve the focused learning of neutral transport physics. If $C_{f_{n_0}}$ is increased any higher, one risks finding trivial solutions at a higher occurrence. Generalizing the optimizers to adaptively update training coefficients \cite{wang2020understanding} is an important pathway for future investigation. All loss functions are trained with mini-batch sampling where $N_0 = N_f = 1000$ using the L-BFGS algorithm---a quasi-Newton optimization algorithm \cite{10.5555/3112655.3112866}. Also, points found to have difficulty converging (e.g. optimizer becomes stuck in local minima) were removed from subsequent training stages to improve learning in remaining regions of the spatiotemporal domain analyzed. In the end, the multi-network framework trains on only 8 (radial) $\times$ 38 (vertical) pixels over 39 frames imaged by the fast camera.

By embedding $f(n_e,T_e)$ in Figure \ref{network_structure_f}, the emissivity predictions by the networks are forced to satisfy CR theory. Similarly, the ionization rate per neutral, $n_e S_{CR}$, is encoded in Eq. \eqref{eq:loss_fn0}. This ensures that the unobserved $n_e$, $T_e$, and $n_0$ being learnt are in agreement with the experimentally measured brightness while trying to satisfy the neutral transport physics for HeI which self-consistently includes time-dependent ionization in the presence of plasma turbulence. The repeated differentiation and summation of networks to represent every term in the ascribed loss functions resultantly constructs a far deeper computation graph representing the collective constraints beyond the 8 hidden layers in each dynamical variable's individual network. The cumulative graph is therefore a truly deep approximation of the physics governing the observed 587.6 nm line emission by the fast camera in Alcator C-Mod.

Due to the stochastic nature of the initialization and multi-task training, learned solutions for $n_e$, $T_e$, $n_0$, and $C$ vary each time an individual optimization is run. This may arise due to a unique solution not necessarily existing given the above optimization constraints. Therefore, an ensemble of realizations are run and it is this collection of runs considered which roughly follow Gaussian statistics. Based upon testing within the optimization framework, the necessary criteria for convergence in normalized units are set to $\mathcal{L}_{GPI} < 10^{2.5}$, $\mathcal{L}_{corr} < -10^3$, and $\mathcal{L}_{f_{n_0}} < 10^{-3}$. Checks for spurious gradients, trivial solutions, and a low number of training iterations were additionally investigated for downselecting converged realizations. For analysis of C-Mod discharge 1120711021, there were 800 runs with 50 sufficiently converging within this present analysis. The scatter in learned turbulent fluctuations among these realizations is used to quantify uncertainty intervals associated with the optimization framework, and as an example, the distribution of inferred measurements at a particular spatial and temporal point are plotted in Figure \ref{learned_1D_turbulence_histogram}. It is also important to note that the loss functions never truly go to zero either and act to quantify potential discrepancies involved in modelling the physical system with deep networks, e.g. $\mathcal{L}_{f_{n_0}}$ can be understood as the outstanding error in approximating the neutral transport theory. For reference when performing future GPI analysis, of these converged runs, the normalized mean loss functions at the end of training for the collection of realizations were found to be $\mathcal{L}_{GPI} = (1.44 \pm 0.42) \times 10^2$, $\mathcal{L}_{corr} = (-4.51 \pm 0.20) \times 10^{3}$, $\mathcal{L}_{relcorr} = 6.75 \pm 0.03$, and $\mathcal{L}_{f_{n_0}} = (3.57 \pm 3.79) \times 10^{-5}$. Simply put, finite values of the loss metrics indicate the degree to which the framework satisfies the collective training conditions. Identifying these errors allows for their iterative improvement, while identifying even better loss functions is an open area for future research. 

 

\section{\label{sec:level4.4}Uncovering plasma-neutral dynamics in experimental turbulence imaging in Alcator C-Mod}


\begin{figure*}[ht]
\centering
\includegraphics[width=1.0\linewidth]{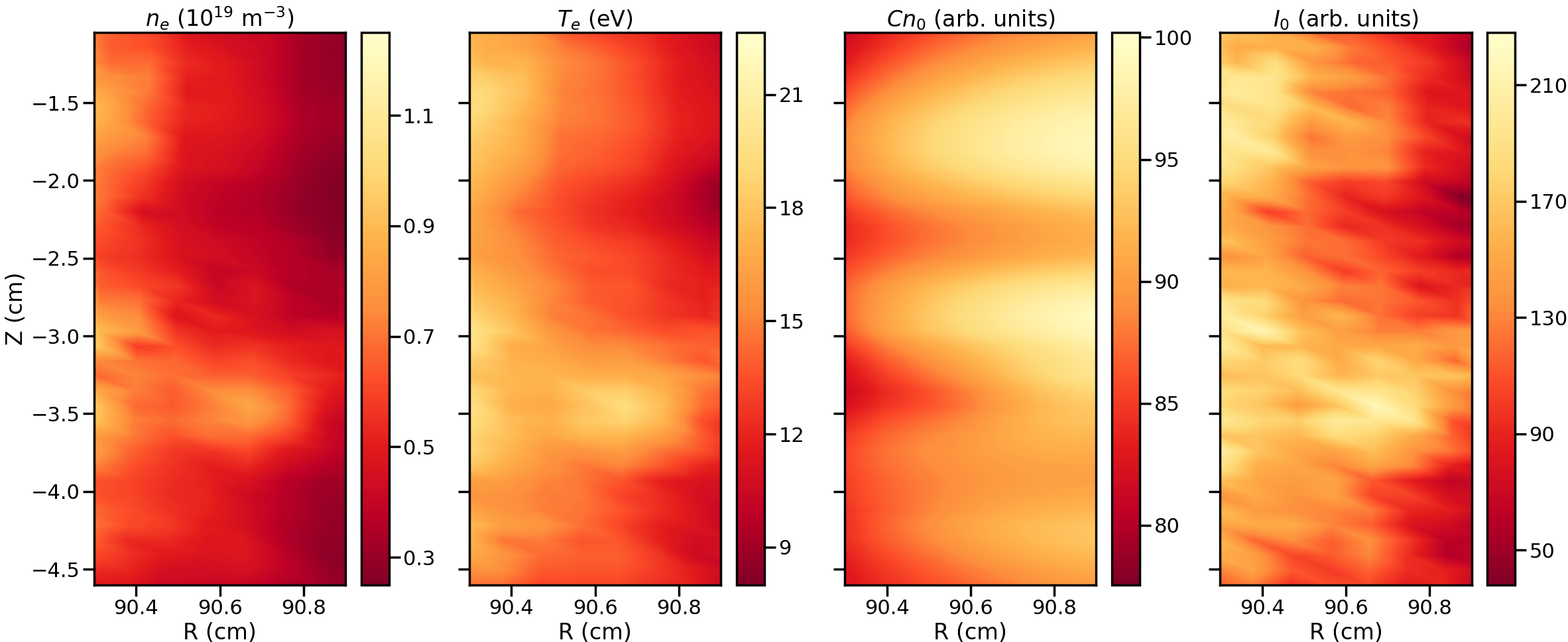}
\caption{\label{learned_2D_turbulence1}The learned 2-dimensional $n_e$, $T_e$, and $Cn_0$ for plasma discharge 1120711021 along with the experimentally observed 587.6 nm photon emission at $t = 1.312815$ s. The learned measurements are based upon the collective predictions within the deep learning framework training against the neutral transport physics and $N_P = 1$ CR theory constraints.}
\end{figure*}

\begin{figure*}[ht]
\centering
\includegraphics[width=1.0\linewidth]{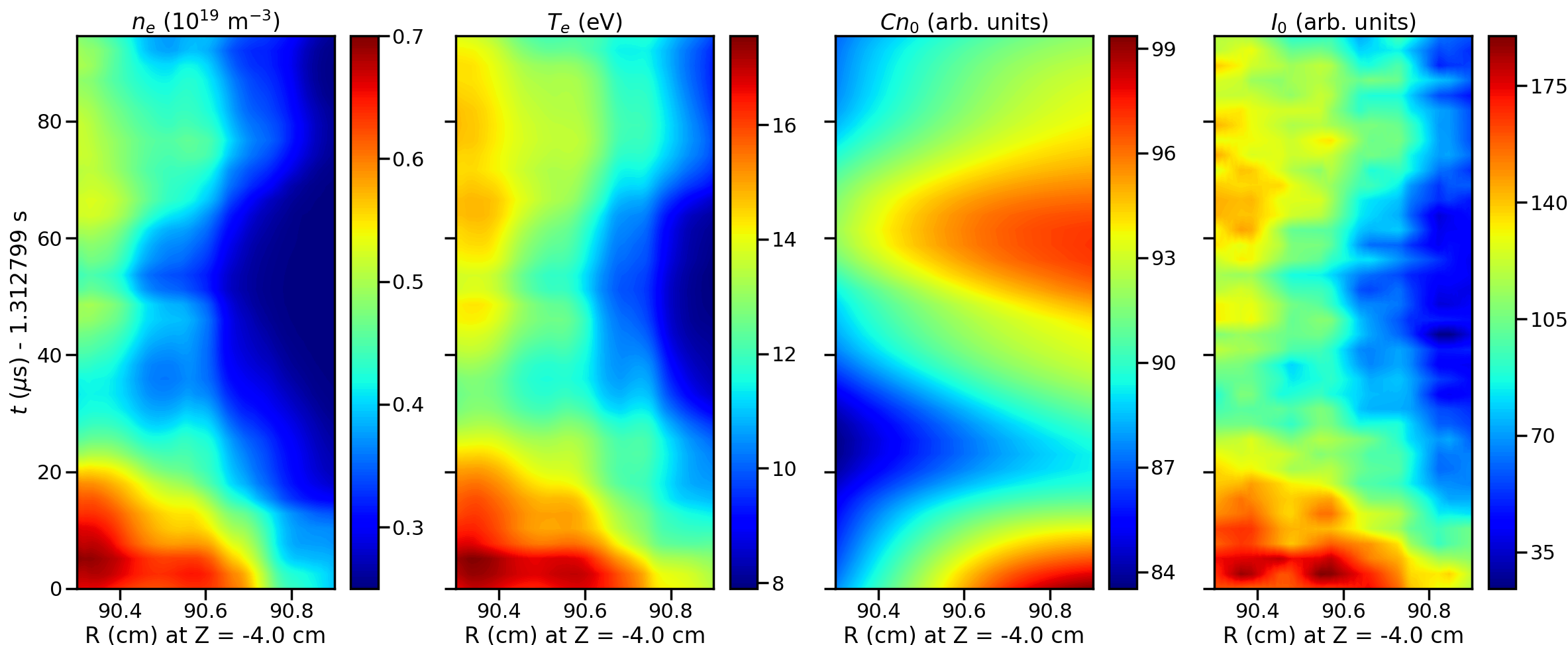}
\caption{\label{GPI_predvexp_1d_1120711021_rvt}The learned $n_e$, $T_e$, and $Cn_0$ along with the experimentally observed 587.6 nm photon brightness for plasma discharge 1120711021 at $Z = -4.0 \text{ cm}$. These quantities are plotted as a function of radius and time.}
\end{figure*}

\begin{figure*}[ht]
\includegraphics[width=1.0\linewidth]{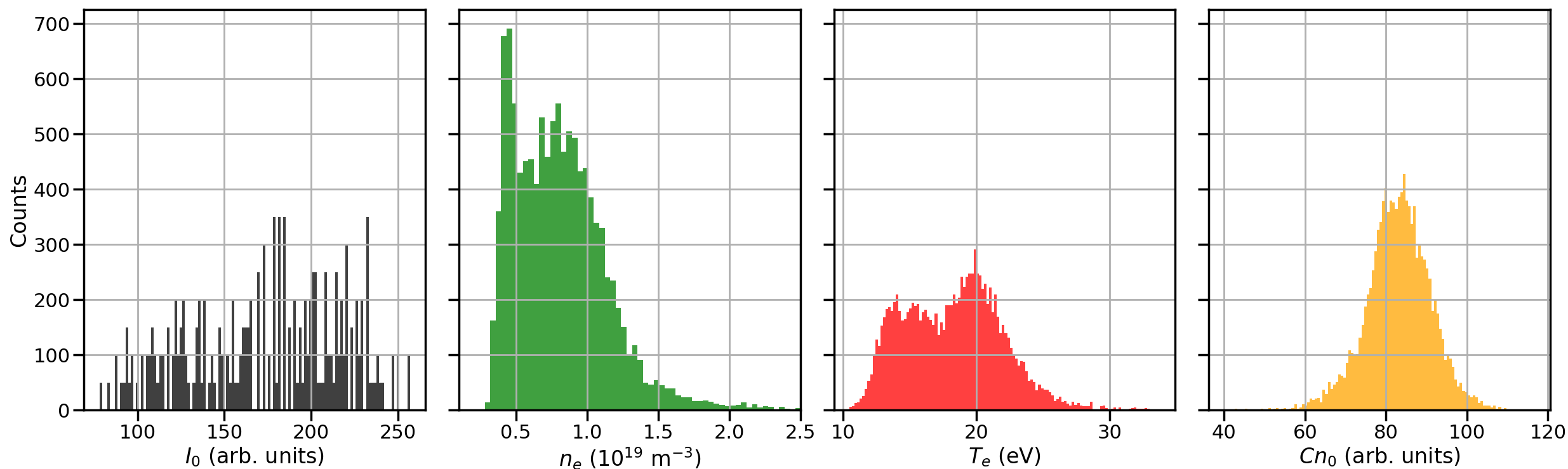}
\caption{\label{turbulence_histogram}Histograms displaying the distribution of turbulent $n_e$, $T_e$, and $Cn_0$ at [$R$ = 90.3 cm, $-4.6 < Z \ \text{(cm)} < -1.0$, $1.312799 < t_{GPI} \ \text{(s)} < 1.312896$] along with the experimentally observed 587.6 nm line intensity.}
\end{figure*}

The learned turbulent $n_e$, $T_e$, and $Cn_0$ from the time-dependent analysis of fast camera imaging for plasma discharge 1120711021 using an ensemble of 50 optimizers are visualized in 2-dimensional space along with experimentally observed GPI measurements in Figure \ref{learned_2D_turbulence1}. The positive fluctuations in brightness are largely correlated with $n_e$ and $T_e$, and these regions tend to have depressed values of $n_0$ as the ionization rate is elevated. This results in a ``shadowing effect'' in atomic helium trajectories arising from increased ionization in regions of positive density fluctuation. The autocorrelation time of $n_0$ also decreases with radius, while it increases for $n_e$ and $T_e$. Temporal variation with radius is visualized in Figure \ref{GPI_predvexp_1d_1120711021_rvt} where, considering a 1-dimensional slice of Figure \ref{learned_2D_turbulence1}, the same physical quantities are plotted at $Z = -4.0$ cm. While correlations vary poloidally and radially, and precise dependencies across the turbulent variables change as $n_e$ and $T_e$ increase, the observed line emission is found to be strongly correlated with electron density and temperature. The atomic helium density fluctuations do not vary directly proportional to $I_0$ in this far edge region on open field lines near the gas tubes. There is instead a weak negative correlation over this narrow radial extent arising from the largest brightness fluctuations corresponding to trajectories with elevated ionization rates causing a depletion, or shadowing, of HeI. A correlation matrix for the normalized relative fluctuations in 2-dimensional space over the roughly 100 $\mu$s time window analyzed are displayed in Table \ref{table_correlation_matrix5}. The maximal $n_0$ fluctuation amplitudes tend to be roughly 30--40\% from peak-to-trough in this far edge region which sits away from the LCFS, where sharper equilibrium gradients and smaller relative fluctuation levels may result in different correlations. And while relative fluctuations may be correlated from $90.3 < R \ \text{(cm)} < 90.9$ as in Table \ref{table_correlation_matrix5}, connections between the turbulent quantities are nonlinear. To better visualize their interdependence, Figure \ref{turbulence_histogram} displays histograms for $n_e$, $T_e$, and $Cn_0$ vertically along $R = 90.3$ cm. The fluctuations follow different statistical distributions and cannot necessarily be linearly mapped from the observed noisy HeI line intensity experimentally measured by the fast camera.

\begin{table}
\centering
\renewcommand{\arraystretch}{1.}
\begin{NiceTabular}{ p{0.7cm}|p{1.1cm}|p{1.1cm}|p{1.1cm}|p{1.1cm}|>{\arraybackslash}p{1.1cm}| }[]
 {} & $\{n_e\}$ & $\{T_e\}$ & $\{n_0\}$ &$\{I^*\}$ &$\{I_0\}$\\ \cline{1-6}
 $\{n_e\}$ & 1.000 & 0.887 & -0.325 & 0.843 & 0.822
\\ \cline{1-6}
 $\{T_e\}$ & 0.887 & 1.000 & -0.307 & 0.925 & 0.902
\\ \cline{1-6}
 $\{n_0\}$ & -0.325 & -0.307 & 1.000 & -0.051 & -0.059
\\ \cline{1-6}
 $\{I^*\}$ & 0.843 & 0.925 & -0.051 & 1.000 & 0.971
\\ \cline{1-6}
 $\{I_0\}$ & 0.822 & 0.902 & -0.059 & 0.971 & 1.000
\end{NiceTabular}
\caption{\label{table_correlation_matrix5}A correlation matrix of the turbulent measurements inferred and observed experimentally in plasma discharge 1120711021. For reference, $I^*$ is the predicted emissivity given by Eq. \eqref{eq:emissivity_GPI}, and $I_0$ is the experimentally observed brightness of the 587.6 nm line. Each quantity's normalized fluctuation amplitude, i.e. $\{X\} = (X - \langle X \rangle)/\langle X \rangle$, is based upon measurements over $90.3 < R \ \text{(cm)} < 90.9$, $-4.6 < Z \ \text{(cm)} < -1.0$, and $1.312799 < t_{GPI} \ \text{(s)} < 1.312896$.}
\end{table}

\begin{figure*}[ht]
\centering
\includegraphics[width=1.0\linewidth]{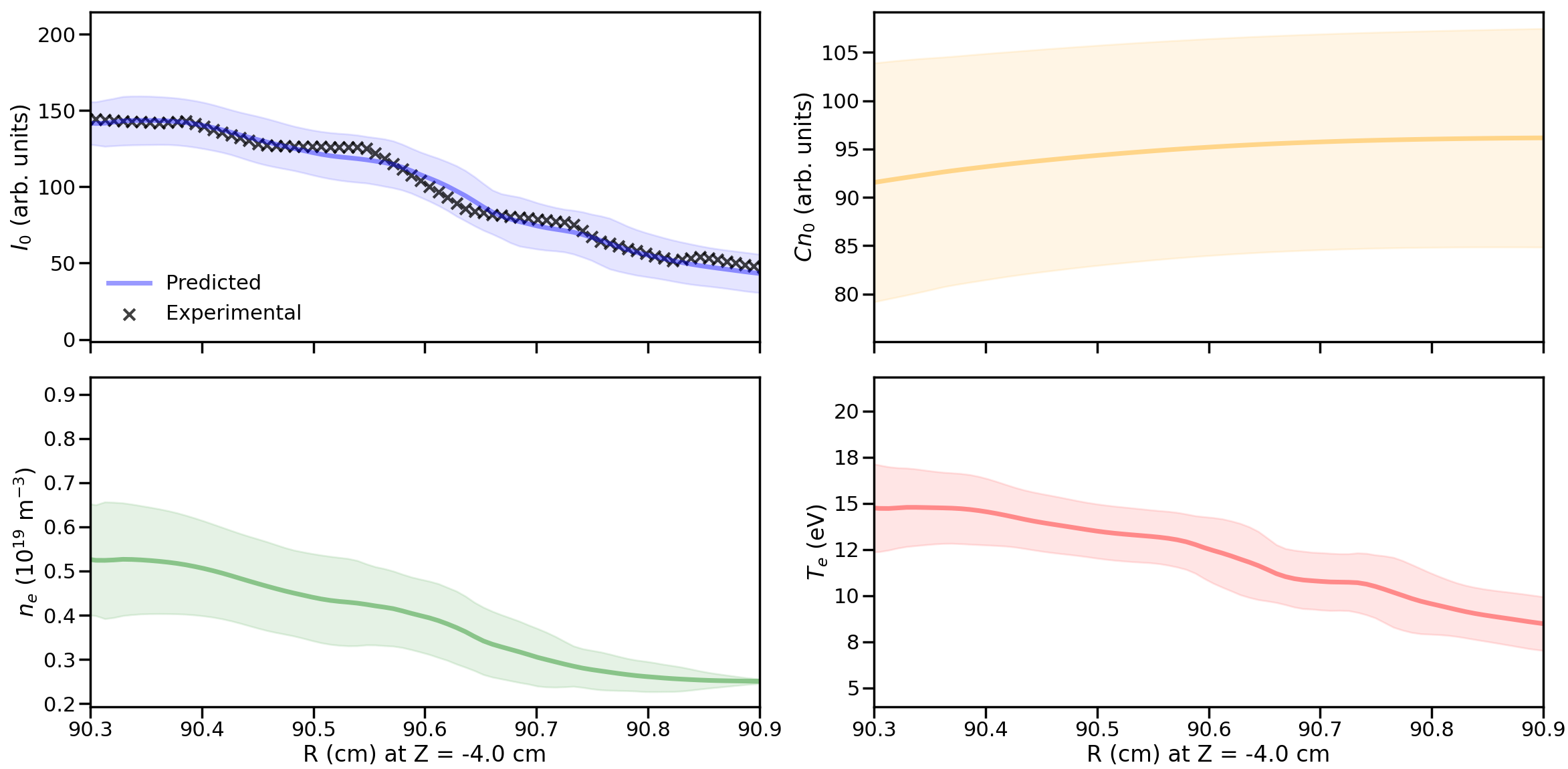}
\caption{\label{predicted_1D}Radial profiles of the inferred turbulent $n_e$, $T_e$, and $Cn_0$ at $[Z = -4.0 \text{ cm}, t = 1.312866 \text{ s}]$ along with a trace of the experimentally observed and predicted GPI intensity profiles. The computed line emission is based upon the deep learning framework following Eq. \eqref{eq:emissivity_GPI}. The dark line in each plot corresponds to the average output of the ensemble of realizations, while the shaded uncertainty intervals correspond to scatter ($\pm 2\sigma$) arising from the independently trained networks.}
\end{figure*}


\begin{figure}[ht]
\centering
\includegraphics[width=1.0\linewidth]{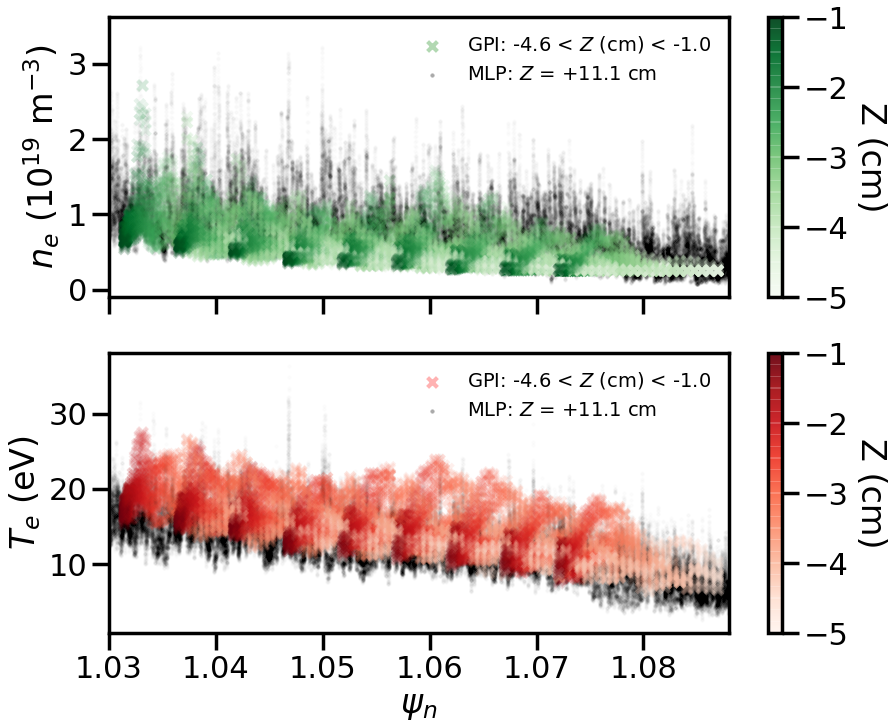}
\caption{\label{GPIvprobes}For comparison, the inferred $n_e$ and $T_e$ are plotted over the roughly 100 $\mu$s time window analyzed from the experimental GPI. The independent mirror Langmuir probe is located at $Z = +11.1$ cm with a radially moving probe head. The MLP scan in the plotted region lasts roughly 6000 $\mu$s (i.e. 60$\times$ longer than the duration of the GPI analysis). All measurements are mapped to normalized poloidal magnetic flux coordinates, $\psi_n$, and none of the data displayed is time-averaged.}
\end{figure}


One should note that these learned $n_e$, $T_e$, and $Cn_0$ are consistent solutions with the collisional radiative and optimization constraints being trained upon, but not necessarily unique solutions. Accordingly, in Figure \ref{predicted_1D}, the predicted light emission from the ensemble of realizations is displayed against the fast camera's measurements. The mean outputs and uncertainty intervals for the turbulent $n_e$, $T_e$, and $Cn_0$ associated with the scatter of running an ensemble of stochastic realizations are also plotted. There is no temporal averaging of the profiles in Figure \ref{predicted_1D}. For GPI on Alcator C-Mod, sharp features exist in the experimental data potentially associated with noise, while the learned line intensity from the collection of networks is smoother and consistent in both magnitude and shape with the observed brightness. These measurements enable novel research pathways into the 2-dimensional tracking of experimental parameters (e.g. turbulent particle and heat fluxes both radially and poloidally) and calculation of fluctuating fields for model validation \cite{Mathews2021PRE,Mathews2021PoP}. They further provide the first quantitative estimates of 2-dimensional time-dependent structure for neutrals on turbulent scales in an experimental fusion plasma. 


To further examine the validity of these results, the turbulent $n_e$ and $T_e$ from the GPI measurements of the single spectral line are juxtaposed against an independent MLP with four electrodes in Figure \ref{GPIvprobes}. This scanning probe is plunged at times overlapping with the gas puff analysis, i.e. $1.287631 < t_{MLP} \ \text{(s)} < 1.317671$ versus $1.312799 < t_{GPI} \ \text{(s)} < 1.312896$, although located at different positions toroidally and vertically as discussed in Section \ref{sec:level4.2}. The MLP measures fluctuations in time as it scans through the edge plasma to construct the resultant radial profile. For the purpose of comparison, turbulent measurements at different $Z$-locations are not time-averaged to compare the $n_e$ and $T_e$ fluctuations. While the measurement regions spatially spanned by the two independent diagnostics are magnetically disconnected, the data are mapped to common poloidal magnetic flux coordinates based upon magnetohydrodynamic equilibrium reconstruction in the tokamak plasma using the EFIT code \cite{Lao_1985}. Deconstructing the GPI fluctuations into the turbulent $n_e$, $T_e$, and $n_0$ instead of the raw brightness from atomic emission largely resolves diagnostic misalignment challenges. Namely, in contrast with past analysis of discharge 1120711021 \cite{Russell_probevGPI}, there is no radial shift applied to align the turbulent fluctuation profiles from these two independent experimental diagnostics. For GPI measurements in this 2-dimensional spatial domain spanning approximately $100 \ \mu$s, peak $n_e$ and $T_e$ fluctuations do not far exceed $3.0 \times 10^{19} \ \text{m}^{-3}$ and 30 eV, respectively, which are roughly consistent with the MLP in the far SOL. When evaluating the two sets of measurements side-by-side in Figure \ref{GPIvprobes}, excellent agreement is found in magnitude and structure between the $T_e$ measurements. The MLP $n_e$ data are slightly elevated on average although still quantitatively consistent within the measurement bounds of the four electrodes. A potential contributing factor to this observed difference in $n_e$ peaks could be natural variations in the poloidal structure of the intermittent fluctuations over the narrow time window analyzed (i.e. 100 $\mu$s for GPI, 6000 $\mu$s for MLP). The two diagnostics are magnetically disconnected and separated vertically by about 10 -- 15 cm. Fluctuation amplitudes measured by either diagnostic for both $n_e$ and $T_e$ are still in the range of 10 -- 100\%. One should also remember that, beyond the diagnostics viewing different spatiotemporal locations, systematic uncertainties extant in both the GPI and probe measurements can cause discrepancies left to be reconciled \cite{hutchinson_2002,probe_review}. For example, the MLP is intrinsically perturbative to local conditions and experimental analysis of the probe edge sheath assumes electrons can be described by a single Maxwellian velocity distribution \cite{kuang_thesis}. Additionally, while the optimization attempts to find consistent solutions within the applied framework, questions of uniqueness and generalized constraints are still being explored for better convergence.







\section{\label{sec:level4.5}Conclusion}

In summary, this chapter has developed a novel time-dependent deep learning framework for uncovering the turbulent fluctuations of both the plasma quantities, $n_e$ and $T_e$, as well as the neutrals underlying experimental imaging of HeI line radiation. Significantly, this allows determination of 2-dimensional fluctuation maps in the plasma boundary, revealing detailed spatiotemporal structure and nonlinear dynamics. It thereby extends the usefulness of the gas puff imaging method. The computational technique intrinsically constrains solutions via collisional radiative theory and trains networks against neutral transport physics. This advancement has allowed for the first estimates of the 2-dimensional $n_e$, $T_e$, and $n_0$ on turbulent scales which reveal fluctuation-induced ionization effects in a fusion plasma based upon optical imaging of just the 587.6 nm line. While the analysis is demonstrated on the edge of the Alcator C-Mod tokamak with quantitative agreement found with independent probe measurements, this technique is generalizable to ionized gases in a wide variety of conditions and geometries (e.g. stellarators, spheromaks, magneto-inertial fusion).

A number of opportunities for future development exist. One key outstanding question is the identification of underlying numerical and physical factors contributing to non-uniqueness in outputs during optimization. From experimental noise to the chaotic properties of the turbulent system, finding sufficient conditions for precise convergence is the focus of ongoing research. Future extensions of the framework also include expanding the radial domain of coverage towards closed flux surfaces, which will require widening the queried bounds on $n_e$, $T_e$, $n_0$, and improving the overall training paradigm via adaptive training and architecture structures \cite{wang2020understanding}. For example, neutral density amplitudes can vary over orders of magnitude with steep shapes in background equilibrium profiles. Tactfully embedding this information during training of the networks can aid with the overall physical modelling via optimization. In this way, better experimental constraints from 1-dimensional data may help uncover further dynamics not otherwise directly probed by edge diagnostics. 

Adaptation to other experiments is a logical next step, and translating this present technique to contemporary experimental devices using helium beams is a pathway that can be explored immediately for regions that are traditionally difficult to probe (e.g. X-point). This deep learning framework can also be extended in principle to 3-dimensional geometries to account for integrated light emission along the camera's lines-of-sight. Further, this global turbulence imaging technique provides new ways to diagnose high pressure plasma events, e.g. disruptive instabilities such as edge localized modes that can be destructive to plasma facing components. Translating the framework for direct analysis of deuterium instead of helium is also possible with a few modifications, but requires investigation of relevant CR physics \cite{Greenland_full} where charge exchange and molecular effects are no longer necessarily negligible \cite{AMJUEL}. One prospect is to couple the turbulent $n_e$ and $T_e$ learned by the framework with Monte Carlo neutral transport codes \cite{STOTLER_time_dependent_DEGAS2}, potentially allowing recovery of 2-dimensional time-dependent estimates of atomic and molecular deuterium density and its emissivity, e.g. through the ultraviolet Ly$_\alpha$ line. These could be compared directly to experimental measurements of line emission from deuterium \cite{Lyalpha0,Lyalpha1}. Such extended comparisons will be important in the testing of reduced edge plasma turbulence models \cite{Mathews2021PRE}.

\chapter{Initial estimates of the turbulent electric field by drift-reduced Braginskii theory in experiment}

\setlength{\epigraphwidth}{0.6\textwidth}
\epigraph{Happy is he who gets to know the reasons for things.}{Virgil (70-19 BCE), Roman poet}

Reduced turbulence models are, by definition, simplified descriptions of chaotic physical systems. Arguments for scale separation and geometric approximations are often undertaken in the pursuit of expedient yet reasonable estimates, but their precise effects on nonlinear processes in turbulence calculations are not always fully understood. As boundary plasmas in magnetic confinement fusion are governed by a vast range of spatiotemporal dynamics, model approximations are inevitable even using modern computing, but may only be weakly valid (if at all). To directly quantify their impacts, this chapter uses the experimental electron density and temperature inferences from Chapter 4 to compute the 2-dimensional turbulent electric field consistent with electrostatic drift-reduced Braginskii fluid theory under the assumption of axisymmetry with a purely toroidal field. The physics-informed deep learning technique outlined in Chapter 2 is used. In this present calculation, neutral deuterium sources and poloidal effects associated with parallel flows are neglected. Chapter 3 found that modelling low-$\beta$ plasmas under these exact assumptions led to excellent agreement when comparing the two-fluid theory's turbulent electric field against electromagnetic gyrokinetics. As an important first test towards translating the computational technique to experiment and directly testing reduced turbulence models, this chapter explores these approximations for an edge plasma in the Alcator C-Mod tokamak. All neglected physics can be re-inserted as a part of future work to ascertain their individual impacts, and as an initial step, the inclusion of helium gas (which is locally puffed in the experiment) is tested to gauge perturbative effects of injected neutral atoms, e.g. via the GPI diagnostic. Past simulations \cite{Thrysoe_2018,Zholobenko_2021_validation} and experiments \cite{exp0neut,exp1neut,exp2neut} have investigated the role of neutrals on edge turbulent fields, although results are at times mixed and/or inconclusive. The particle and energy sources associated with time-dependent ionization of HeI in the numerical framework are found to cause broadening in the computed turbulent electric field amplitudes along with an enhancement in correlation with the electron pressure that is not otherwise extant in plasmas without such neutral dynamics. This intensification of fields, which is due to the plasma-neutral interactions, reveals stronger ${\bf E \times B}$ flows and elevated average shearing rates on turbulent scales than expected in fully ionized gases \cite{mathews2022ErwHeI}.

\section{\label{sec:level5.1}The experimental calculation}

The focus of the present analysis will be plasma discharge 1120711021 from Alcator C-Mod as described in Chapter 4. This turbulent electric field calculation framework assumes the 2-dimensional experimental estimates of the electron density and temperature are parallel to the background magnetic field as developed in Chapter 2, but since GPI on Alcator C-Mod views the edge plasma in the $(R,Z)$-plane \cite{mathews2022deep}, an approximation of a purely toroidal magnetic geometry is the result. The plasma is further assumed to be magnetized, collisional, and quasineutral with the perpendicular fluid velocity given by ${\bf E \times B}$, diamagnetic, and ion polarization drifts. This chapter follows the prescription developed in Chapter 2 which utilizes just field-aligned turbulent $n_e$ and $T_e$ measurements along with Eqs. \eqref{eq:nDotGDBH} and \eqref{eq:TeDotGDBH} to calculate $\phi$ consistent with drift-reduced Braginskii theory. The equations are cast in a full-$f$ representation where fluctuations and global profiles evolve together \cite{francisquez_thesis}.

No boundary nor initial conditions are explicitly assumed within the physics-informed deep learning framework. All analytic terms encoded in these continuum equations are computed exactly by the neural networks without any approximation as this machine learning framework uses a continuous spatiotemporal domain (e.g. no linearization nor discretization). Hyperdiffusion, which is ordinarily applied for stability in numerical codes, is set to zero. Density sources and energy sinks associated with time-dependent ionization of the local helium gas based upon collisional radiative modelling are outlined in Chapter 4. The sources and sinks are given by $\nSrcN = n_0 n_e S_{CR}$ and $S_{E,e} = -E_{HeI} \nSrcN$, where $n_0$ is the atomic helium density, $S_{CR}$ corresponds to the ionization rate coefficient, and $E_{HeI} = 24.587$ eV is the ionization energy of HeI. The 2-dimensional turbulent $n_e$ and $T_e$ in experiment come from an $(R,Z)$-aligned plane on open field lines with a rectangular cross-section that roughly spans $[90.3 < R \ \text{(cm)} < 90.9, -4.6 < Z \ \text{(cm)} < -1.0]$ over a duration of $1.312799 < t_{GPI} \ \text{(s)} < 1.312896$. By assuming $\bhatZ$ to be parallel to these measurements, the plasma turbulence model essentially neglects the poloidal component of the field lines present in Alcator C-Mod. For physical orientation, when viewed from above the machine, the toroidal magnetic field and plasma current in this discharge run clockwise. This results in the local magnetic field lines being pointed towards the imaging system and $\bf{B} \times \nabla B$ being directed downwards. Moreover, in keeping with the parallel uniformity approximation in Chapter 2, gradients along this field-aligned (nominally toroidal) direction are assumed to be small, i.e. $\nabla_\parallel \rightarrow 0$. Accordingly, an orthogonal right-handed geometry is employed for modelling whereby $x \equiv R$ is the radial coordinate, the parallel coordinate $\bhatZ = +{\bf z}$ is purely toroidal, and the binormal (nominally vertical) direction is $y \equiv Z$. The plasma theory consists of electrons and deuterium ions with real electron-ion mass ratio, i.e. $m_i = 3.34 \times 10^{-27} \text{ kg}$ and $m_e = 9.11\times 10^{-31} \text{ kg}$. Beyond the inclusion of appropriate sources and collisional drifts,  this technique \cite{Mathews2021PRE} to calculate the turbulent electric field is applicable even if multiple ions and impurities are present in the experimental plasma due to quasi-neutrality underlying the electron fluid theory in the machine learning framework \cite{multi_species}.

\begin{figure}
\centering
\includegraphics[width=1.0\linewidth]{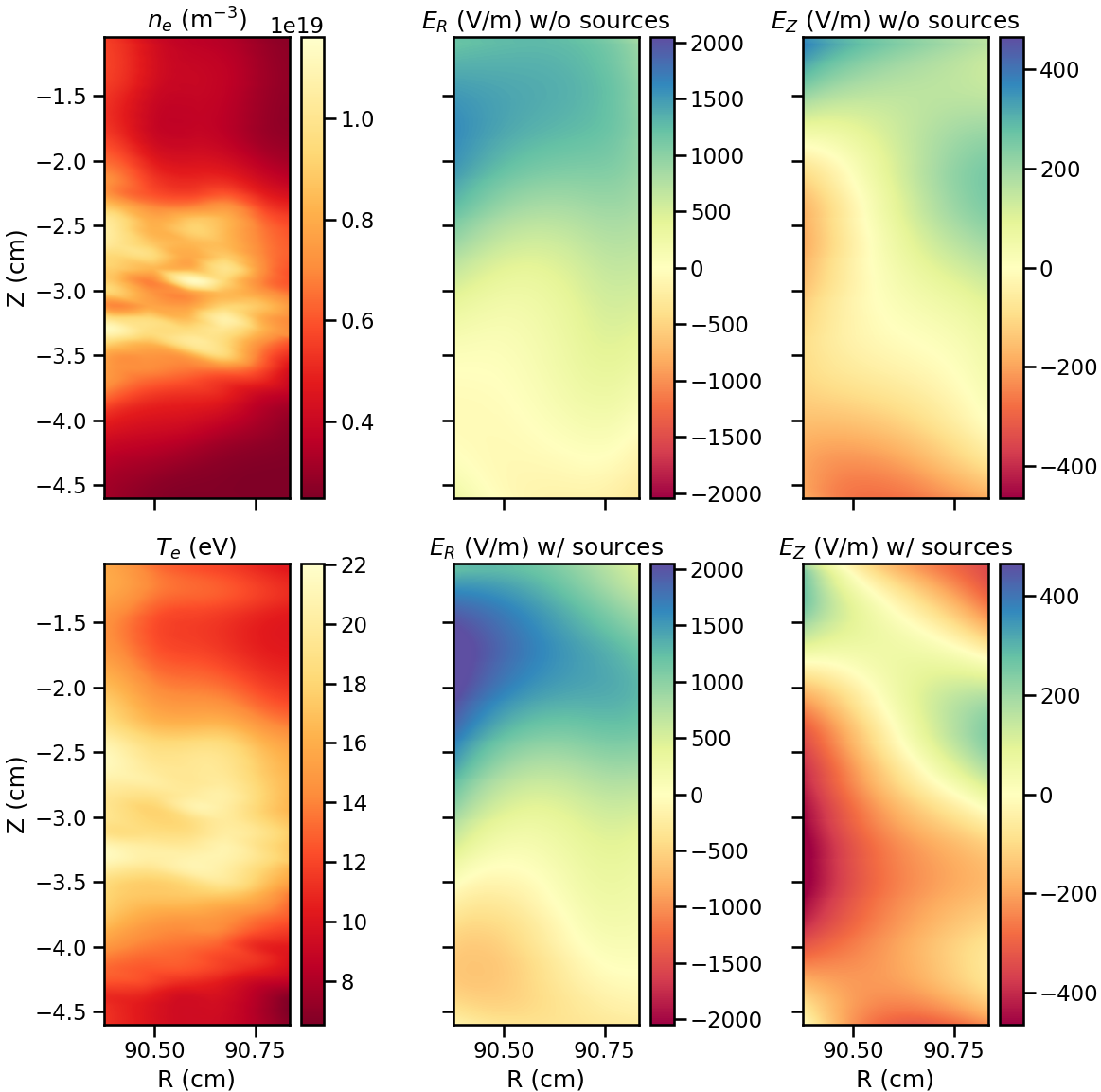}
\caption{\label{observed_neTe_predicted_Er}The 2-dimensional $n_e$ and $T_e$ (top row) are computed from experimental GPI measurements from plasma discharge 1120711021 on Alcator C-Mod at $t = 1.312886$ s \cite{mathews2022deep}. The $E_R$ and $E_Z$ are inferred from drift-reduced Braginskii theory using these experimental $n_e$ and $T_e$ according to the deep learning framework outlined in Chapter 2 in the limiting cases of with (i.e. scaling factor of $n_0^* = 10^{19}$ m$^{-3}$) and without (i.e. $n_0^* = 0$) atomic helium sources.}
\end{figure}

\begin{table}
\centering
\renewcommand{\arraystretch}{1.}
\begin{NiceTabular}{ p{0.6cm}|p{1.1cm}|p{1.1cm}|p{1.1cm}|p{1.1cm}|p{1.1cm}|>{\arraybackslash}p{1.2cm}| }[]
 {} & $n_e$ & $T_e$ & $E_R$&$E_R^{n_0}$ & $E_Z$ & $E_Z^{n_0}$\\ \cline{1-7}
 $n_e$ & 1.000 & 0.971 & -0.001 & 0.141 & 0.001 & -0.203
\\ \cline{1-7}
 $T_e$ & 0.971 & 1.000 & 0.016 & 0.154 & -0.014 & -0.203
\\ \cline{1-7} 
 $E_R$ & -0.001 & 0.016 & 1.000 & 0.850 & 0.343 & 0.268
\\ \cline{1-7}
 $E_R^{n_0}$ & 0.141 & 0.154 & 0.850 & 1.000 & 0.347 & 0.287
\\ \cline{1-7}
 $E_Z$ & 0.001 & -0.014 & 0.343 & 0.347 & 1.000 & 0.882
\\ \cline{1-7}
 $E_Z^{n_0}$ & -0.203 & -0.203 & 0.268 & 0.287 & 0.882 & 1.000
\\
\end{NiceTabular}
\caption[A correlation matrix of the turbulent fluctuations where $n_e$ and $T_e$ are inferred from plasma discharge 1120711021 based upon experimental GPI measurements over $90.3 < R \ \text{(cm)} < 90.9$, $-4.6 < Z \ \text{(cm)} < -1.0$, and $1.312799 < t_{GPI} \ \text{(s)} < 1.312896$.]{\label{table_correlation_matrix}A correlation matrix of the turbulent fluctuations where $n_e$ and $T_e$ are inferred from plasma discharge 1120711021 based upon experimental GPI measurements over $90.3 < R \ \text{(cm)} < 90.9$, $-4.6 < Z \ \text{(cm)} < -1.0$, and $1.312799 < t_{GPI} \ \text{(s)} < 1.312896$. The quantities $E_R^{n_0}$ and $E_R$ ($E_Z^{n_0}$ and $E_Z$) in this table correspond to the radial (vertical) turbulent electric fields predicted by drift-reduced Braginskii theory with and without HeI sources, respectively.}
\end{table}

\begin{figure}[ht]
\includegraphics[width=1.0\linewidth]{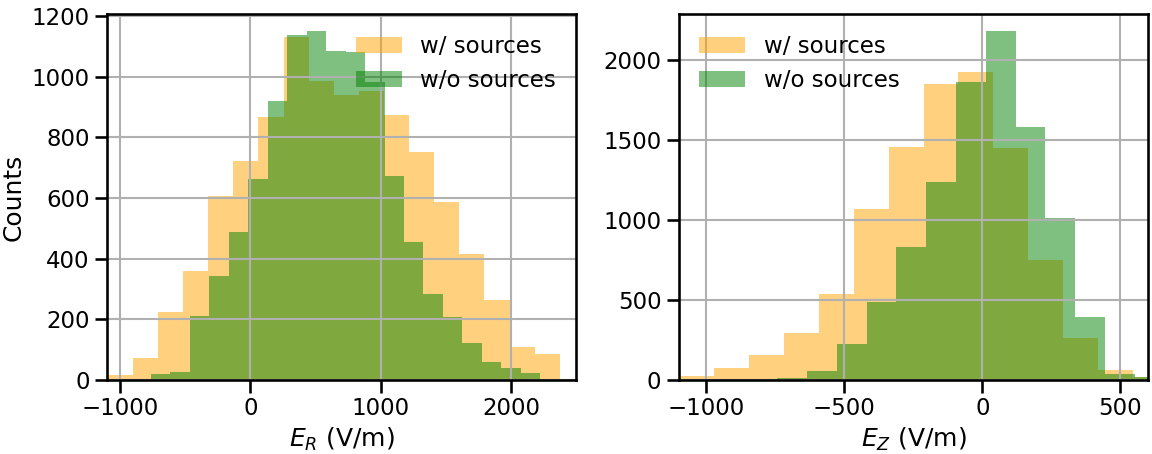}
\caption{\label{histogram_Er}Histograms of $E_R$ and $E_Z$ consistent with drift-reduced Braginskii theory in a toroidal axisymmetric geometry evaluated at the GPI pixels from $90.3 < R \ \text{(cm)} < 90.9$, $-4.6 < Z \ \text{(cm)} < -1.0$, and $1.312799 < t_{GPI} \ \text{(s)} < 1.312896$ in plasma discharge 1120711021 from the Alcator C-Mod tokamak.}
\end{figure}

\begin{figure}[ht]
\includegraphics[width=1.0\linewidth]{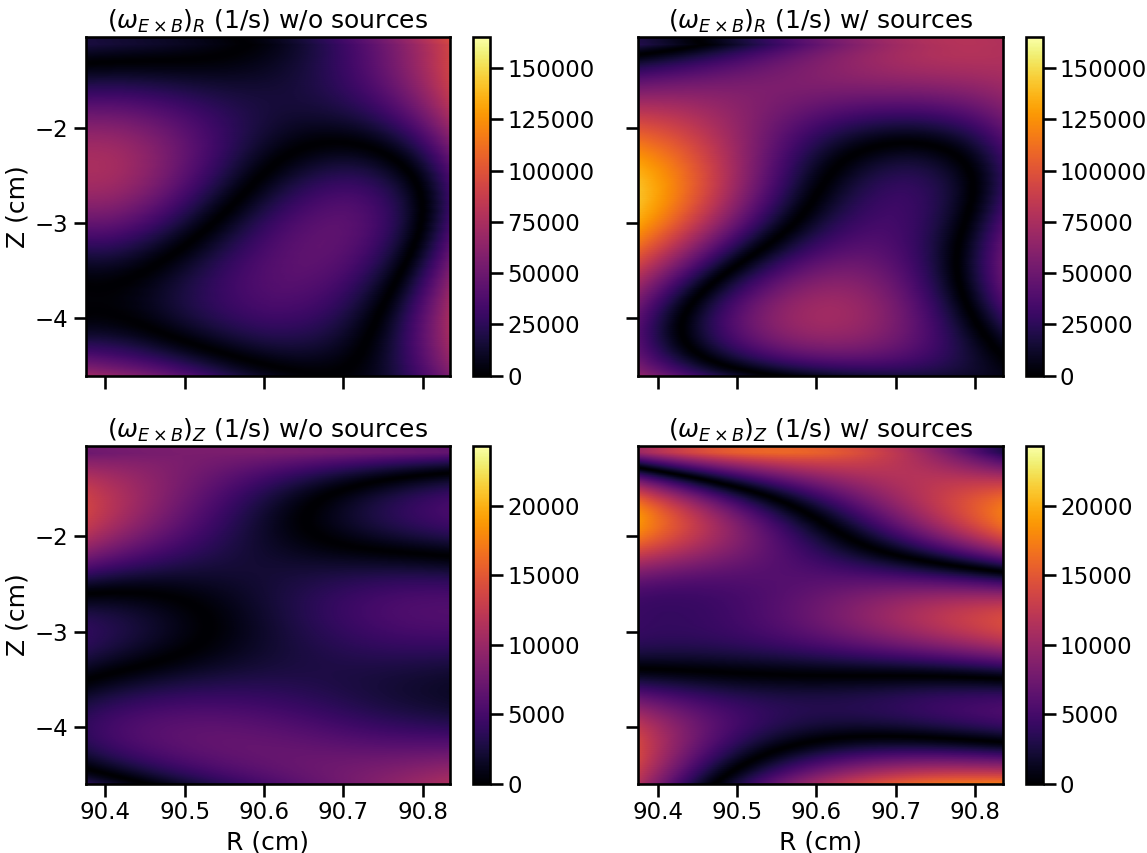}
\caption{\label{ExB_velocity_shear}Visualizations of the radial and vertical turbulence shearing rates predicted by drift-reduced Braginskii theory in discharge 1120711021 at $t = 1.312886$ s on Alcator C-Mod under the assumption of axisymmetry with a purely toroidal magnetic field. The plots consider no sources (left) and, alternatively, neutral sources to account for time-dependent ionization of atomic helium (right).}
\end{figure}

As visualized in Figure \ref{observed_neTe_predicted_Er}, using the 2-dimensional $(R,Z)$-aligned experimentally-inferred $n_e$ and $T_e$ measurements from the helium GPI diagnostic, the time-dependent turbulent electric field predicted by the ascribed drift-reduced Braginskii theory is computed in the limits of (i) no sources and (ii) source effects due to time-dependent ionization of HeI. Since only the relative (and not absolute) brightness of the line emission across the field-of-view of the GPI is known for this plasma discharge, only the structure of the experimental turbulent profile of $n_0$ can be inferred. Nevertheless, as a conservative lower bound on $n_0$ \cite{Terry_private} in the model calculations to test the impacts of neutral dynamics on turbulent fields, the atomic helium density is scaled to an amplitude of approximately $10^{19}$ m$^{-3}$. Much larger scaling factors for $n_0$ (e.g. $10^{20}$ m$^{-3}$) were found to lead to numerical instability in the optimization, which suggests mathematical terms (e.g. poloidal flows) are missing in the reduced turbulence model's equations and/or that such high $n_0$ are unphysical. A matrix of correlation coefficients for these fluctuations is given in Table \ref{table_correlation_matrix}. The correlations between the turbulent electric field and $n_e$ and $T_e$ predicted by the plasma theory in fully ionized conditions are found to be nearly zero. This nonlinear connection changes with the inclusion of plasma-neutral interactions: time-dependent ionization effects due to atomic helium induce a positive (negative) dependence of the computed $E_R$ ($E_Z$) on $n_e$ and $T_e$. If the experimental $n_0$ is truly larger in magnitude, the reported correlations between these dynamical variables are expected to be even stronger.

Further, the addition of neutral helium dynamics to drift-reduced Braginskii theory are found to broaden the distribution of turbulent field magnitudes over the 2-dimensional spatial domain as displayed in Figure \ref{histogram_Er}. This leads to amplified electric field fluctuations with sharper radial variation in the electric potential structure, and manifests as larger ${\bf E \times B}$ flows on turbulent scales in the boundary plasma. Intuitively, this all arises since the observed spatiotemporal evolution of $n_e$ and $T_e$ is not solely due to transport, but instead the self-consistent turbulent ${\bf E \times B}$ flows have to be mathematically balanced in Eqs. \eqref{eq:L_f_n_DotGDBH} and \eqref{eq:L_f_Te_DotGDBH} with sources and sinks. Experimentally, such effects are important for turbulence spreading \cite{Grenfell_2018} and material interactions since even small drifts can compete with flows perpendicular to surfaces at the plasma-sheath interface \cite{edge_flows}. Additionally, the radial and vertical turbulence shearing rates, $({\omega_{\bf E \times B}})_R = \lvert \partial({v_{\bf E \times B}})_Z/\partial R \rvert$ and $({\omega_{\bf E \times B}})_Z = \lvert \partial({v_{\bf E \times B}})_R/\partial Z \rvert$, are elevated on average when atomic helium is present in the edge compared to the case with no time-dependent ionization, i.e. $\langle ({\omega_{\bf E \times B}})_R \rangle$ rises from $4.74 \times 10^4$ to $5.38 \times 10^4$ s$^{-1}$ and $\langle ({\omega_{\bf E \times B}})_Z \rangle$ increases from $9.08 \times 10^3$ to $1.27 \times 10^4$ s$^{-1}$. At intermediate $n_0$ densities, $\langle ({\omega_{\bf E \times B}})_R \rangle$ and $\langle ({\omega_{\bf E \times B}})_Z \rangle$ still increase with $n_0$, although the trend is not strictly linear, as displayed in Table \ref{table_shear}. The modified shearing rates on turbulent scales visualized in Figure \ref{ExB_velocity_shear} can impact shear flow stabilization and cross-field transport of coherent structures. Not including time-dependent neutral dynamics in nonlinear simulations can accordingly mask these effects in edge profile predictions. The amplification of fields due to atomic helium and presence of correlations not present in fully ionized gases demonstrates the importance of neutrals on turbulent scales. They should thus be accounted in experimental tests to precisely validate reduced edge turbulence models, otherwise such errors in predicted fields due to plasma-neutral interactions that scale nonlinearly with $n_0$ will exist.

\begin{table}[ht]
\centering
\renewcommand{\arraystretch}{1.}
\begin{NiceTabular}{ p{2.5cm}|p{1.2cm}|p{1.2cm}|>{\arraybackslash}p{1.2cm} }[]
 {$n_0^* \ (10^{19}$ m$^{-3})$} & $1/4$ & $1/2$&$1$\\ \cline{1-4}
 $\Delta \langle ({\omega_{\bf E \times B}})_R \rangle$ & 0.42\% & 1.48\% & 13.50\%
\\ \cline{1-4}
 $\Delta \langle ({\omega_{\bf E \times B}})_Z \rangle$ & 6.50\% & 13.44\% & 39.87\%
\end{NiceTabular}
\caption{\label{table_shear}Change in nonlinear turbulence shearing rates computed at varying $n_0$ and averaged over the spatiotemporal domain spanned by the camera frames. These calculations of relative change in turbulence shearing rate are with respect to the case with no sources where $\langle ({\omega_{\bf E \times B}})_R \rangle = 4.74 \times 10^4$ s$^{-1}$ and $\langle ({\omega_{\bf E \times B}})_Z \rangle = 9.08 \times 10^3$ s$^{-1}$.}
\end{table}

\section{\label{sec:level5.2}Present limitations and upcoming extensions} 
Due to the axisymmetric toroidal geometry assumed within the drift-reduced Braginskii model, direct comparisons with independent experimental diagnostics from Alcator C-Mod are not yet possible. But going forward, there are several extensions possible in translating these calculations towards empirical testing in magnetic confinement fusion devices to uncover new physics. For example, once flows and geometric effects arising from the poloidal magnetic field \cite{LaBombard_2005_flows} are inserted into the deep learning framework (and validated using modern 3-dimensional codes \cite{3D_geometry_1,3D_geometry_2,giacomin2021gbs}), the predictions from drift-reduced Braginskii theory can be directly compared to available experimental poloidal electric field measurements \cite{mccarthy_thesis}. Such experimental information can then even be used to invert the computational technique to potentially begin learning missing or misrepresented physical terms (e.g. transport coefficients, source functions). The development of edge diagnostics with wide coverage, e.g. probe arrays \cite{TCV_probe,Shao_2018}, capable of measuring radial and poloidal electric fields can thus significantly aid validation efforts especially if magnetically connected with the GPI emission cloud \cite{mathews2022deep}. It is important to underline that the presently used experimental inferences of $n_e$ and $T_e$ come from a 2-dimensional $(R,Z)$-aligned plane. If the vertically-stacked gas tubes utilized for GPI were oriented with the pitch angle of the local magnetic field, then the existing deep learning methodology, which assumes field-aligned 2D observations of $n_e$ and $T_e$, could be directly applied to better approximate the tokamak geometry with the reduced turbulence model. While such diagnostic adjustments are no longer possible on the retired Alcator C-Mod, they can be enacted on existing and upcoming fusion devices. Also, the time-dependent 2-dimensional $n_e$ and $T_e$ are based upon generalized collisional radiative constraints that are agnostic to any turbulence model. This permits the self-consistent learning of time-dependent 2-dimensional profiles for neutral species such as atomic and molecular deuterium \cite{GBS_3fluid_kineticneutrals} via application of existing Monte Carlo transport codes \cite{STOTLER_time_dependent_DEGAS2}, which could be playing a considerable role---as exemplified above by ionization of atomic helium---and can be added into the computational framework akin to that used here for helium. By isolating these effects such as the broadening of turbulent field amplitudes and shearing rates due to atomic helium, essential physics in the development of effective reduced turbulence models can be quantitatively identified. Overall, these initial calculations illustrate a novel pathway towards uncovering unobserved dynamics in experimental fusion plasmas which are conventionally difficult to diagnose. Further, by making no explicit assumptions on boundary conditions or the initializations for turbulent fields within the physics-informed deep learning framework, the nonlinear impacts of approximations (e.g. neglecting time-dependent neutrals) in these chaotic systems can be quantified on turbulent scales.

\chapter{Final conclusions}

Predicting edge plasma profiles is one of the greatest uncertainties in the design of fusion energy devices. To begin reducing uncertainty in turbulence models, by focusing on the defining trait of any nonlinear theory---the connections between dynamical variables---this thesis has demonstrated an original deep learning framework to start examining the quantitative accuracy of turbulent electric field predictions by the widely applied drift-reduced Braginskii model in the edge of fusion devices.


A brief summary of the most important results of this thesis are as follows: First, a novel physics-informed machine learning system was created in Chapter 2 to uncover the unobserved turbulent electric field consistent with drift-reduced Braginskii theory from just partial 2-dimensional observations of the $n_e$ and $T_e$ in a synthetic plasma. This is not otherwise possible using conventional equilibrium models such as the Boltzmann relation or ion pressure balance. Moreover, this computational technique is robust to noisy measurements, which enhances its experimental applicability. 

In Chapter 3, this deep learning technique was utilized to compare the reduced fluid turbulence model against higher fidelity full-$f$ simulations. It demonstrates the first ever direct comparisons of nonlinear fields between two distinct global turbulence models: electrostatic drift-reduced Braginskii theory and long-wavelength electromagnetic gyrokinetics. Good quantitative agreement was confirmed between the independent models in helical plasmas similar to the edge of NSTX. At artificially elevated $\beta$, significant discrepancies in electric fields were not only observed but quantified to demonstrate that the two turbulence models were definitively inconsistent. 

In Chapter 4, a path was embarked on to start translating these techniques to the highest fidelity plasma of all: experiment. For this task, an entirely new optimization scheme based upon a multi-network physics-integrated framework was used to convert brightness measurements of HeI line radiation into local plasma and neutral fluctuations via the merging of transport physics and collisional radiative modelling for the $3^3 D - 2^3 P$ transition in atomic helium. This analysis for ionized gases is highly transferable to both magnetized and unmagnetized environments with arbitrary geometries. This work extends the gas puff imaging approach applied around the globe in fusion plasmas where conventional diagnostics are unable to provide such extended coverage of fluctuations. With this technique, based upon fast camera data on the Alcator C-Mod tokamak, the first 2-dimensional time-dependent experimental measurements of the $n_e$, $T_e$, and $n_0$ on turbulent scales are presented revealing shadowing effects in a fusion plasma using a single spectral line. 

The previous chapters' results are then collectively utilized in Chapter 5 to estimate the 2-dimensional turbulent electric field consistent with (i) drift-reduced Braginskii theory under the framework of an axisymmetric fusion plasma with purely toroidal field and (ii) experimental $n_e$ and $T_e$ measurements via gas puff imaging on Alcator C-Mod. The inclusion of atomic helium effects on particle and energy sources within the reduced turbulence model are found to strengthen correlations between the electric field and plasma pressure. The neutrals are also associated with an observed broadening of the turbulent field amplitudes and increased ${\bf E \times B}$ shearing rates.

Overall, while the goal of improving confidence in predictions of edge $n_e$ and $T_e$ profiles still remains, this thesis has begun the development of a new physics-informed deep learning framework to quantitatively test a commonly used reduced edge turbulence model. In particular, by examining the relationship between $n_e$ and $T_e$ with $\phi$ on turbulent scales, good agreement is found between drift-reduced Braginskii theory and gyrokinetics at conditions relevant to modern tokamaks and novel pathways are opened for precise comparisons with experimental plasmas. This work thus emphasizes the importance of improving aspects of edge codes beyond the equations such as boundary conditions and initialization of simulations. Further, for full confidence in edge $n_e$ and $T_e$ predictions by reduced models, the channels examined need to be extended to all dynamical variables (e.g. $j_{||}$, $T_i$) beyond just $\phi$. Nevertheless, this thesis presents an important and necessary step towards this goal. 

\section{\label{sec:level6.1}Future works}

From a computational physics perspective, there are numerous open questions. For starters, there are uncertainties arising from the intrinsic scatter associated with the stochastic optimization techniques utilized within the deep learning frameworks described in Chapters 2 to 5 to perform these turbulence calculations. While switching from first-order gradient-based methods (e.g. Adam) to approximately second-order (e.g. L-BFGS) aided training, safeguarding convergence in solutions remains a major area for improvement. Novel approaches (e.g. proximal gradient optimization algorithms, generalized loss functions) could help. The observed nonuniqueness may be a natural consequence arising from the sensitivity of the chaotic systems themselves being represented by the networks, but work is ongoing to identify the reasons---both numerical and physical---for this range and to improve precision. To better statistically capture errors, fully propagating experimental uncertainties is to be explored, although this thesis indicates a level of robustness exists to noisy $n_e$ and $T_e$ measurements. And extending the framework to infer relationships between all dynamical variables in the multi-field model with neutrals \cite{GBS_3fluid_kineticneutrals} is required for full testing.


Experimentally, there are several future directions created by this thesis. As noted in Chapter 5, if the gas tubes associated with GPI are aligned with the local magnetic field, then improved reconstructions of the turbulent electric field can be captured using the framework presently outlined in Chapter 2 even in the presence of significant poloidal field. Alternatively, extending the technique to innately handle 3-dimensional geometries and/or training against 2-dimensional $n_e$ and $T_e$ data from highly realistic drift-reduced Braginskii simulations employing proper tokamak geometry and sources as in \cite{giacomin2021gbs,GBS_3fluid_kineticneutrals,Zholobenko_2021_validation} could enable this work to use the existing GPI data for comparison of turbulent electric fields with available probe measurements. These increasingly sophisticated full device simulations include self-consistent kinetic neutrals (i.e. D$_2$, D) along with charged molecular species (i.e. D$_2^+$). Testing with these effects accounted will improve the overall applicability of this model to extract the turbulent electric field on new fusion devices. As an immediate extension, by coupling the learned $n_e$ and $T_e$ from Chapter 4---which uses a technique completely oblivious to plasma turbulence beyond the criteria outlined---with Monte Carlo neutral transport codes \cite{STOTLER_time_dependent_DEGAS2}, this can recover 2-dimensional time-dependent estimates of atomic and molecular species in experiment on turbulent scales. Emissivity predictions, e.g. of Ly$_\alpha$ radiation, could then be compared against experimental measurements of deuterium line emission as a secondary test of validity \cite{Lyalpha0,Lyalpha1}. These neutral profiles could also be applied in the experimental testing of reduced turbulence models.

Further, the framework outlined in Chapter 4 yields new experimental pathways altogether for the GPI diagnostic. Efforts to transfer this analysis technique to existing devices with variable geometry (e.g. W7-X, TCV, DIII-D) can be tackled right away. Widening the domain towards the confined plasma to analyze pedestal dynamics will likely require significant advancements to the computational framework to simultaneously handle the multiscale behaviour in fluctuations and equilibrium profiles \cite{Fourier_multirate,wang2020understanding,wang2020eigenvector,wang2022causality}. As an example, Appendix B outlines a deep Gaussian process capable of learning transient and steep features (e.g. formation of pedestals across transitions between confinement regimes) in otherwise slowly evolving profiles. Bayesian integration of networks with such experimental information on macroscopic scales from independent plasma diagnostics could help augment their ability to learn edge dynamics. In addition, if not generalizing the GPI analysis technique to 3 dimensions, applying highly collimated HeI beams is sought for good reconstruction of the turbulent $n_e$ and $T_e$. On this point, the encoded velocity closure for neutral transport should be carefully examined---or perhaps even learned by networks---in novel experimental scenarios. A spectrometer with sufficiently high spectral resolution could also experimentally measure the Doppler shift. Finally, while the deep learning framework for GPI analysis can technically utilize deuterium line emission instead of puffing HeI, all the criteria listed in Chapter 4 would need to be re-visited and suitably developed to account for effects such as charge-exchange and recombination.

\section{\label{sec:level6.2}Last remarks}

While we began to understand the structure and potential of atomic nuclei just about one century ago, we are only now starting to delve into the structure and potential of neural networks. The practical ability for these artificial circuits to represent physical systems such as ionized gases permits this thesis to no longer use power laws but partial differential equations as regression constraints when training on spatiotemporally evolving (simulation or experimental) data on turbulent scales beyond 0-dimensional quantities. But unlocking their potential may not always be simple. Turbulence is fundamentally complex, and so may be the tools used to model it. Nevertheless, these tools can provide useful representations and insights not otherwise easy to uncover to view plasma dynamics in altogether new ways. This thesis itself represents just part of the beginning of a powerful technique with computational graphs that can solve essential problems in turbulence and the confinement of fusion energy systems to help continue providing breakthroughs in advancing human knowledge and technology.

\setlength{\epigraphwidth}{0.6\textwidth}
\epigraph{The hardest problems we have to face do not come from philosophical questions about whether brains are machines or not. There is not the slightest reason to doubt that brains are anything other than machines with enormous numbers of parts that work in perfect accord with physical laws. As far as anyone can tell, our minds are merely complex processes. The serious problems come from our having had so little experience with machines of such complexity that we are not yet prepared to think effectively about them.}{Marvin Minsky}

\newpage

\setlength{\epigraphwidth}{0.6\textwidth}
\epigraph{Let us return for a moment to Lady Lovelace’s objection, which stated that the machine can only do what we tell it to do. One could say that a man can ``inject'' an idea into the machine, and that it will respond to a certain extent and then drop into quiescence, like a piano string struck by a hammer. Another simile would be an atomic pile of less than critical size: an injected idea is to correspond to a neutron entering the pile from without. Each such neutron will cause a certain disturbance which eventually dies away. If, however, the size of the pile is sufficiently increased, the disturbance caused by such an incoming neutron will very likely go on and on increasing until the whole pile is destroyed. Is there a corresponding phenomenon for minds, and is there one for machines? There does seem to be one for the human mind. The majority of them seem to be ``sub critical,'' i.e. to correspond in this analogy to piles of sub-critical size. An idea presented to such a mind will on average give rise to less than one idea in reply. A smallish proportion are supercritical. An idea presented to such a mind may give rise to a whole ``theory'' consisting of secondary, tertiary and more remote ideas. Animals’ minds seem to be very definitely sub-critical. Adhering to this analogy we ask, ``Can a machine be made to be super-critical?''}{Alan Turing}






\newpage\phantom{blabla}
\appendix
\chapter{Normalization of drift reduced Braginskii fluid theory}

\setlength{\epigraphwidth}{0.7\textwidth}
\epigraph{Keep computations to the lowest level of the multiplication table.}{David Hilbert}

Converting the drift-reduced Braginskii equations from physical units to a normalized form is useful to numerically solve the model with both finite difference schema and physics-informed machine learning codes. For completeness, the full normalization procedure is shown below. The physical variables and all associated quantities are transformed according to \cite{francisquez2020fluid}
\begin{align} \label{eq:normSMT1}
\begin{split}
  n &\leftarrow n/\nRef, \\
  \phi &\leftarrow \phi/\phi_0, \\
\end{split}
\quad
\begin{split}
  T_{s} &\leftarrow T_{s}/T_{s0}, \\
  v_{\parallel s} &\leftarrow v_{\parallel s}/c_{s0}, \\
\end{split}
\end{align}
where $n_0 = 5 \times 10^{19} \text{ m}^{-3}$, $T_{s0} = 25 \text{ eV}$, $c_{s0}^2 = T_{s0}/m_i$, $\phi_0 = \BRef a_0^2/c\tRef$, and $\tRef = \sqrt{a_0 R_c/2}/c_{e0}$ is the interchange-like reference timescale. To match the simulation with experimental edge parameters of the Alcator C-Mod tokamak, $\BRef = B_{axis} R_0 /(R_0 + a_0)$ and $R_c = R_0 + a_0$. This in turn defines the following dimensionless constants

\begin{eqnal}
\!\begin{aligned}
  \epa &= \frac{2a}{R_c}, \\
  \epv &= \frac{c_{e0}\tRef}{R_c}, \\
  \tau &= \frac{\TiRef}{\TeRef},
\end{aligned}
\qquad
\!\begin{aligned}
  \kappai &= 3.9\frac{2}{3}\frac{\TiRef\tRef}{m_i R_c^2\nu_{i0}}, \\
  \eta &= 0.51\nu_{e0}\tRef, \\
  \kappae &= 3.2\frac{2}{3}\frac{\TeRef\tRef}{\nu_{e0}m_e R_c^2},
\end{aligned}
\qquad
\!\begin{aligned}
  \alphad &= \frac{\TeRef c\tRef}{e\BRef a^2}, \\
  \epg &= \frac{0.08\tau}{\nu_{i0} t_0}, \\
  \epge &= \frac{0.73}{12\nu_{e0}t_0}
\end{aligned}
\end{eqnal}
where $c$ and $\nu_{s 0}$ denote the speed of light and collision rate \cite{Huba2013}, respectively. The spatiotemporal coordinates are normalized by the following conversions
\begin{align} \label{eq:normXYZT}
\begin{split}
  x &\leftarrow x/a_0, \\
  z &\leftarrow z/R_0, \\
\end{split}
\quad
\begin{split}
  y &\leftarrow y/a_0, \\
  t &\leftarrow t/t_0. \\
\end{split}
\end{align}
With these transformations, the unitless equations numerically solved are
\begin{eqnal}\label{eq:normlnDotGDBH}
\d{^e\ln n}{t} = -\epa\left[\curv{\phi}-\alphad\frac{\curv{p_e}}{n}\right] -\epv\delpar{\vpe}+\frac{1}{n}\nSrcN+\mathcal{D}_{\ln n}
\end{eqnal}
\begin{eqnal}\label{eq:normwDotGDBH}
\pd{\gvort}{t} &= \curv{p_e}+\tau\curv{p_i}+\frac{\epv}{\alphad\epa}\delpar{\cur}-\epg\curv{G_i} \\
&\quad-\div{\lbrace\frac{n}{B^3}\left[\phi,\gradperp{\phi}+\tau\alphad\frac{\gradperp{p_i}}{n}\right]+ \\
&\quad\sqrt{\tau}\epv\frac{n}{B^2}\vpi\delpar{\left(\gradperp{\phi}+\tau\alphad\frac{\gradperp{ p_i}}{n}\right)}\rbrace}+\mathcal{D}_{\gvort}
\end{eqnal}
\begin{eqnal}\label{eq:normvpeDotGDBH}
\d{^e\vpe}{t} &= \frac{m_i}{m_e}\epv\left(\frac{1}{\alphad}\delpar{\phi}-\frac{\delpar{p_e}}{n}-0.71\delpar{T_e}\right) \\ &\quad +4\epv\epge\frac{m_i}{m_e}\frac{\delpar{G_e}}{n} +\epa\alphad T_e\curv{\vpe} +\eta\frac{\cur}{T_e^{3/2}} +\momSrce +\mathcal{D}_{\vpi}
\end{eqnal}
\begin{eqnal}\label{eq:normvpiDotGDBH}
\d{^i\vpi}{t} &= -\frac{\epv}{\sqrt{\tau}}\left(\frac{1}{\alphad}\delpar{\phi}+\tau\frac{\delpar{p_i}}{n}-0.71\delpar{T_e}\right) \\
&\quad +\frac{4\epv\epg}{\sqrt{\tau}}\frac{\delpar{G_i}}{n} -\epa\tau\alphad T_i\curv{\vpi} -\frac{m_e}{m_i}\frac{\eta}{\sqrt{\tau}}\frac{\cur}{T_e^{3/2}} +\momSrci+\mathcal{D}_{\vpi}
\end{eqnal}
\begin{eqnal}\label{eq:normlTeDotGDBH}
\d{^e\ln T_e}{t} &= \frac{5}{3}\epa\alphad\curv{T_e}+\frac{\kappa^e}{p_e}\delpar{T_e^{7/2}\delpar{\ln T_e}} -\frac{2}{3}\epv\delpar{\vpe} \\  &+\frac{2}{3n}\left[0.71\epv\left(\delpar{\cur}-\cur\delpar{\ln T_e}\right) +\frac{m_e}{m_i}\eta\frac{\cur^2}{T_e^{5/2}}\right] \\ &-\frac{2}{3}\epa\left[\curv{\phi}-\alphad\frac{\curv{p_e}}{n}\right] +\frac{2}{3}\frac{1}{p_e}\enerSrceN +\mathcal{D}_{\ln T_e}
\end{eqnal}
\begin{eqnal}\label{eq:normlTiDotGDBH}
\d{^i\ln T_i}{t} &= -\frac{5}{3}\tau\epa\alphad \curv{T_i}+\frac{\kappa^i}{p_i}\delpar{T_i^{7/2}\delpar{\ln T_i}} +\frac{2}{3}\frac{1}{p_i}\enerSrciN+\mathcal{D}_{\ln T_i}\\ &+\frac{2}{3}\left\lbrace -\epa\left[\curv{\phi}-\alphad\frac{\curv{p_e}}{n}\right] -\sqrt{\tau}\epv\delpar{\vpi} +\epv\frac{\delpar{\cur}}{n}\right\rbrace,
\end{eqnal}
where the normalized diffusivities applied for all dynamical variables in only Chapter 2 are $\chi_x = -4.54 \times 10^{-10}$, $\chi_y = -1.89 \times 10^{-9}$, and $\chi_z = -8.91 \times 10^{-3}$. The normalized evolution equations given by \eqref{eq:normlnDotGDBH} and \eqref{eq:normlTeDotGDBH} are the physical model constraints learnt in the machine learning framework employed in Chapters 2, 3, and 5. 

A few subtle yet importance differences exist between the physical theory posed and the construction of the synthetic plasma in Chapter 2. One deviation between the theorized plasma and the one produced computationally is that the numerical code actually evolves the logarithmic form of $n$, $T_e$, and $T_i$ to enforce positivity and the high order diffusion operators act on these logarithmic quantities, too. While equivalent analytically, this choice numerically forces the drift-reduced Braginskii equations to be posed and solved in non-conservative form by the finite difference solver. Consequent errors due to numerical approximation can manifest as unexpected artificial sources or sinks in the simulation domain \cite{francisquez2020fluid}. In addition, simulation boundaries applied in practice only approximately satisfy the zero flux conditions when employing even- and odd-symmetry conditions on a cell-centered discretized grid \cite{francisquez2020fluid}. These computational discrepancies can cause potential misalignment between inferred dynamics using idealized theory and numerical modelling of the synthetic plasma's turbulent fields. Physics-informed deep learning can overcome these numerical limitations when representing plasma theory since positivity can be intrinsically encoded in the network. Further, it employs a continuous spatiotemporal domain and the nonlinear continuum equations represented by \eqref{eq:f_n_DotGDBH} and \eqref{eq:f_Te_DotGDBH} are consequently evaluated exactly up to computer precision \cite{Raissi_JMLR}. Unphysical numerical dissipation in observational data can therefore present deviations from reflecting the sought theory, but reasonable agreement is nevertheless found when analyzing the synthetic measurements with the partial differential equations embedded in the machine learning framework.
\clearpage
\newpage

\chapter{Quantifying experimental profile evolution via multidimensional Gaussian process regression}

\setlength{\epigraphwidth}{0.875\textwidth}
\epigraph{I remember my friend Johnny von Neumann used to say, `with four parameters I can fit an elephant and with five I can make him wiggle his trunk.'}{Enrico Fermi, as quoted by Freeman Dyson}

The edge density and temperature profiles of tokamak plasmas are strongly correlated with energy and particle confinement and resolving these profiles is fundamental to understanding edge dynamics. These quantities exhibit behaviours ranging from sharp plasma gradients and fast transient phenomena (e.g. transitions between low and high confinement regimes) to nominal stationary phases. Analysis of experimental edge measurements motivates robust fitting techniques to capture dynamic spatiotemporal evolution. Additionally, fusion plasma diagnostics have intrinsic measurement errors and data analysis requires a statistical framework to accurately quantify uncertainties. Appendix B outlines a generalized multidimensional adaptive Gaussian process routine capable of automatically handling noisy data and spatiotemporal correlations. This technique focuses on the edge-pedestal region in order to underline advancements in quantifying time-dependent plasma structures including transport barrier formation on the Alcator C-Mod tokamak. Outputs from this regression can be used to physically inform and prime neural networks about background profiles.

Automatically generating accurate pedestal plasma density and temperature profiles requires handling large quantities of noisy observations. Current fitting routines in the fusion community involve a range of methods including nonlinear least squares via modified hyperbolic tangent functions \cite{Groebner_2001}, cubic splines \cite{Diallo_2011}, and various Bayesian techniques \cite{Fischer_2010}. Past Gaussian process (GP) regression codes typically fixed covariance function length scales \cite{Chilenski_2015} and generally permitted only one-dimensional scenarios to build radial profiles. This requires filtering or averaging temporal variation in data which can be limiting in the edge especially when analyzing transient phenomena such as spontaneous transitions of confinement regimes and transport barrier formation \cite{Mathews_2019}. Capability to capture both mean spatial and temporal variations of edge plasma profiles and associated gradients is therefore sought. Towards this task a deep multidimensional heteroscedastic GP routine was developed to provide automated fitting and uncertainty estimates from the Thomson scattering diagnostic on the Alcator C-Mod tokamak. Evolution of both plasma density and temperature is tracked across the edge-pedestal region with varying length scales typical of experimental data including the formation of both particle and energy transport barriers. This technique has the capability to routinely process thousands of discharges automatically to yield profile statistics and be run across novel experiments. The methodology and accompanying mathematical proofs are provided along with demonstrations on experimental data from Alcator C-Mod exhibiting transport barriers.

\section{Method}
The technique applied for reconstructing edge-pedestal plasma profiles is an adaptive heteroscedastic multidimensional GP routine. Each of these terms are individually defined and outlined below to introduce the scope of this method. 
\subsection{Gaussian process}
A GP is a supervised learning method capable of solving classification and regression problems. The capability of fitting profiles via nonlinear probabilistic regression is the main focus of this appendix. In particular, the underlying assumption is that the variable (e.g. plasma density) being predicted at a certain location is normally distributed and spatiotemporally correlated with neighbouring points, indicating that partial observations provide information at nearby locations for conditioning future predictions. The function space definition of a GP is that any finite collection of the random variables modelled follow a joint Gaussian distribution, i.e. $(y_i,y_j) \sim \mathcal{N}(\mu,\Sigma)$ and any subset is given by $y_i \sim f({\bf x}_i) + \sigma_n\mathcal{N}(0,I)$ \cite{Rasmussen_2005}, where $\sigma^2_n$ is the noise variance. A GP is specified entirely up to its second-order statistics as denoted by the mean, $\mu({\bf x}_i)$, and covariance, $\Sigma({\bf x}_i,{\bf x}_j)$. Consequently, conditional predictions at ${\bf x}_b$ based upon observations at ${\bf x}_a$ are analytically given by \cite{Rasmussen_2005}:
\begin{align}
\label{GPR_machinery1}
{\boldsymbol\mu}_{{ \bf y}_b \vert { \bf y}_a} = \boldsymbol\mu_b+\Sigma_{b,a}{\Sigma_{a,a}}^{-1}({\boldsymbol y_a}-\boldsymbol\mu_a)
\end{align}
\begin{align}
\label{GPR_machinery2}
{\Sigma}_{{ \bf y}_b \vert { \bf y}_a} = \Sigma_{b,b}-\Sigma_{b,a}{\Sigma_{a,a}}^{-1}\Sigma_{a,b}
\end{align}
To provide a brief proof of (\ref{GPR_machinery1}) and (\ref{GPR_machinery2}), one should note that a normally distributed random variable, ${\bf y}$, in $N$-dimensions is modelled by \cite{Rasmussen_2005}
\begin{equation}
P({ \bf y} \vert {\boldsymbol\mu}, { \bf \Sigma}) = \frac {1}{2\pi^{N/2} \lvert { \bf  \Sigma} \rvert^{1/2}} \exp[-\frac{1}{2}({ \bf y}-{\boldsymbol\mu})^T { \bf \Sigma}^{-1} ({ \bf y}-{\boldsymbol\mu})],
\end{equation}
where ${ \bf \Sigma}$ is a positive semi-definite covariance matrix, ${\boldsymbol\mu} = [\mu_1, ..., \mu_N]$, ${ \bf y} = [y_1, ..., y_N]$, and ${\bf \Sigma}_{i,j} = \mathbb{E}[(y_i - \mu_i)(y_j - \mu_j)]$. In this formalism, the conditional probability of predicting a new point (or set of points), ${\bf y}_b$, can be ascertained from an observed point (or set of points), ${ \bf y}_a$, through the posterior distribution: ${ \bf y}_b \vert { \bf y}_a \sim \mathcal{N}({\boldsymbol\mu}_{{ \bf y}_b \vert { \bf y}_a},{\Sigma}_{{ \bf y}_b \vert { \bf y}_a})$. 

 Following the treatment in \cite{Tso2010}, \eqref{GPR_machinery1} and \eqref{GPR_machinery2} can be derived by defining ${\bf z} \equiv {\bf y}_b + {\bf A} {\bf y}_a $ where ${\bf A} \equiv -\Sigma_{b,a} \Sigma^{-1}_{a,a}$, implying 
\begin{align} 
{\rm cov}({\bf z}, {\bf y}_a) &= {\rm cov}( {\bf y}_{b}, {\bf y}_a ) + 
{\rm cov}({\bf A}{\bf y}_a, {\bf y}_a) \nonumber \\
&= \Sigma_{b,a} + {\bf A} {\rm var}({\bf y}_a) \nonumber \\
&= \Sigma_{b,a} - \Sigma_{b,a} \Sigma^{-1}_{a,a} \Sigma_{a,a} \nonumber \\
&= 0
\end{align}
Consequently, ${\bf z}$ and ${\bf y}_a$ are uncorrelated and, since they are assumed jointly normal in GP regression, they are independent. It is evident $\mathbb{E}[{\bf z}] = {\boldsymbol \mu}_b + {\bf A}  {\boldsymbol \mu}_a$, and it follows that the conditional mean can be expressed as
\begin{align}
\mathbb{E}[{\bf y}_b | {\bf y}_a] &= \mathbb{E}[ {\bf z} - {\bf A} {\bf y}_a | {\bf y}_b] \nonumber \\
& = \mathbb{E}[{\bf z}|{\bf y}_a] -  \mathbb{E}[{\bf A}{\bf y}_a|{\bf y}_a] \nonumber \\
& = \mathbb{E}[{\bf z}] - {\bf A}{\bf y}_a \nonumber \\
& = {\boldsymbol \mu}_b + {\bf A}  ({\boldsymbol \mu}_a - {\bf y}_a) \nonumber \\
& = {\boldsymbol \mu}_b + \Sigma_{b,a} \Sigma^{-1}_{a,a} ({\bf y}_a- {\boldsymbol \mu}_a)
\end{align}
which proves the conditional mean \eqref{GPR_machinery1}, and 
\begin{align}
{\rm var}({\bf x}-{\bf D}{\bf y}) &\equiv {\rm var}({\bf x}) + {\bf D}{\rm var}({\bf y}){\bf D}^T \nonumber - {\rm cov}({\bf x},{\bf y}){\bf D}^T - {\bf D}{\rm cov}({\bf y},{\bf x})
\end{align}
implies
\begin{align}
{\rm var}({\bf y}_b|{\bf y}_a) &= {\rm var}({\bf z} - {\bf A} {\bf y}_a | {\bf y}_a) \nonumber \\
&= {\rm var}({\bf z}|{\bf y}_a) + {\rm var}({\bf A} {\bf y}_a | {\bf y}_a) \nonumber - {\bf A}{\rm cov}({\bf y}_a, {\bf z}) - {\rm cov}({\bf z}, {\bf y}_a) {\bf A}^T \nonumber \\
&= {\rm var}({\bf z}|{\bf y}_a) = {\rm var}({\bf z})
\end{align}
Plugging the above result into the conditional variance yields \eqref{GPR_machinery2},
\begin{align}
{\rm var}({\bf y}_b|{\bf y}_a) &= {\rm var}( {\bf y}_b + {\bf A} {\bf y}_a ) \nonumber \\
&= {\rm var}( {\bf y}_b ) + {\bf A} {\rm var}( {\bf y}_a ) {\bf A}^T + {\bf A} {\rm cov}({\bf y}_b,{\bf y}_a) + {\rm cov}({\bf y}_a,{\bf y}_b) {\bf A}^T \nonumber \\
&= \Sigma_{b,b} +\Sigma_{a,b} \Sigma^{-1}_{a,a} \Sigma_{a,a}\Sigma^{-1}_{a,a}\Sigma_{a,b} - 2 \Sigma_{b,a} \Sigma_{a,a}^{-1} \Sigma_{a,b} \nonumber \\
&= \Sigma_{b,b} +\Sigma_{b,a} \Sigma^{-1}_{a,a}\Sigma_{a,b}
\nonumber - \indent 2 \Sigma_{b,a} \Sigma_{a,a}^{-1} \Sigma_{a,b} \nonumber \\ 
&= \Sigma_{b,b} -\Sigma_{b,a} \Sigma^{-1}_{a,a}\Sigma_{a,b}
\end{align}
Therefore, all that is required for the GP machinery are priors over ${\bf y}_a$ and ${\bf y}_b$ to obtain $\boldsymbol\mu_a$ and $\boldsymbol\mu_b$, respectively, and a covariance function which is defined in this paper by a heteroscedastic uncertainty component, $\epsilon$, along with an adaptive kernel function, $k({\bf x}_i,{\bf x}_j)$. The resulting covariance, $\Sigma_{i,j} = k({\bf x}_i,{\bf x}_j) + \epsilon({\bf x}_i)  \delta_{i,j}$, describes similarity between data points and accounts for spatiotemporal correlations in the data since ${\bf x}$ represents the independent variables (e.g. $\psi$ and $t$). The hyperparameters of the prescribed adaptive kernel function are optimized through maximum {\it a posteriori} estimation on each individual discharge's observed data. Practically, given an experimental set of noisy measurements at arbitrary positions and times, this formalism allows inference of expected values across the spatiotemporal domain. 

Suitably selecting the kernel function, $k({\bf x}_i,{\bf x}_j)$, representing the full covariance is critical to the GP since it specifies the correlation between any pairs of random variables. Constraining kernels will consequently limit the range of behaviour that can be captured by the GP which may only be physically warranted in certain scenarios. To remain robust to tracking a wide range of spatiotemporal behaviour, an adaptive heteroscedastic kernel is optimized against experimental data. Utilizing alternative distributions (e.g. log-normal) and latent variable transformations can also permit scenarios with non-Gaussian residuals \cite{wang2012gaussian}, although Gaussianity is assumed here.

\subsection{Adaptivity}
GP regression is a nonparametric method without an explicit functional form. Nonparametric in this context means that there are no fixed number of constraining model parameters but instead the fitting routine becomes increasingly constrained as training data increases. Correlations between observed data points are based upon the prescribed covariance function. Various kernels have been proposed to embed this structure ranging from a Gaussian function (for expectedly smooth behaviour) to periodic functions (for expectedly cyclic behaviour) to combinations of multiple kernels (e.g. for automatic relevance determination) \cite{DD_thesis}. Despite the generic regression technique lacking a strictly fixed functional form, the optimized hyperparameters defining the kernel are typically constrained themselves. For example, a standard stationary isotropic Mat\'ern kernel is defined by
\begin{equation}
    k({\bf x}_i,{\bf x}_j) = \sigma^2_k\frac{2^{1-\nu}}{\Gamma(\nu)}\Bigg(\sqrt{2\nu}\frac{|{\bf x}_i - {\bf x}_j| }{\rho}\Bigg)^\nu K_\nu\Bigg(\sqrt{2\nu}\frac{|{\bf x}_i - {\bf x}_j|}{\rho}\Bigg)
\end{equation}
where $\Gamma$ is the gamma function, $K_\nu$ is the modified Bessel function of the second kind, and $\rho$ and $\nu$ are non-negative globally constant hyperparameters which control spatial range and smoothness, respectively. A Mat\'ern kernel is $\nu - 1$ times differentiable and reduces to a Gaussian kernel in the limit $\nu \rightarrow \infty$ while becoming an exponential kernel when $\nu = 1/2$ \cite{Genton_2002}. It resultantly covers a wide class of kernels. Nevertheless, the hyperparameters are quite restrictive if simply constants \cite{heinonen16,2d-gpr-tomography}. Therefore, a version of the generalized nonstationary Mat\'ern kernel is encoded as \cite{Paciorek_2006,Plagemann_2008}:
\begin{equation}
\label{adaptive_kernel}
    k({\bf x}_i,{\bf x}_j) = 
    \sigma^2_k \frac{2^{1-\nu}}{\Gamma(\nu)}
    \frac{|\rho_i|^{1/2}|\rho_j|^{1/2}}{\sqrt{\frac{1}{2}\rho^2_i + \frac{1}{2}\rho^2_j}} 
    \Bigg(2\sqrt{\frac{2\nu |{\bf x}_i - {\bf x}_j|^2}{\rho^2_i + \rho^2_j}}\Bigg)^\nu K_\nu\Bigg(2\sqrt{\frac{2\nu |{\bf x}_i - {\bf x}_j|^2}{\rho^2_i + \rho^2_j}}\Bigg)
\end{equation}
where $\rho$ varies across the entire multidimensional domain and adapts to optimize the length scale based upon the experimental data being trained upon in each individual plasma discharge. The kernel accomplishes learning point estimates of local smoothness by representing the primary GP's locally isotropic length scale hyperparameter by a secondary GP with radial basis function (RBF) kernel that allows global variation of $\rho$ across the spatiotemporal grid. It is this second-level GP which introduces the notion of a deep process and adaptivity to the overall regression technique. A stationary kernel is purely a function of $\bf{x}_i - \bf{x}_j$, while additional local dependence exists in (\ref{adaptive_kernel}) through $\rho$ which introduces nonstationary behaviour \cite{Plagemann_2008}. A k-means clustering algorithm is used for training of the secondary GP which parametrizes the nonstationary covariances with latent length-scales \cite{Plagemann_2008}.

\begin{figure}
\centering
\includegraphics[width=0.495\linewidth]{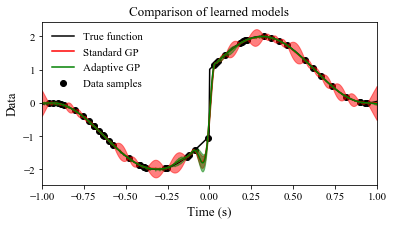}
\includegraphics[width=0.49\linewidth]{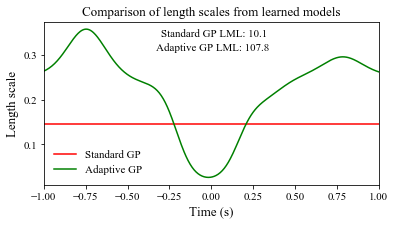}
\caption{Comparison of two separate GPs simply training on 1-dimensional data: one employs a standard Mat\'ern kernel while the other includes an adaptive length scale. Fits to the original data samples (left) and computed length scales (right) are displayed. Figure courtesy of J.H. Metzen.}
\label{fig:1dlls}
\end{figure}
\begin{figure} 
\centering
\includegraphics[width=0.48\linewidth]{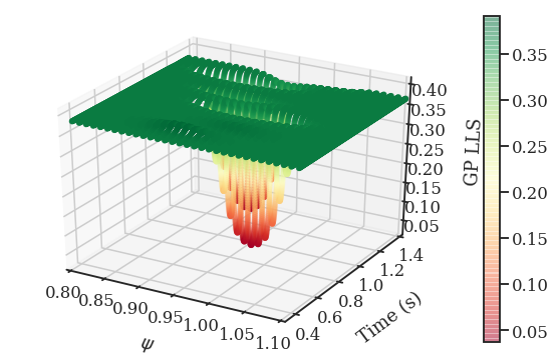}
\includegraphics[width=0.48\linewidth]{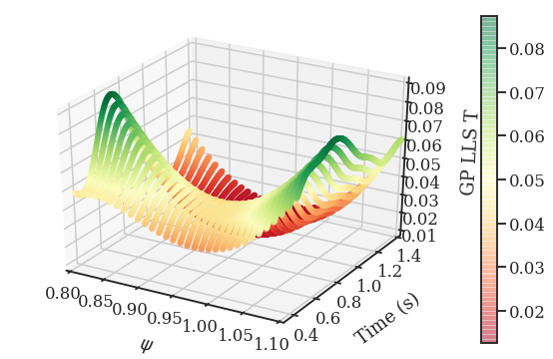}

\caption{Two examples of learning length scales that vary across the spatiotemporal domain by training identical GP models on experimentally measured electron density (left) and temperature (right) from the Thomson scattering diagnostic.}
\label{fig:2dlls}
\end{figure}

Figure \ref{fig:1dlls} demonstrates a basic 1-dimensional example of a sinusoidal function with imposed discontinuity. The advantage conferred by the adaptive length scale can be quantitatively observed by comparing the log marginal likelihood (LML) \cite{Rasmussen_2005} between a standard Mat\'ern kernel and one with locally adaptive length scale. The order of magnitude improvement of LML (which is a logarithmic quantity) indicated in Figure \ref{fig:1dlls} occurs because a stationary Mat\'ern kernel is forced to decrease its constant global length scale considerably while an adaptive length scale permits reducing its value locally only near the discontinuity. The adaptive length scales not only provide the capability to better capture singular or transient phenomena on otherwise slowly-varying profiles but importantly improves uncertainty estimates across the domain. The code sets $\nu = 3/2$, which allows for potentially stiff behaviour and this value can be modified, if sought. It can also be kept as a variable for optimization to capture an entire spectrum of kernel functions. The user can freely specify upper and lower bounds on length scales to be learned across the grid which are given uniform prior distributions during training. 
As a preview for the application of these methods, the length scales can be extended to multidimensional scenarios as depicted in Figure \ref{fig:2dlls} where the learned input length scale varies across the entire spatial and temporal domain. The exact same kernel was initialized for both the data sets used in Figure \ref{fig:2dlls}, but since the datasets were themselves different, the locally learned length scales vary. A custom stochastic optimizer based on differential evolution is used in these examples for GP hyperparameter-tuning and finally polished off with gradient-based descent. Finding a global optimum in the likelihood function is not guaranteed, therefore this is helpful because the loss function may be multimodal and challenging for simple gradient-based methods acting on non-convex problems. 


    
\subsection{Heteroscedasticity}
\begin{figure}
\centering
\includegraphics[width=0.47\linewidth]{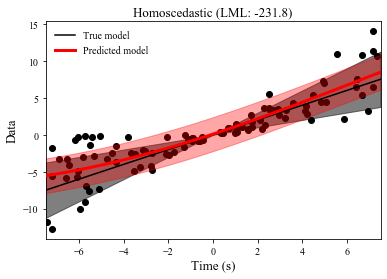} 
\includegraphics[width=0.47\linewidth]{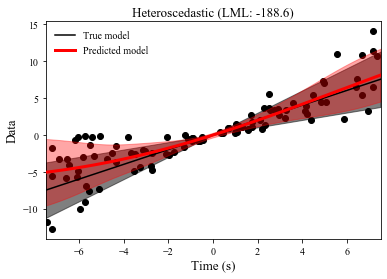}
\caption{Comparison of two separate GPs on 1-dimensional data both using an RBF kernel. The difference is that one utilizes an additional homoscedastic component (left) while the other GP uses a heteroscedastic component (right) in the full covariance structure. Figure courtesy of J.H. Metzen.}
\label{hetero_figure}
\end{figure}

Heteroscedasticity refers to learning intrinsic scatter associated with the noisy variable, $y$, and introducing non-constant variances. The full covariance function applied in the GP can be broken down into an adaptive kernel and noisy variance component:

\begin{equation}
    \Sigma({\bf x}_i,{\bf x}_j) = \underbrace{k({\bf x}_i,{\bf x}_j)}_{\text{adaptive}} + \underbrace{\sigma^2_n({\bf x}_i) \delta_{ij}}_{\text{heteroscedastic}}
\end{equation}
and heteroscedasticity is mathematically defined by $\sigma_n({\bf x}_i)$ having an explicit dependence on points in the input space. To contrast, homoscedasticity would entail a globally constant $\sigma_n$. To more vividly demonstrate the benefit of heteroscedasticity, a 1-dimensional example is displayed in Figure \ref{hetero_figure} by applying both homoscedastic and heteroscedastic components. The function to be learned is a linear relationship with variance growing quadratically along the abscissa. In this scenario with a homoscedastic noise model, the GP is forced to learn a constant intrinsic scatter in the underlying data commonly represented by white noise functions of constant amplitude. It is evident that enabling a heteroscedastic covariance function better captures the distribution of observed data. The mean estimates of both models are equivalent, but the predicted variances across the domain are markedly different and the heteroscedastic example obtains a larger LML and better captures intrinsic scatter in the data. The non-constant variances are learned across the domain using a k-means algorithm \cite{LIKAS2003451} to once again identify clustering centers to provide characteristic noise estimates even when explicit error bars are absent. These prototype values are then extended across the domain by using a pairwise RBF kernel. Both the adaptive length scale kernel and heteroscedastic components are combined to significantly improve overall fitting and stability of the numerical optimization. For example, they avoid zero variances while training. This heteroscedastic term in the full data-driven covariance function is modular to an extent and can be optionally subtracted away to output confidence intervals ($\mathbb{V}[f_*]$) instead of prediction intervals ($\mathbb{V}[y_*]$) which are wider and account for intrinsic scatter in observations. (Note that $\mathbb{E}[f_*] \equiv \mathbb{E}[y_*]$.) 

\subsection{Multidimensional}

GPs do not require a fixed discretization scheme over the physical domain and hence easily handle spatially and temporally scattered diagnostic measurements involved in nonlinear regression. The generalized kernel above is encoded to handle data spanning any number of dimensions. Due to the highly non-convex optimization problem associated with finding optimal hyperparameters across the multidimensional space, the training applies stochastic optimization (Storn algorithm \cite{Storn_alg}) with gradient descent at the end to help avoid becoming trapped in local minima. An associated error checking routine has been developed and available in the code on GitHub to automatically identify regions of the domain, if any, over which the GP did not successfully converge during training and requiring further optimization. Generally, multidimensional sampling of the GP is performed by applying 
\begin{equation}
    f_* = \mu_* + B\mathcal{N}(0,I)
\end{equation}
where $B B^T = \Sigma_*$ and $B$ is a triangular matrix known as the Cholesky decomposition of the covariance matrix. To account for constraints such as monotonicity or positivity with respect to the mean in sampled profiles, modified or truncated normal distributions can be enabled in the code to permit constrained GP sampling to yield physically relevant results. Finally, the technique's flexibility allows it to automatically handle vast data sets (e.g. thousands of discharges fitted in parallel) to efficiently construct large multidimensional profile databases with minimal user input.

\begin{figure}
\begin{center}
    \includegraphics[width=0.975\linewidth]{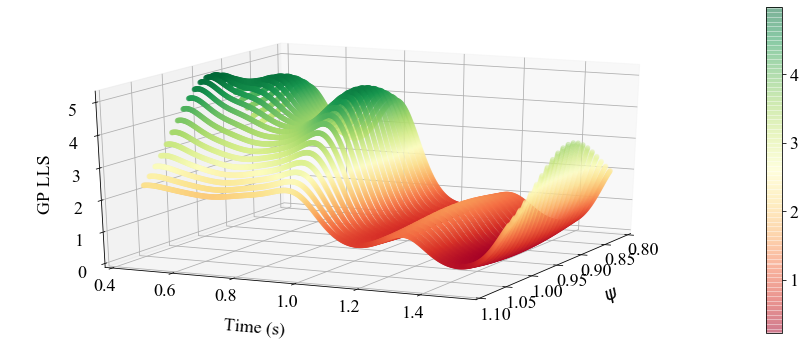}  
\caption{Corresponding length scales, $\rho$, learned across the spatial and temporal domain based on training of the adaptive heteroscedastic GP.} 
\label{lls_n_2d}
\end{center}
\end{figure} 

\section{Application to experimental data}
The aforementioned adaptive heteroscedastic GP is now directly applied to experimental data originating from the Thomson scattering diagnostic on Alcator C-Mod which consists of two Nd:YAG lasers, each pulsing at 30 Hz. The measurements have an approximate spatial resolution of 1.0 cm and 1.3 mm in the core and edge, respectively \cite{JWHughes-TS-diagnostic}. It is noted that the GP itself is a continuous regression model and can provide fittings at arbitrary spatial and temporal resolution. Discharge number 1091016033 is analyzed which exhibits L-, H-, and I-mode behaviours within this single discharge as detailed in \cite{Whyte_2010}. Ion cyclotron range of frequencies (ICRF) heating of up to 5 MW is applied in the experiment with power primarily deposited in the core on the hydrogen minority species. The on-axis magnetic field is 5.6 T with a plasma current of approximately 1.2 MA. There is a wide range of plasma behaviour associated with time-varying density and temperature pedestal structure even in this single discharge including transport barrier formation and confinement mode transitions necessitating a suitably robust regression method. The tools outlined above have been demonstrated on discharge 1091016033, and can be easily automated for edge data analyses across any set of plasma discharges on Alcator C-Mod, or on other tokamaks with sufficiently resolved kinetic profiles.

\begin{figure}
\centering
\includegraphics[width=0.48\linewidth]{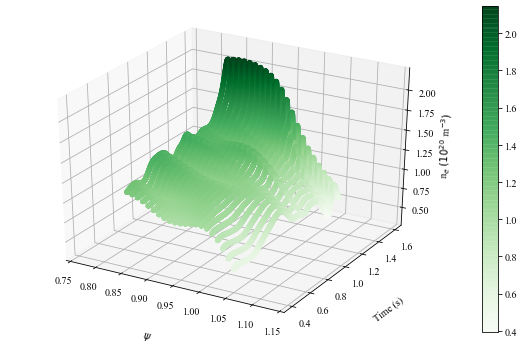} 
\includegraphics[width=0.48\linewidth]{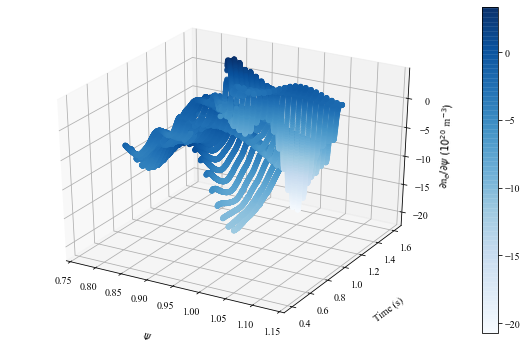}
\caption{Electron density and corresponding spatial gradients produced by the adaptive heteroscedastic GP. This technique accounts for both spatial and temporal evolution of experimental data across the entire discharge over the edge-pedestal region (i.e. $0.85 < \psi < 1.05$, which covers data on both open and closed field lines).}  
\label{2D-GPR}
\end{figure} 

\begin{center}
\begin{figure*}  
\includegraphics[width=0.325\linewidth]{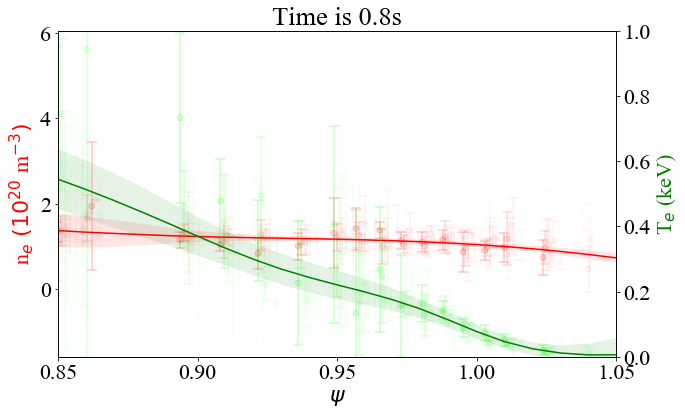} 
\includegraphics[width=0.325\linewidth]{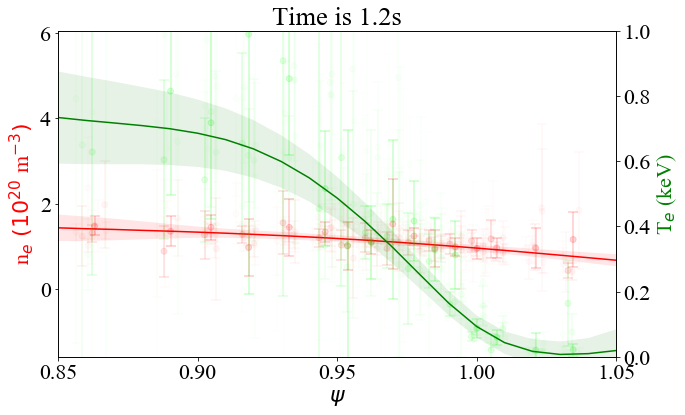}
\includegraphics[width=0.325\linewidth]{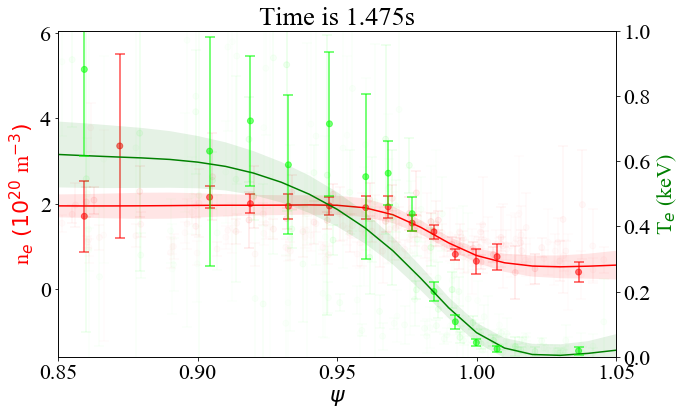}
\\ 
\includegraphics[width=0.325\linewidth]{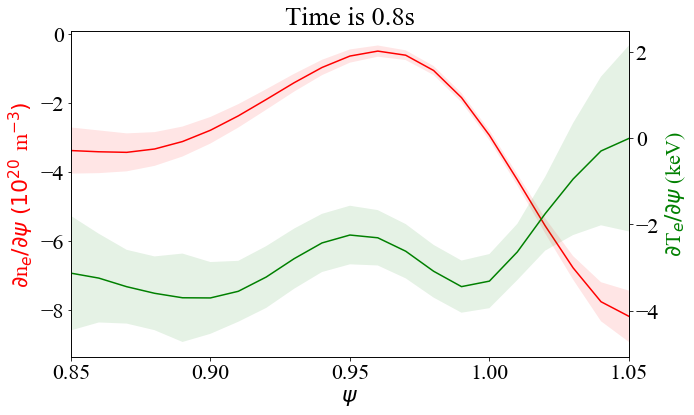} 
\includegraphics[width=0.325\linewidth]{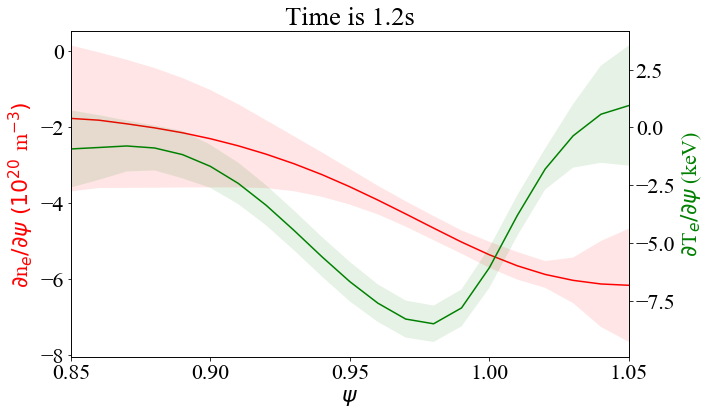} 
\includegraphics[width=0.325\linewidth]{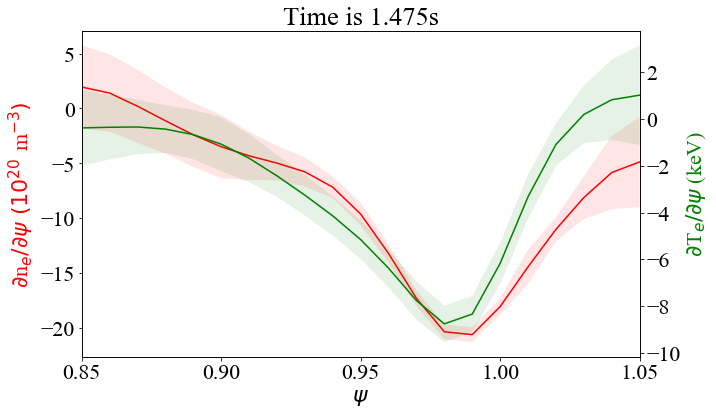}
\caption{Electron density and temperature measurements during  L- (left), I- (middle), and H-modes (right) fitted by the adaptive heteroscedastic GP without time-averaging experimental data. Proximity of experimental data to the time at which the GP is predicting is indicated by transparency of the data points. For simple demonstrative purposes, 95\% prediction intervals are displayed in the top three plots while 95\% confidence intervals are applied for the bottom.}
\label{1D-GPR}
\end{figure*}  
\end{center}

To begin, the trained GPs can compute expected mean density and temperature along with corresponding uncertainties across the entire spatiotemporal domain covered by the plasma diagnostic. Spatial gradients (or time derivatives) can be directly produced and are displayed in Figure \ref{2D-GPR} for electron density across the grid as well. The key advantage of applying the adaptive GP for edge profile fitting is its ability to learn spatiotemporal correlations across the entire domain without any temporal averaging nor spatial filtering of data. This freedom in training is evident in the learned variable length scales for density across the discharge as visualized in Figure \ref{lls_n_2d}. Particular time slices can also be evaluated by the GPs. Fitted L-, I-, and H-mode profiles from discharge 1091016033 are displayed in Figure \ref{1D-GPR} without employing any time-averaging or filtering of experimental data as required when repeatedly applying a modified tanh function, which can miss important profile variation even within a confinement regime (e.g. while ramping up auxiliary heating). Density and temperature gradients along with uncertainties for all quantities are available across the experiment during L-, H-, and I-mode phases at times of 800, 1200, and 1475 milliseconds, respectively \cite{Whyte_2010}. 

A major benefit derived from applying the adaptive heteroscedastic GP is its ability to provide defining features (e.g. spatial and temporal gradients) of experimental profiles automatically across entire discharges. In past classification exercises to develop large confinement regime databases \cite{Mathews_2019}, individual time slices needed to be manually reviewed for the presence of density and/or temperature pedestals to identify windows containing different regimes (e.g. L-, H-, or I-modes). This is a highly time-intensive and arguably subjective route to develop meaningful confinement regime databases since discretely classifying plasma behaviour can miss underlying nuances (e.g. staircase pedestals, profile hollowing). Applying this multidimensional GP regression method can automate characterization of discharges based upon quantitative profile characteristics such as gradient scale lengths. It can handle variation across the edge-pedestal region which connects the high temperature core with the colder SOL upon crossing the separatrix. Different varieties of stationary confinement regimes may have quite different profile dynamics and structure for temperature and density necessitating adaptive fitting capability. Additionally, outputting the time-dependent evolution of plasma profiles and gradients with corresponding uncertainties helps capture subtleties in edge structures which are essential to better understand how confinement regimes across the entire discharge. Resolving these features helps improve systematic reconstructions of experimental measurements for further large scale analysis, for example in running stability codes or scientific machine learning, or in empirical database studies, e.g. characterizing upstream conditions of background plasmas for divertor heat flux studies \cite{Silvagni_2020}. For example, the multidimensional GP can provide key profile information into plasma simulations (e.g. inputs for global gyrokinetic codes or comparisons with EPED \cite{Snyder_2009}) which may require sampling inputs such as gradient profiles to output sufficient statistics. Additionally, the denoised equilibrium plasma dynamics can be useful observational constraints in scientific learning applications such as in Chapter 4.

Overall, the outlined adaptive multidimensional GP regression routine can automate fitting and uncertainty estimation of edge-pedestal measurements from the Thomson scattering diagnostic on the Alcator C-Mod tokamak while being robust to edge-pedestal phenomena. Spatiotemporal evolution of plasma density and temperature is tracked with varying length scales extant in experimental data including the formation of both particle and energy transport barriers. Structure imposed on learned edge gradient profiles is minimized by using a data-driven kernel function which can be critical to model plasmas near sensitive instability boundaries. The application is focused on edge measurements of tokamak plasmas with relevance to numerical analysis of the pedestal, but these techniques extend beyond analysis of the edge-pedestal region and can be suitably adapted to novel scenarios exhibiting singular transient events. The GP introduced provides an automated tool to tackle nonlinear multidimensional regression tasks and helps resolve measurements of equilibrium profiles with dynamics spanning a wide range of physical scales. 

\chapter{Code and data availability}
\section*{Chapter 2}
\noindent All relevant data files and codes for constructing the deep networks can be found on Github at: \url{https://github.com/AbhilashMathews/PlasmaPINNs}.
\section*{Chapter 3}
\noindent All relevant data files and codes can be found on Github at \url{https://github.com/AbhilashMathews/PlasmaPINNtheory} and \url{https://github.com/ammarhakim/gky\\l-paper-inp/tree/master/2021_PoP_GK_Fluid_PINN}. Full documentation on the gyrokinetic simulation framework with the \texttt{Gkeyll} code is located at \url{https://gkeyll.readthedocs.io/en/latest/}.
\section*{Chapter 4} 
The codes applied to analyze experimental GPI data are available on the MIT Engaging cluster at the following directory: \texttt{/home/mathewsa/PINNs/GPI}.

\noindent The scripts there for training the deep learning framework realizations are: 

\noindent \texttt{GPI\_test\_HeI\_DEGAS2\_probe\_time\_2012\_best\_const\_priming\_rep\_v0.py} -- \\ \texttt{GPI\_test\_HeI\_DEGAS2\_probe\_time\_2012\_best\_const\_priming\_rep\_v799.py}. 

\noindent After training is complete, one can utilize the following scripts for plotting: \\ \texttt{GPI\_theory\_2D\_paper\_best\_prime\_1120711021\_probe\_time\_interval.py} and \\ \texttt{GPIpaper\_bestprime\_figures\_plot+save.py} (running certain functions in \\these codes will require access to the MFE workstations located at the PSFC).
\\ \\
\noindent The HeI collisional radiative code \cite{GOTO_2003,Zholobenko_thesis} is written in C and utilizes the GNU Scientific Library (GSL) for numerical calculations. It is located on Engaging at: \\ \texttt{/net/eofe-data005/psfclab001/mathewsa/hecrmodel\_WZmodified\_NAG\_compatible}

\noindent To compute all population and rate coefficients, users should prepare their own main function according to their purposes. Example makefiles (e.g. \texttt{Makefile\_PEC\_loop}) are included in the directory for this purpose. The energy levels are numbered in the code with $1^1$S, $2^1$S, $2^3$S, $2^1$P, $2^3$P, $3^1$S, $3^3$S, ... are indexed as 0, 1, 2, 3, 4, 5, 6, ..., respectively. Consequently, all singlet levels except $1^1$S have odd numbers and all triplet levels have even numbers. Labels like \texttt{S3P}, which means ``singlet 3P'', can be used to designate energy levels instead of the numbers. The correspondence between the numbers and the labels are defined in \texttt{hecrm.h}. The input parameters are the magnetic field strength (T) which is used for the singlet-triplet wavefunction mixing calculations, the electron temperature (eV) and the electron density (cm$^{-3}$). They are set in the \texttt{crmodel.prm structure}. See \texttt{hecrm.h} and \texttt{run.c}. The results are stored in the three structures, i.e., rate coefficient, population coefficient and \texttt{cr\_rate\_coefficient}, after calling the \texttt{hecrmodel} function. See \texttt{hecrm.h} for details.

\noindent The population of a level $p$, $n(p)$ is expressed with the population coefficients as
\begin{equation}
n(p) = r_0(p)n_en_i + r_1(p)n_en(1^1S) + r_2(p)n_en(2^1S) + r_3(p)n_en(2^3S)
\end{equation}

\noindent in Formulation I, and as
\begin{equation}
n(p) = R_0(p)n_en_i + R_1(p)n_en(1^1S)
\end{equation}

\noindent in Formulation II, which is the one utilized in this thesis. The population coefficients in Formulation I and in Formulation II are stored respectively in the vectors \texttt{r0}, \texttt{r1}, \texttt{r2}, and \texttt{r3} and in the vectors \texttt{rr0} and \texttt{rr1} of the population coefficient structure. When the population coefficient structure is declared as \texttt{popcoe} like
in the sample code \texttt{PEC\_run\_loop\_dens.c}, $R_1(3^3D)$ is found in \texttt{popcoe[i].rr1[T3D]}, where \texttt{i} is the index for the \texttt{ne} array. The spontaneous transition probabilities, i.e. Einstein A coefficients, are given in the matrix \texttt{a} of the rate coefficient structure. When the structure is declared as \texttt{rate}, the Einstein A coefficient for the transition $3^3D \rightarrow 2^3P$, for example, is taken as \texttt{gsl\_matrix\_get(rate.a, T3D, T2P)} or \texttt{MG(rate.a,
T3D, T2P)}. Here, \texttt{MG} is the short form of \texttt{gsl\_matrix\_get} as defined in \texttt{hecrm.h}. Additional coefficients calculated in the code are stored in \texttt{cr\_rate\_coefficient} \cite{Fujimoto1979ACM,GOTO_2003}\footnote{Note: the electron temperature for the results in Fig. 7 of \cite{GOTO_2003} is 1 eV, not 10 eV.}.
\\ \\
\noindent Files for computing Greenland's criteria and their outputs including $\tau_Q$ are located in the directory:
\texttt{/net/eofe-data005/psfclab001/mathewsa/hecrmodel\_WZmodified\_\\NAG\_compatible/NAG/nll6i271bl/hecrmodel\_WZmodified}. Note that running these codes require usage of the commercial package \texttt{NAG Library (Mark 27.1)}. 

\section*{Chapter 5}
The scripts for training the deep learning framework realizations are:\\ \texttt{CMod\_ES\_DRB\_exp\_Er\_1120711021bestprime\_no\_sources\_rep\_v0.py} --\\ \texttt{CMod\_ES\_DRB\_exp\_Er\_1120711021bestprime\_no\_sources\_rep\_v99.py} and\\
\texttt{CMod\_ES\_DRB\_exp\_Er\_1120711021bestprime\_fix\_sources\_rep\_v0.py} --\\ \texttt{CMod\_ES\_DRB\_exp\_Er\_1120711021bestprime\_fix\_sources\_rep\_v99.py} with\\
\texttt{plotting\_final\_paper+chapter\_plots.py} for plotting.

\section*{Appendix B} 
All relevant data files and codes can be found on Github at \url{https://github.com/AbhilashMathews/gp_extras_applications}, which are based upon the \texttt{gp\_extras} package by J.H. Metzen at \url{https://github.com/jmetzen/gp\_extras}.



\begin{singlespace}
\bibliography{main}
\bibliographystyle{plain}
\end{singlespace}

\end{document}